\documentclass{aastex631}

\newcommand{\e}[1]{\ensuremath{\times 10^{#1}}}
\newcommand{\um}{$\mu$m }

\received{2021 Dec 30}
\revised{2022 Apr 20}
\accepted{2022 Apr 21}
\submitjournal{PSJ}

\shorttitle{Water Ice On Miranda}
\shortauthors{DeColibus et al.}

\graphicspath{{./}{figures/}}

\begin{document}

\title{Longitudinal Variation of H$_2$O Ice Absorption on Miranda}

\correspondingauthor{Riley DeColibus}
\email{decolib@nmsu.edu}

\author[0000-0002-1647-2358]{Riley A. DeColibus}
\affiliation{Department of Astronomy, New Mexico State University \\
Box 30001, Dept.4500 \\
Las Cruces, NM 88003, USA}

\author[0000-0002-9984-4670]{Nancy J. Chanover}
\affiliation{Department of Astronomy, New Mexico State University \\
Box 30001, Dept.4500 \\
Las Cruces, NM 88003, USA}

\author[0000-0002-6886-6009]{Richard J. Cartwright}
\altaffiliation{Visiting Astronomer at the Infrared Telescope Facility, which is operated by the University of Hawaii under contract 80HQTR19D0030 with the National Aeronautics and Space Administration.}
\affiliation{The Carl Sagan Center at the SETI Institute\\
189 Bernardo Ave., Suite 200\\
Mountain View, CA 94043, USA}

\begin{abstract}

Many tidally locked icy satellites in the outer Solar System show leading/trailing hemispherical asymmetries in the strength of near-infrared (NIR) H$_2$O ice absorption bands, in which the absorption bands are stronger on the leading hemisphere. This is often attributed to a combination of magnetospheric irradiation effects and impact gardening, which can modify grain size, expose fresh ice, and produce dark contaminating compounds that reduce the strength of absorption features. Previous research identified this leading/trailing asymmetry on the four largest classical Uranian satellites but did not find a clear leading/trailing asymmetry on Miranda, the smallest and innermost classical moon. We undertook an extensive observational campaign to investigate variations of the NIR spectral signature of H$_2$O ice with longitude on Miranda's northern hemisphere. We acquired 22 new spectra with the TripleSpec spectrograph on the ARC 3.5m telescope and 4 new spectra with GNIRS on Gemini North. Our analysis also includes 3 unpublished and 7 previously published spectra taken with SpeX on the 3m IRTF. We confirm that Miranda has no substantial leading/trailing hemispherical asymmetry in the strength of its H$_2$O ice absorption features. We additionally find evidence for an anti-Uranus/sub-Uranus asymmetry in the strength of the 1.5-\um H$_2$O ice band that is not seen on the other Uranian satellites, suggesting that additional endogenic or exogenic processes influence the longitudinal distribution of H$_2$O ice band strengths on Miranda.
\end{abstract}


\section{Introduction} \label{sec:intro}

The five classical Uranian satellites -- Miranda, Ariel, Umbriel, Titania, and Oberon -- represent a distinctly different population of icy bodies compared to the moons of Jupiter and Saturn. They are smaller than the nearly planet-sized Jovian moons and colder than the Saturnian satellites. While the Uranian moons are subject to significant seasonal variations in surface temperature due to the extreme axial tilt of Uranus ($\sim$98\degr), they are warmer than Triton and the distant trans-Neptunian objects (TNOs). Only one spacecraft has visited the Uranian system, with the \textit{Voyager 2} flyby in January 1986 \citep{Stone1986,Smith1986}. During its short visit, \textit{Voyager} observed the southern hemispheres of the Uranian moons (sub-solar latitude $\sim$82$\degr$S) and discovered a wealth of geologic features on the Uranian moons, as well as strong evidence for geologically recent activity on Miranda ($<$1 Ga) and Ariel (1 -- 2 Ga) \citep[e.g.][]{Smith1986,Greenberg1991,Schenk1991,Zahnle2003,Kirchoff2022}. For reference, we include the physical, orbital, and photometric properties of the five classical Uranian satellites in Table \ref{tab:MoonProperties}.

The extent of our knowledge of the Uranian moons is largely comparable to that of the Saturnian satellites prior to the arrival of the \textit{Cassini-Huygens} mission. The \textit{Voyager} spacecraft did not carry a near-IR mapping spectrometer, therefore little information on the surface compositions of the Uranian moons was obtained during the \textit{Voyager 2} flyby. With no other spacecraft missions to the Uranian system currently planned or in development, our knowledge of their surface compositions is therefore limited to disk-integrated spectra obtained from ground-based and space-based telescopes. The surface compositions of airless Solar System bodies like the Uranian moons are often studied with near-infrared (near-IR) spectroscopy of reflected sunlight. This region of the electromagnetic spectrum contains several diagnostic absorption features for H$_2$O ice-rich bodies, including broad absorption bands centered near 1.52, 1.65, and 2.02 \um \citep{Mastrapa2008}. 

Miranda, the smallest and innermost classical moon, has an apparent visual magnitude of $m_V \sim 16.5$. This makes Miranda difficult to observe from Earth due to its proximity to Uranus, and previous near-IR studies have only achieved limited longitudinal coverage in comparison to the other Uranian satellites, hindering systematic studies of Miranda's surface composition. The other Uranian moons show a leading/trailing asymmetry in the strength of near-IR H$_2$O ice bands, but Miranda does not seem to follow the same spatial trends in surface composition that are seen on the other Uranian satellites \citep{Cart2018,Cart2020IRAC}. The apparent discrepancy between expectations and observations motivated an observational campaign to study the spectral properties of H$_2$O ice as a function of sub-observer longitude on Miranda. We acquired 24 new near-IR spectra of Miranda's northern hemisphere (sub-solar latitudes 40--55\degr N). The northern hemispheres of the Uranian satellites were not observed by \textit{Voyager 2}, as these latitudes were in winter darkness at the time of the \textit{Voyager} flyby.

In the following subsections we describe the state of knowledge for the surface composition of Miranda and the other classical Uranian moons. We discuss our observations of Miranda in Section \ref{sec:obs} and our analysis of integrated band areas and depths of the H$_2$O absorption bands in Section \ref{sec:analysis}. We summarize our results in Section \ref{sec:results}, and discuss the implications of our findings in Section \ref{sec:discussion}.

\begin{deluxetable*}{lllcllllcll}
\tablecaption{Properties of the classical Uranian satellites\label{tab:MoonProperties}}
\tablehead{
\colhead{Moon} & \colhead{Radius} & \colhead{Mass} & \colhead{$\rho$} &
\colhead{$P_{orb}$} & \colhead{$a$} & \colhead{$i$} & \colhead{$e$} & \colhead{$m_{K,opp}$} & \colhead{$A_{geom,K}$} & \colhead{$A_{Bond}$}\\
\colhead{} & \colhead{km} & \colhead{kg} & \colhead{g cm$^{-3}$} & \colhead{days} &
\colhead{km} & \colhead{\degr} & \colhead{} & \colhead{mag} & \colhead{} & \colhead{}
}
\startdata
Miranda & $\sim$236 & 6.590\e{19} & 1.20 & 1.413 & 129,800 & 4.22 & 0.0027 & 14.6 & 0.25$\pm$0.03 & 0.20$\pm$0.03 \\
Ariel & $\sim$579 & 1.353\e{21} & 1.67 & 2.520 & 191,200 & 0.31 & 0.0034 & 12.4 & 0.31$\pm$0.025 & 0.23$\pm$0.025 \\
Umbriel & 584.7 & 1.172\e{21} & 1.40 & 4.144 & 266,000 & 0.36 & 0.0050 & 13.1 & 0.20$\pm$0.01 & 0.10$\pm$0.01 \\
Titania & 788.4 & 3.527\e{21} & 1.71 & 8.706 & 435,800 & 0.10 & 0.0022 & 12.0 & 0.23$\pm$0.015 & 0.17$\pm$0.015 \\
Oberon & 761.4 & 3.014\e{21} & 1.63 & 13.463 & 582,600 & 0.10 & 0.0008 & 12.2 & 0.22$\pm$0.015 & 0.14$\pm$0.015 \\
\enddata
\tablecomments{\footnotesize Physical, orbital, and photometric properties of the classical Uranian satellites. The radii listed are the average value for each moon. Miranda's radius is 240.4$\times$234.2$\times$232.9 km, while Ariel's radius is 581.1$\times$577.9$\times$577.7 km. $\rho$ is the bulk density of the moon, $P_{orb}$ is the orbital period, $a$ is the orbital semimajor axis, $i$ is the inclination, $e$ is the eccentricity, $m_{K,opp}$ is the K-band apparent magnitude at opposition, $A_{geom,K}$ is the geometric albedo in K-band, and $A_{Bond}$ is the bolometric Bond albedo. All five moons are tidally locked with Uranus, so their orbital periods are equivalent to their rotational periods. Physical and orbital values are from the JPL Horizons database, while photometric properties are derived from \citet{Kark2001}. }
\end{deluxetable*}

\subsection{Surface Composition}\label{ssec:surfcomp}
The first near-IR spectrophotometric studies of the classical Uranian satellites determined that their surface compositions are dominated by H$_2$O ice \citep{Cruikshank1980,CruikshankBrown1981,BrownCruikshank1983,BrownClark1984}. However, the Uranian satellites have lower albedos and weaker near-IR H$_2$O absorption bands compared to similarly sized H$_2$O-rich Saturnian moons. The surfaces of the Uranian moons could be mixed or coated with a `dirty' but spectrally neutral component to account for this darkening and weakening of the H$_2$O ice bands. This component is still unidentified, but it is likely to be carbonaceous or silicate-rich in nature \citep{Brown1983,ClarkLucey1984}. Previous studies have modeled this component as amorphous carbon \citep{Bauer2002,Cart2015,Cart2018} or a synthetic spectrally neutral absorber \citep{Grundy1999}, while \citet{Sharkey2021} found that the Neptunian satellite Nereid had similar spectral properties to the Uranian satellites, and that the dark absorber could be modeled with magnetite. As with many bodies in the outer Solar System, near-IR spectra of the Uranian satellites show strong evidence for the 1.65-\um absorption feature that is associated with low-temperature H$_2$O ice in a crystalline phase \citep{Grundy1999,Bauer2002}. 

The trailing hemispheres of Ariel, Umbriel, Titania, and Oberon also show evidence for three narrow absorption features at 1.966, 2.012, and 2.070 \micron, which have been attributed to ``pure'' CO$_2$ ice deposits \citep{Grundy2003,Grundy2006,Cart2015}. The spatial distribution of this CO$_2$ ice appears to be controlled by exogenic processes (Section \ref{ssec:trends}). CO$_2$ ice has not been detected on Miranda \citep{Gourgeot2014,Cart2018,Cart2020IRAC}. 

Finally, the presence of a weak absorption feature near 2.2 \um on Miranda has been attributed to ammonia (NH$_3$)-bearing species, hinting at geologically recent exposure, as these compounds are not expected to survive over geological timescales \citep{Bauer2002,Moore2007}. This feature appears qualitatively similar to the well-studied 2.21-\um feature on Charon and other bodies in the Pluto system \citep[e.g.][]{BrownCalvin2000,Holler2017,Cook2007,Cook2018}. The detection of a 2.2-\um band on Miranda has neither been confirmed nor ruled out by subsequent studies \citep{Gourgeot2014,Cart2018}. However, low level spectral features in the 2.2 \um region on the other Uranian moons hint that this feature may be present across the Uranian satellite system \citep{Cart2018}. On Ariel, these absorption features have been attributed to NH$_3$- and NH$_4$-bearing species \citep{Cart2020Ariel}, but the exact compound(s) responsible have not been determined.

The photometric colors of the Uranian moons are mostly neutral, with the exception of minor hemisphere-dependent reddening \citep{Buratti1991,Cart2018}. The smaller, outer irregular moons of Uranus are much redder, and icy bodies farther from the Sun, such as Centaurs and TNOs, often have significantly reddened surfaces attributed to modification of carbon compounds by irradiation \citep[e.g.][]{Sagan1979,Thompson1987,Barucci2008,Bennett2013,DalleOre2015}. However, the colors of the giant planet irregular satellites are not as red as many TNOs, so additional processes may have an influence on their surface coloration \citep{Graykowski2018}. The Uranian satellites also show strong, narrow opposition surge effects, with sharp increases in brightness near zero phase angle \citep{Kark2001}. Polarimetric studies have identified negative-branch polarization characteristic of highly porous, `fluffy' surfaces with an abundance of $<$1 \um grains, which could be more analogous to mid-sized, H$_2$O-ice rich TNOs than the icy moons of Jupiter and Saturn \citep{Afanasiev2014}.  The surface temperatures of the Uranian moons are thought to average 70--85 K, depending on albedo, and are subject to significant seasonal and latitudinal variation \citep{Hanel1986,Grundy1999,Grundy2006}, although thermal modeling has suggested that the temperature range may be as wide as 30--90 K \citep{Sori2017}. Although it is difficult to directly compare the evolutionary history of icy satellites and TNOs, the major Uranian satellites are at what has been described as a ``heliocentric crossroads'' between the warmer Saturnian system and the more distant and colder TNOs \citep{Cart2018}. Furthermore, the Uranian satellites surface compositions are dominated by H$_2$O ice mixed with a dark, spectrally neutral absorber, similar to Charon, Orcus, and other mid-sized, H$_2$O-bearing TNOs. Volatile species such as the CO, CH$_4$, and N$_2$ ices seen on Triton and Pluto \citep{Cruikshank1993,Owen1993} are not stable on the Uranian satellites given their sizes and heliocentric distance, and the Uranian satellites lack the ultra-red coloration of some TNOs (e.g. the `RR-type' TNOs of \citet{Barucci2005}). With their ``crossroads'' of characteristics between warmer icy satellites and TNOs, an improved understanding of the characteristics of H$_2$O ice in the Uranian system can therefore enhance our understanding of icy bodies throughout the outer Solar System.

\subsection{Compositional Trends on the Uranian Satellites\label{ssec:trends}}
\begin{figure}[ht!]
\centering
\makebox[\textwidth][c]{\includegraphics[width=0.8\textwidth]{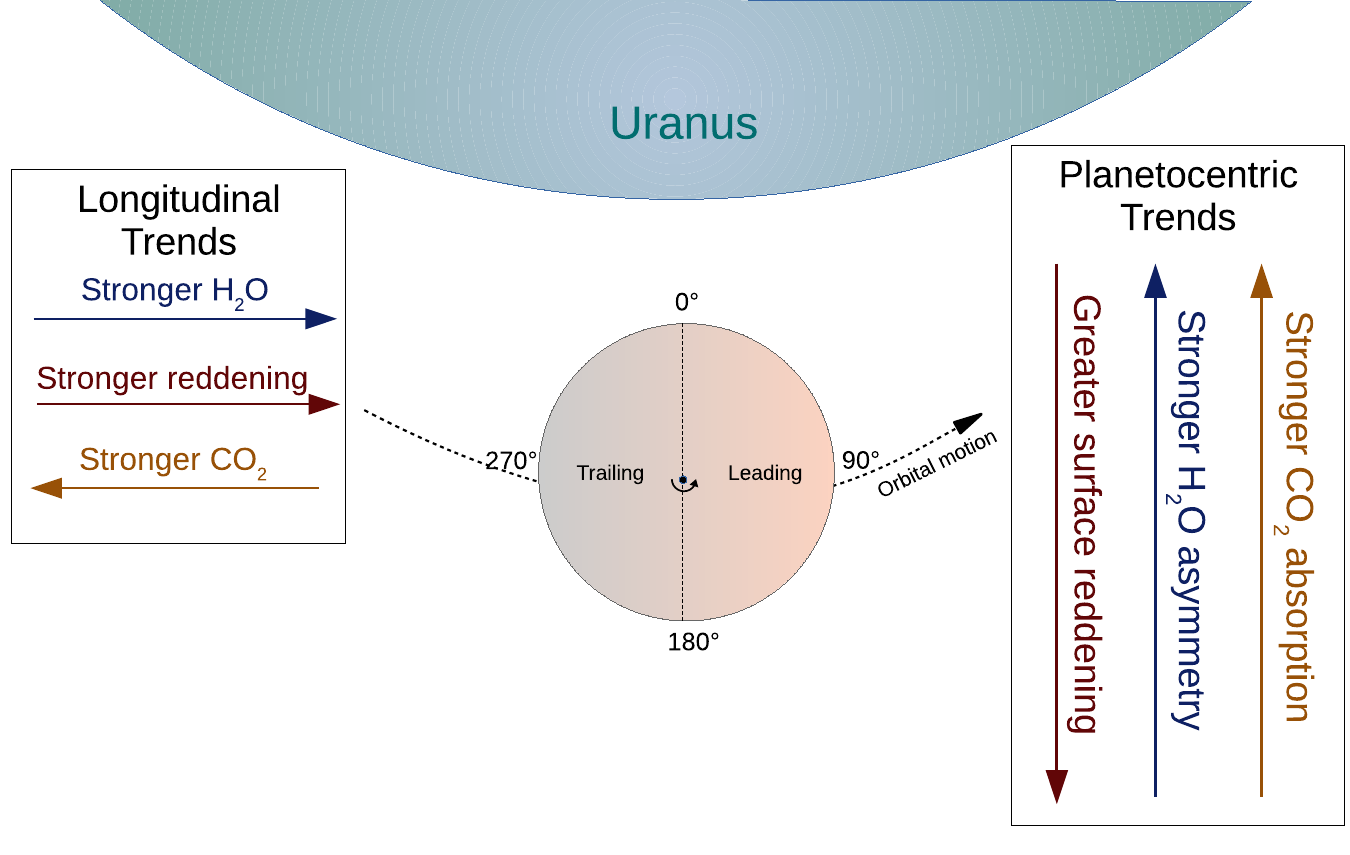}}
\vspace{-15pt}
\caption{\footnotesize A diagram summarizing the previously observed planetocentric and longitudinal trends in the surface compositions of the classical Uranian moons, excluding Miranda \citep{Buratti1991,Grundy2003,Grundy2006,Cart2015,Cart2018}.}
\vspace{-5pt}
\label{fig:trends}
\end{figure}

In systems of natural satellites around giant planets like Uranus, multiple processes can modify the surface composition of the moons, including solar UV irradiation and magnetospheric particle irradiation, sublimation of volatiles, dust impacts and regolith overturn, and endogenic geologic activity \citep[e.g.][and references therein]{ClarkBrown1984,Lanz1987,Cassidy2010,Bennett2013,Cart2015}. The surface compositions of the Uranian satellites show spatial trends, varying with longitude on the satellites and with distance from Uranus. These trends in surface composition have been identified by systematic studies with near-IR reflectance spectroscopy (Figure \ref{fig:trends}).

For example, the flux (and kinetic energy) of heliocentric impactors should be highest close to Uranus due to the planet's gravity \citep{Zahnle2001,Zahnle2003}. The leading hemispheres of the orbiting moons are more likely to experience impacts from this heliocentric population, in much the same way that a moving vehicle receives more raindrops on the front windshield compared to the back window.
Large and small (micrometeorite) impacts could overturn the regolith and expose fresh H$_2$O ice, thereby enhancing H$_2$O ice bands on the leading hemispheres of tidally-locked moons. However, heliocentric impacts could also break down the regolith into smaller ice grains, which would result in shallower absorption bands \citep{Clark1983}. The larger population of heliocentric impactors near the planet, moving at faster relative velocities, implies that the asymmetry in impactor flux between the leading and trailing hemispheres should also be larger closer to the planet \citep{Zahnle2001,Zahnle2003,Grundy2006,Bennett2013}. 

In addition to heliocentric impacts, the trailing hemispheres of tidally-locked satellites tend to receive higher levels of irradiation from charged particles embedded in the host planet's magnetosphere. The giant planets and their magnetic field lines rotate more quickly than the orbital velocities of their satellites. The magnetic field lines and their embedded charged particles should thus `sweep' over the trailing hemispheres of the satellites, leading to a higher radiation dose on the trailing hemisphere compared to the leading hemisphere. This simplified model is incomplete, however, as a planetary magnetosphere is populated with a variety of electrons, protons, and heavy ions of different species and different energies. These can take differing trajectories along magnetospheric field lines, bombard differing regions of satellite surfaces, penetrate to different depths, and have differing efficiencies in modifying surface chemistry \citep[e.g.][]{Lanz1987,Cassidy2013,Hendrix2018}. Uranus' magnetic field is also offset from its rotation axis \citep{Ness1986}, and moon-magnetosphere interactions may be more complicated in the Uranian system. 

This charged particle irradiation can interact with surface composition in complex ways. It should cause preferential sputtering and removal of small H$_2$O ice grains, leaving larger grains behind and strengthening absorption bands. However, radiolysis of an icy satellite's surface can produce dark contaminants and modify the lattice structure of crystalline H$_2$O ice, converting it into amorphous H$_2$O ice \citep[e.g.][]{Kouchi1990,Loeffler2020}. This can instead reduce the strength of H$_2$O ice absorption bands, especially the 1.65-\um feature strongly associated with crystalline H$_2$O ice \citep{Mastrapa2008,Mastrapa2013}. The net result of magnetospheric bombardment could therefore enhance leading/trailing asymmetries in composition. This asymmetry between leading and trailing hemispheres is likely strongest closer to the planet, where plasma densities are usually higher, and decreases with planetocentric distance.

In the Jovian system, Europa, Ganymede, and Callisto tend to show hemispherical asymmetries in the strengths of H$_2$O ice absorption, where the 1.5-\um and 2.0-\um bands are less deep on their trailing hemispheres. The 1.04-\um band is stronger on the trailing hemispheres of all three moons, from which we can infer larger H$_2$O ice grain size \citep{ClarkBrown1984,Calvin1995}. The apparent asymmetry is stronger on Europa and Ganymede than it is on Callisto, but the presence of dark contaminants on Callisto, Europa's young surface, and Ganymede's intrinsic magnetic field complicate this simple picture. Nonetheless, these spatial trends are consistent with a combination of the magnetospheric bombardment and micrometeorite hypotheses \citep{Wolff1983,ClarkBrown1984,Calvin1995,DelitskyLane1998,Cassidy2013}. Thermal sintering of smaller grains into larger ones on geologically short timescales is likely to also play a role in controlling grain size in the Galilean system, but this process should be less effective at Saturn and negligible at the surface temperatures of the moons of Uranus \citep{Clark1983}. 

For the mid-sized icy moons in the Saturnian system (Mimas, Enceladus, Tethys, Dione, and Rhea), the leading/trailing asymmetries in H$_2$O ice absorption features are similar to those on the Jovian icy satellites. The 1.5-\um and 2.0-\um bands appear to be less deep on the trailing hemispheres of all five of these moons \citep{ClarkBrown1984,Emery2005,Verb2006}. \citet{Emery2005} found that the leading hemispheres of Dione, Tethys, and Rhea have similar grain sizes as the trailing hemispheres, but a lower abundance of dark contaminants, consistent with the magnetospheric interactions hypothesis \citep{ClarkBrown1984,Grundy1999,DelitskyLane2002,Emery2005,Verb2006,Hendrix2018}. Mimas appears to be an outlier, as it seems to have larger grains and more contaminants on the leading hemisphere, but this may be due to preferential mantling of the trailing hemisphere by material from Saturn's E-ring \citep{Fila2012,Hendrix2018}. Enceladus also shows slightly deeper absorption bands on its leading hemisphere, but plume activity appears to have mantled the entire moon in a layer of H$_2$O ice frost \citep{Verb2006}. Deposition of E-ring material throughout the rest of the Saturnian system is complex, but it generally mantles the leading hemispheres of Tethys, Dione, and Rhea, although perhaps with a significant sub-Saturnian/anti-Saturnian component \citep{Schenk2011,Fila2012,Hendrix2018,Kempf2018}.

The strength of the H$_2$O ice absorption bands on the four largest Uranian satellites are also stronger on their leading hemispheres compared to their trailing hemispheres. This longitudinal asymmetry in H$_2$O band strength decreases with increasing distance from Uranus, with the strongest asymmetry on Ariel and the weakest on the outermost classical moon Oberon \citep{Grundy2006,Cart2015,Cart2018}. 
The exact mechanism or combination of mechanisms responsible for these spatial trends remains uncertain, but the observed longitudinal and planetocentric trends observed on the Uranian moons are consistent with both the heliocentric impactor and the magnetospheric `sweeping' hypotheses.

Additionally, the leading hemispheres of the Uranian moons are spectrally redder than their trailing hemispheres, with the reddening increasing with distance from Uranus. This has been attributed to planetocentric dust grains sourced from Uranus' retrograde irregular satellites \citep{Buratti1991,Tamayo2013,Cart2018}. Mantling of the leading hemisphere with this non-icy material should decrease leading hemisphere H$_2$O ice band strengths, and therefore diminish the longitudinal asymmetry. However, most of this planetocentric dust should be swept up by the outer moons Titania and Oberon, and little of this dust should collide with the innermost major moon Miranda \citep{Tamayo2013}. 

The four largest Uranian moons also show evidence of CO$_2$ ice deposits on their trailing hemispheres. These CO$_2$ ice absorption features are stronger on the moons closer to Uranus, in particular Ariel \citep{Grundy2003,Grundy2006,Cart2015}. The distribution of this CO$_2$ ice is consistent with radiolytic production from native H$_2$O ice and carbonaceous material on the surfaces of the Uranian moons, which likely migrates to low-latitude cold traps on their trailing hemispheres \citep{Grundy2006,Sori2017,Cart2022}. As CO$_2$ ice is not stable at the peak surface temperatures of the Uranian satellites over geological timescales, ongoing radiolytic production is the most likely explanation for its presence (and longitudinal trends) on the larger Uranian satellites.

Miranda, however, does not appear to follow the compositional trends observed on the other Uranian satellites. Those trends suggest that the leading-trailing asymmetry in H$_2$O ice band strengths should be most prominent closest to Uranus, but previous studies instead indicate that this asymmetry is weak or nonexistent on Miranda \citep{Gourgeot2014,Cart2018,Cart2020IRAC}. Similarly, if the CO$_2$ ice observed on the other Uranian satellites is produced through radiolysis, then radiolytic production of CO$_2$ should be occurring on Miranda as well. However, no CO$_2$ ice has been detected on Miranda's surface, possibly as a consequence of its low escape velocity hindering the retention of CO$_2$ molecules traveling at thermal velocities \citep{Sori2017}. 

\section{Observations} \label{sec:obs}
\begin{deluxetable*}{lccp{2cm}cclcc}
\tabletypesize{\footnotesize}
\tablecaption{Miranda Observations\label{tab:obstable}}
\tablehead{
\colhead{UT Date} & \colhead{Latitude} & \colhead{Mean Longitude} &
\colhead{Longitude Range} & \colhead{Phase} & \colhead{Instrument} & \colhead{Standard(s)} & \colhead{$t_{int}$} & \colhead{PI}\\
\colhead{} & \colhead{Sub-Earth} & \colhead{Sub-Earth} & \colhead{Sub-Earth} &
\colhead{angle} & \colhead{} & \colhead{} & \colhead{} & \colhead{} \\
\colhead{} & \colhead{\degr N} & \colhead{\degr E} & \colhead{\degr E} &
\colhead{\degr} & \colhead{} & \colhead{} & \colhead{s} & \colhead{}
}  
\startdata
2000/09/07 & -35.4 & 36.3 & 33.6 – 39.0 & 1.32 & SpeX SXD & SAO 164 & 2220 & Rivkin\tablenotemark{a} \\
2012/09/25 & 21.4 & 256.1 & 248.1 – 266.7 & 0.20 & SpeX SXD & BD-00 4557 & 5520 & Gourgeot\tablenotemark{b} \\
2012/09/26 & 21.4 & 152.6 & 146.4 – 158.8 & 0.15 & SpeX SXD & BD-00 4557 & 3840 & Gourgeot\tablenotemark{b} \\
2014/11/30 & 24.7 & 279.4 & 268.6 – 289.8 & 2.32 & SpeX SXD & SAO 110201, SAO 109567 & 7200 & Cartwright\tablenotemark{a} \\
2015/09/11 & 30.4 & 236.2 & 235.0 – 237.6 & 1.49 & SpeX PRISM & SAO 110201, BD+08 205 & 960 & Cartwright\tablenotemark{a} \\
2015/09/12 & 30.4 & 92.1 & 89.4 – 94.7 & 1.45 & SpeX PRISM & SAO 110201, BD+08 205 & 1680 & Cartwright\tablenotemark{a} \\
2015/09/17 & 30.2 & 280.3 & 277.6 – 282.9 & 1.23 & SpeX PRISM & SAO 110201, BD+08 205 & 1920 & Cartwright\tablenotemark{a} \\
2017/09/25 & 36.7 & 232.5 & 219.3 – 245.0 & 1.21 & SpeX SXD & SA 93-101 & 8880 & this work \\
2017/09/30 & 36.7 & 80.2 & 73.1 – 87.2 & 0.97 & SpeX PRISM & SA 93-101 & 4800 & this work \\
2017/10/10 & 36.1 & 107.1 & 101.8 – 112.4 & 0.47 & SpeX PRISM & BD+09 213 & 3600 & this work \\
2019/10/13a & 44.6 & 285.7 & 283.6 – 287.9 & 0.77 & TripleSpec & HD 19061 & 1440 & this work \\
2019/10/13b & 44.6 & 307.5 & 305.4 – 309.7 & 0.77 & TripleSpec & HD 16017 & 1440 & this work \\
2019/10/21 & 44.4 & 102.8 & 101.9 – 103.6 & 0.38 & TripleSpec & HD 16017 & 720 & this work \\
2019/10/22a & 44.3 & 37.7 & 33.4 – 42.2 & 0.32 & TripleSpec & BD+15 4915 & 2880 & this work \\
2019/10/22b & 44.3 & 63.6 & 59.3 – 70.3 & 0.32 & TripleSpec & HD 16017, HD 19061 & 2880 & this work \\
2019/10/26 & 44.2 & 308.2 & 300.9 – 316.2 & 0.12 & TripleSpec & HD 19061 & 4320 & this work \\
2019/10/30 & 44.0 & 277.8 & 275.8 – 280.0 & 0.10 & TripleSpec & HD 28099 & 1440 & this work \\
2019/11/04 & 43.8 & 80.5 & 67.0 – 93.8 & 0.35 & TripleSpec & HD 19061, HD 224251 & 7680 & this work \\
2020/01/18 & 42.3 & 70.6 & 67.0 – 75.0 & 2.83 & TripleSpec & HD 28099 & 2520 & this work \\
2020/09/07 & 50.4 & 96.1 & 69.3 – 119.7 & 2.40 & TripleSpec & HD 15942 & 12600 & this work \\
2020/09/12 & 50.3 & 292.4 & 268.4 – 314.3 & 2.24 & TripleSpec & HD 15942 & 10980 & this work \\
2020/09/13 & 50.3 & 184.9 & 159.8 – 208.9 & 2.21 & TripleSpec & HD 15942 & 12060 & this work \\
2020/09/30 & 49.9 & 192.8 & 166.7 – 218.8 & 1.54 & TripleSpec & HD 15942 & 11880 & this work \\
2020/10/07 & 49.7 & 174.2 & 150.7 – 198.5 & 1.23 & TripleSpec & HD 16275 & 11520 & this work \\
2020/10/08 & 49.6 & 109.4 & 102.4 – 117.1 & 1.18 & GNIRS XD & HD 16275 & 4320 & this work \\
2020/10/13 & 49.5 & 262.8 & 238.5 – 288.0 & 0.94 & TripleSpec & HD 15942 & 12960 & this work \\
2020/11/04 & 48.7 & 68.7 & 53.4 – 83.7 & 0.18 & TripleSpec & HD 15942 & 6840 & this work \\
2020/12/06 & 47.6 & 274.6 & 241.5 – 304.8 & 1.72 & TripleSpec & HD 16275 & 12960 & this work \\
2020/12/12 & 47.5 & 356.4 & 348.0 – 6.0 & 1.95 & TripleSpec & HD 16275 & 3600 & this work \\
2020/12/29a & 47.2 & 348.6 & 334.8 – 356.6 & 2.49 & TripleSpec & HD 16275 & 3780 & this work \\
2020/12/29b & 47.2 & 29.7 & 23.4 – 36.8 & 2.49 & TripleSpec & HD 16275 & 3600 & this work \\
2020/12/30 & 47.2 & 320.4 & 311.0 – 330.2 & 2.52 & GNIRS XD & HD 16275 & 3780 & this work \\
2021/01/01 & 47.2 & 51.8 & 19.1 – 84.3 & 2.56 & TripleSpec & HD 16275 & 14760 & this work \\
2021/01/05 & 47.2 & 351.5 & 320.0 – 23.0 & 2.65 & TripleSpec & HD 16275 & 15840 & this work \\
2021/01/15 & 47.2 & 11.8 & 351.0 – 32.7 & 2.79 & TripleSpec & HD 16275 & 10800 & this work \\
2021/11/06 & 53.8 & 355.4 & 346.9 – 3.9 & 0.07 & GNIRS XD & HD 16275 & 4860 & this work \\
2021/11/21 & 53.3 & 203.7 & 194.2 – 212.9 & 0.84 & GNIRS XD & HD 16275 & 4860 & this work \\
\enddata
\tablenotetext{a}{\footnotesize\citet{Cart2018}}
\tablenotetext{b}{\footnotesize\citet{Gourgeot2014}}
\tablecomments{\footnotesize The sub-Earth latitude, mean longitude, range of longitudes, and phase angle} for each spectrum are determined from the JPL Horizons database using the individual timing of each spectral frame. The latitude changes negligibly across a single night. In addition to published SpeX spectra from other authors that we use in our analysis, we also include three SpeX spectra not previously published. An `a' or `b' suffix on the UT date indicates that two separate, non-contiguous spectra were taken on one night and analyzed individually.
\end{deluxetable*}

\begin{figure}[ht!]
\centering
\plottwo{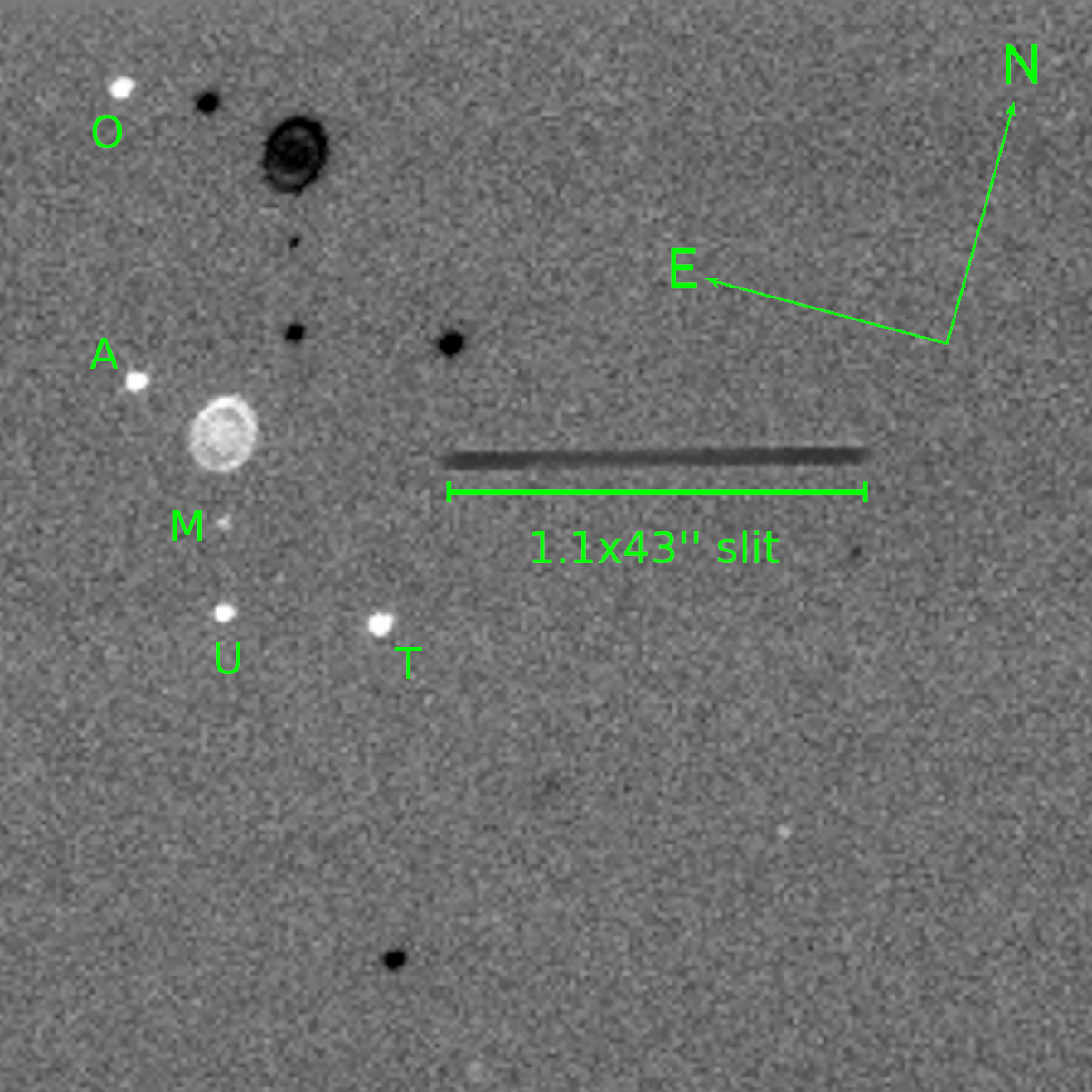}{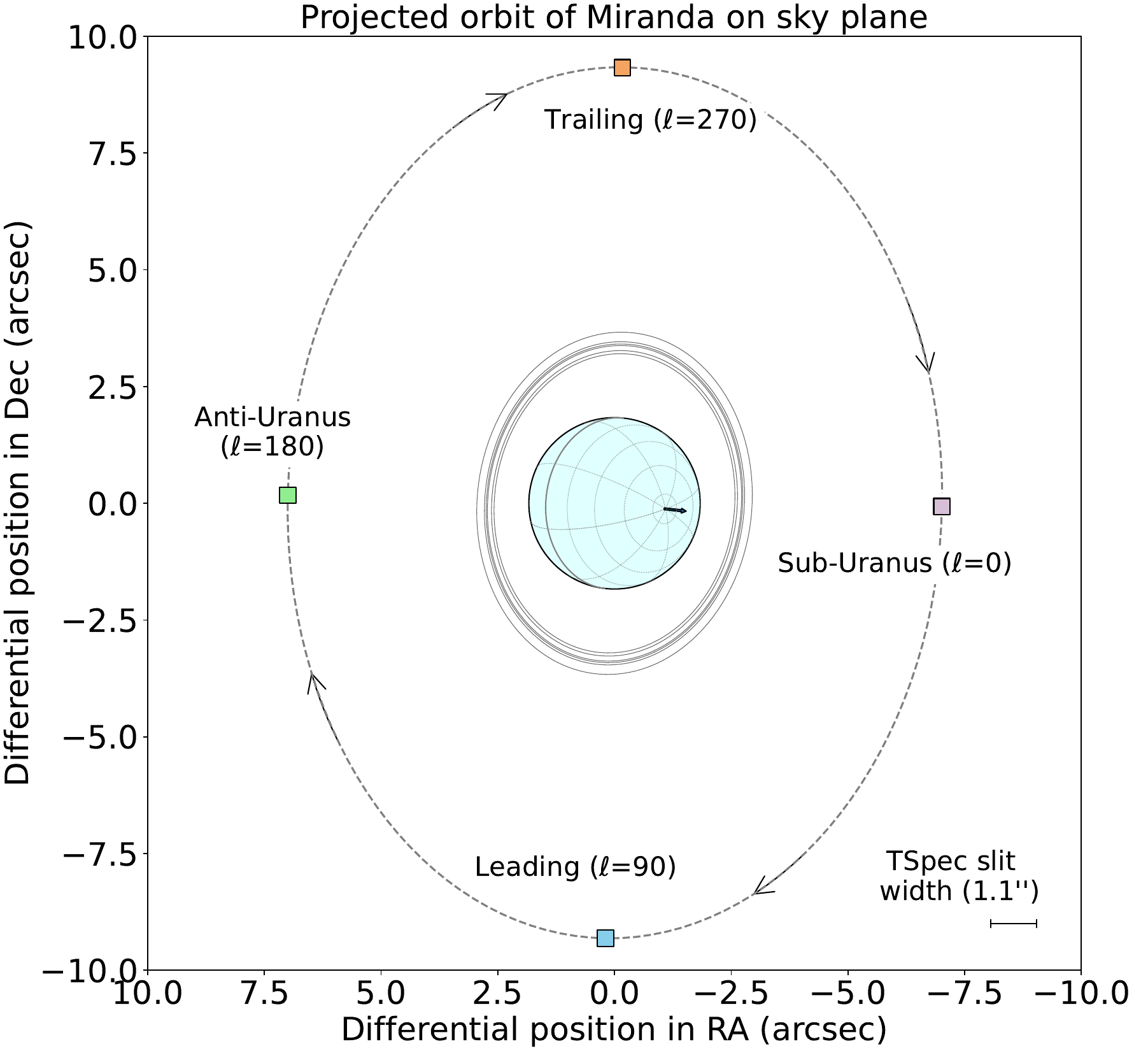}
\caption{\footnotesize \textit{(Left panel):} An annotated image of the Uranian system as observed with the TripleSpec K$_s$-band slit-viewing camera on 2020/09/07 (UT). Uranus, its ring system, and all five classical satellites are clearly visible in this 15 second exposure. The dark negative image above the slit is the result of subtracting a previous slitviewer frame to decrease the sky background. \textit{(Right panel):} The orientation of the inner Uranian system as seen from Earth when the slitviewer image was taken, at the beginning of acquisition of the UT200907 spectrum. The projected orbit of Miranda is shown, with markers denoting Miranda's position when the centers of certain quadrants are facing Earth during each orbit. The axial tilt of the Uranian system is clear from the projected longitude-latitude grid on Uranus, with the north pole pointing towards the viewer. While the sub-observer latitude may differ between Miranda and Uranus by a few degrees, this diagram is generally representative of the typical geometry during our 2019--2021 observations. }
\label{fig:slitviewer}
\end{figure}

We undertook an extensive observational campaign in order to investigate the longitudinal distribution of H$_2$O ice band strengths on Miranda, summarized in Table \ref{tab:obstable}. Although several prior observational studies have investigated Miranda \citep{BrownClark1984,Bauer2002,Gourgeot2014,Cart2018}, it is much less well characterized in the near-IR than the other Uranian moons. Miranda's faintness (m$_{V}\sim$16.5, m$_{K}\sim$15) requires significant observing time with 3 to 4-meter class telescopes to achieve a high signal-to-noise ratio (S/N). Additionally, Miranda's maximum angular proximity to Uranus is on the order of a few arcseconds (Figure \ref{fig:slitviewer}), resulting in significant scattered light in the spectrograph slit and guiding difficulties in the optical regime. Miranda's rapid orbital motion and synchronous rotation (with a period of 1.413 days, $\sim$34 hours) covers roughly 10$\degr$ of sub-observer longitude in about an hour, making it more difficult to obtain high S/N data on a limited range of longitudes. As we required multiple hours on Miranda to obtain a spectrum with reasonable S/N, we accounted for the range of sub-observer longitudes covered over time when planning our observations. Miranda's orbital motion and the proximity of Uranus and the other Uranian moons additionally required frequent adjustment of the spectrograph slit position angle, often incompatible with keeping the slit at a parallactic angle to minimize atmospheric dispersion. 

The dearth of near-IR spectra covering all longitude ranges offered us an opportunity to pursue a sub-observer longitudinal mapping of Miranda at mid-northern sub-observer latitudes, bringing it up to par with the coverage on the other Uranian satellites. In total, we observed Miranda on 24 different occasions. Twenty of these observations were obtained with the Astrophysical Research Consortium (ARC) 3.5m telescope at Apache Point Observatory. Additionally, we report four near-IR spectra of Miranda collected with the GNIRS spectrograph on the 8.1-meter Gemini North telescope, which likely represent some of the highest quality ground-based data ever collected for this moon. Our data set includes some of the first spectra acquired of the sub-Uranian quadrant of Miranda and significant coverage of all other quadrants. ARC 3.5m observations are typically scheduled in half-night blocks, which we refer to as `nights' for convenience. Of the 20 3.5m nights, 13 of them were dedicated solely to Miranda, with the exception of observing telluric standard stars and Uranus for scattered light correction. The spectra of Miranda analyzed in this study are publicly available \citep{MirandaDataset}. This collection includes our newly acquired spectra of Miranda from TripleSpec and GNIRS, along with previously published SpeX spectra from other authors.

\begin{figure}[ht!]
\centering
\makebox[\textwidth][c]{\includegraphics[width=\textwidth]{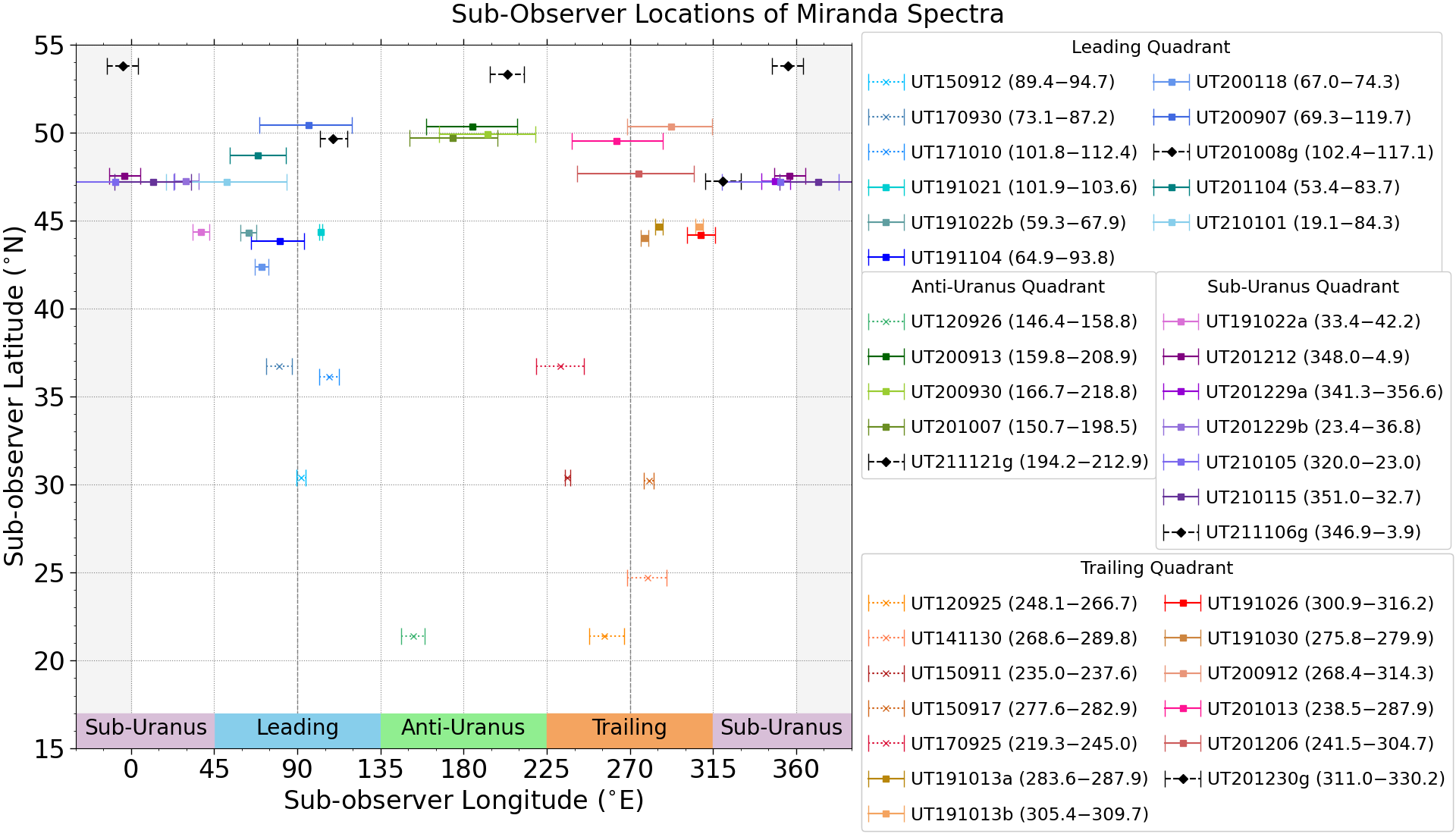}}
\vspace{-10pt}
\caption{\footnotesize Sub-observer latitudes and longitude ranges of the Miranda spectra presented in this work. The width of the horizontal bars attached to each data point represents the sub-observer longitudes at the beginning and end of the set of spectral frames that contributed to a particular final spectrum. Spectra acquired with SpeX are marked with colored X's and dotted errorbars, TripleSpec with colored squares and solid errorbars, and GNIRS with black diamonds and dashed errorbars. The SpeX spectrum from UT000907, our only spectrum of the southern hemisphere, has been omitted in order to better visualize the latitudinal spread of the northern hemisphere spectra. Spectra acquired at latitudes $>$35\degr N are new in this work. Gray shaded regions on either side of the plot show overlap from spectra crossing 0$\degr$ longitude, and spectra in this region are plotted on both the left and right sides of the diagram. In this plot and spectral plots, we maintain a color scheme of blue tones for the leading quadrant, green tones for the anti-Uranus quadrant, orange tones for the trailing quadrant, and purple tones for the sub-Uranus quadrant.}
\vspace{-5pt}
\label{fig:locations}
\end{figure}

Figure \ref{fig:locations} shows the sub-observer (sub-Earth) latitude and longitude range of each of the spectra included in this work. We note that we did not spatially resolve Miranda during any of our observations. Miranda's angular diameter is only 0.035$^{\prime\prime}$ as seen from Earth, therefore all of our spectra are disk-integrated. While a small amount of the opposite hemisphere is visible during our observations at these latitudes (e.g. a fraction of the trailing hemisphere is visible in leading hemisphere spectra, see Uranus in Figure \ref{fig:slitviewer}), the spectrum is still dominated by the contribution from the sub-observer longitude at the center of the target disk. The sub-observer longitude range is defined here as the longitudes spanning between the sub-observer longitude at disk center on Miranda at the time of the first and last individual spectral frames contributing to a given spectrum, calculated with the NASA/JPL Horizons online ephemeris service. This work uses the IAU left-handed, east-positive convention for longitude, where 0$\degr$ is the meridian on Miranda directly facing Uranus, and 90$\degr$ is the center of the leading hemisphere. The sub-observer longitude increases with time as Miranda rotates, while the sub-observer latitude on Miranda's surface does not change appreciably ($<$0.1\degr) within a single night. However, the sub-observer latitude does change over longer timescales due to Miranda's inclined orbit around Uranus, in addition to Uranus's extreme axial tilt slowly moving the sub-solar point northward as the planet approaches its northern summer solstice. We categorize our spectra in four `quadrants' of longitude on Miranda's surface: the leading quadrant (46--135\degr), the anti-Uranus quadrant (136--225\degr), the trailing quadrant (226--315\degr), and the sub-Uranus quadrant (316--45\degr). In some of our analyses, we categorize our spectra into the leading hemisphere (1--180\degr) and the trailing hemisphere (181--360\degr) instead.

\subsection{ARC 3.5m / TripleSpec}
\begin{figure}[ht!]
\centering
\makebox[\textwidth][c]{\includegraphics[width=\textwidth]{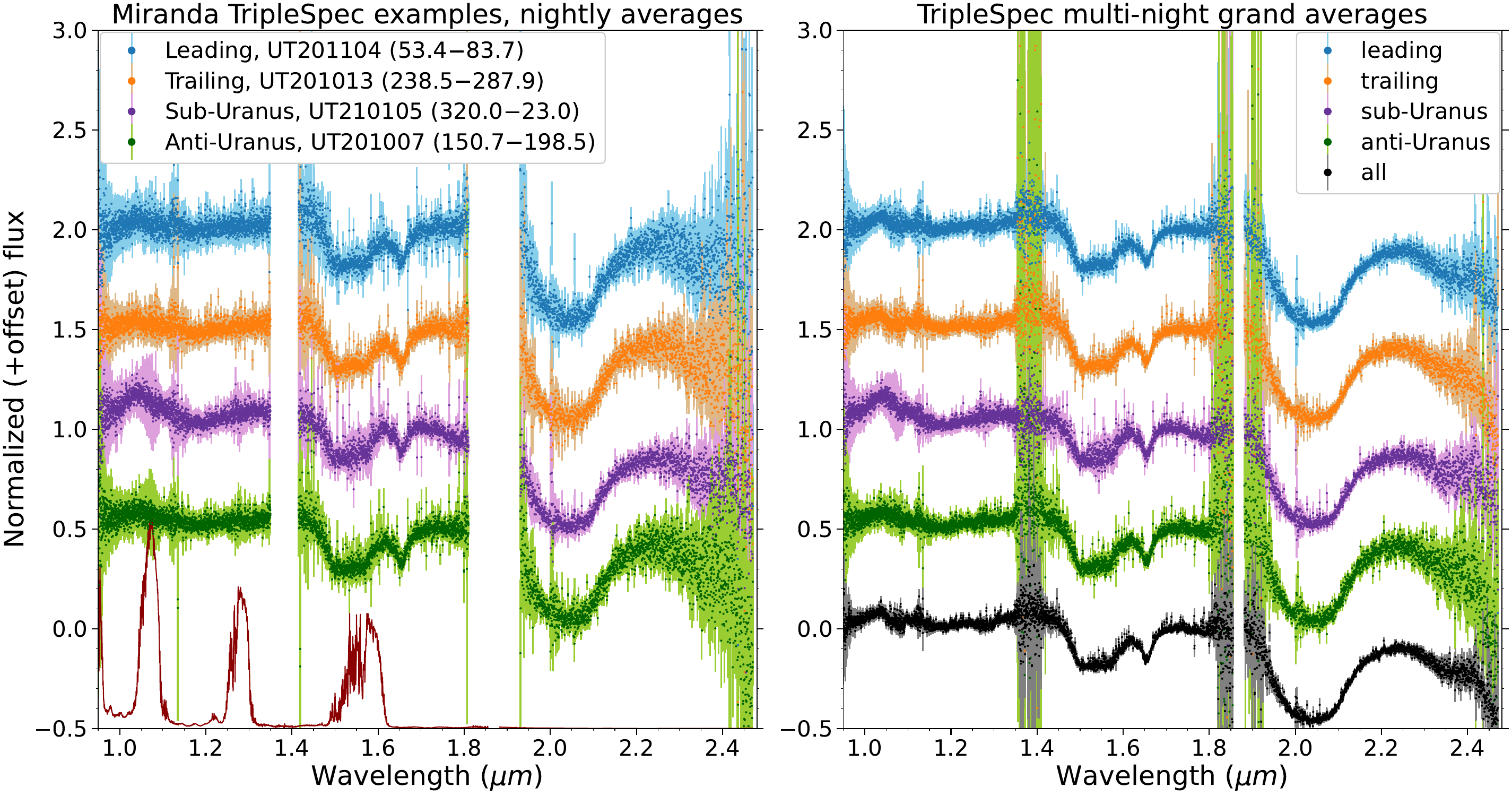}}
\vspace{-15pt}
\caption{\footnotesize \textit{(Left panel):} Examples of four nightly average spectra acquired with TripleSpec. The wavelengths between 1.35--1.42 \um and 1.81--1.93 \um are strongly affected by telluric noise and have been omitted for clarity. The short-wavelength `bump' between 0.95--1.15 \um in the TripleSpec data is an artifact of merging spectral orders. A spectrum of Uranus is also plotted to show which wavelength ranges have the strongest contamination, visible as increased error estimates in certain regions of the spectrum, primarily on the anti-Uranus and sub-Uranus spectra. See Table \ref{tab:obstable} for observation details. \textit{(Right panel):} The `grand average' TripleSpec spectra, constructed from multiple nights of data. Spectra in both panels are normalized to unity at 1.72 \um and offset in increments of 0.5.}
\vspace{-5pt}
\label{fig:TSpecExamples}
\end{figure}
The majority of our observations were carried out using the ARC 3.5-meter telescope, located at Apache Point Observatory (APO) in Sunspot, New Mexico. Examples of our acquired spectra are shown in Figure \ref{fig:TSpecExamples} and \hyperref[adx:collection]{Appendix C}. We used the TripleSpec (TSpec) near-IR spectrograph \citep{Wilson2004}. TripleSpec is a cross-dispersed echellette spectrograph of the same basic design as NIRES on Keck II, ArCoIRIS at CTIO, and TripleSpec at Palomar. A single TripleSpec exposure covers the wavelength range 0.95--2.45 \um simultaneously in five spectral orders at a spectral resolution of R$\sim$3500, with only a small gap between 1.85--1.88 \micron. The 1.1$\times$43$^{\prime\prime}$ slit offers ample space for nodding along the slit. The 3.5m telescope control software can track at non-sidereal rates, but we could not guide on the other Uranian satellites as the guide software is not designed to compensate for guide `stars' moving at non-sidereal rates and Miranda is too faint to guide on. We therefore guided the telescope manually during Miranda observations, which worked well except on particularly windy nights.

Due to the proximity of Uranus and the other Uranian satellites to Miranda's position on the sky, it was often difficult or impossible to maintain the spectrograph slit at the ideal parallactic angle to minimize atmospheric dispersion. We instead adopted a strategy of positioning the slit such that the slit ran perpendicular to a line connecting Miranda and Uranus, or in other words, approximately parallel to Miranda's orbital motion. This choice minimized scattered light by keeping Uranus as far as possible from the slit, and the scattered light that entered the slit ran along the spatial direction and was better accounted for by fitting a background during extraction. We obtained a spectrum of Uranus during every night that we observed Miranda with the 3.5m, and careful selection of the slit position angle sometimes allowed us to observe more than one Uranian satellite in the same exposure.

Our observations were carried out in a manner typical for near-IR observations, involving nodding along the slit in an ABBA pattern. Certain regions of the near-IR spectrum are heavily absorbed by the Earth's atmosphere. These telluric absorption features are one of the principal limitations on near-IR spectroscopy from the ground. In particular, the regions between 1.35--1.43 \micron, 1.80--1.94 \micron, and $>$2.45 \um are strongly absorbed by atmospheric water vapor, which varies dramatically in abundance over short timescales. Two more strong absorption bands between 2.0--2.1 \um are due to atmospheric CO$_2$. 
Near-IR observations typically use A0V stars for correction of telluric absorption by the Earth's atmosphere, due to the ease of modeling their intrinsic spectra to retrieve a clean telluric spectrum \citep{Vacca2003}. However, for Solar System objects observed in reflected sunlight, we can instead observe G stars, as a simple division of the object spectrum by the standard spectrum effectively corrects for the intrinsic stellar spectrum of the Sun, the instrument throughput function, and telluric absorption. We observed early G dwarfs as our standard stars at similar airmasses and times as our targets (described further in \hyperref[adx:telluric]{Appendix B}).

The proximity of Miranda to Uranus led to significant contamination of the Miranda spectra by scattered light from the planet, at a level which would make accurate measurement of the 1.5-\um H$_2$O ice band complex difficult. We attempted correction of these effects on an order-by-order basis by subtracting a scaled spectrum of Uranus. This procedure reliably retrieves the shape of the 1.5-\um H$_2$O ice band complex. Uranus is much fainter at wavelengths longer than 2 \micron, and the K-band was negligibly contaminated by this scattered light. A full discussion of our Uranian scattered light correction procedure is included in \hyperref[adx:uranus]{Appendix A}.

\subsection{Gemini North / GNIRS}
\begin{figure}[ht!]
\centering
\makebox[\textwidth][c]{\includegraphics[width=\textwidth]{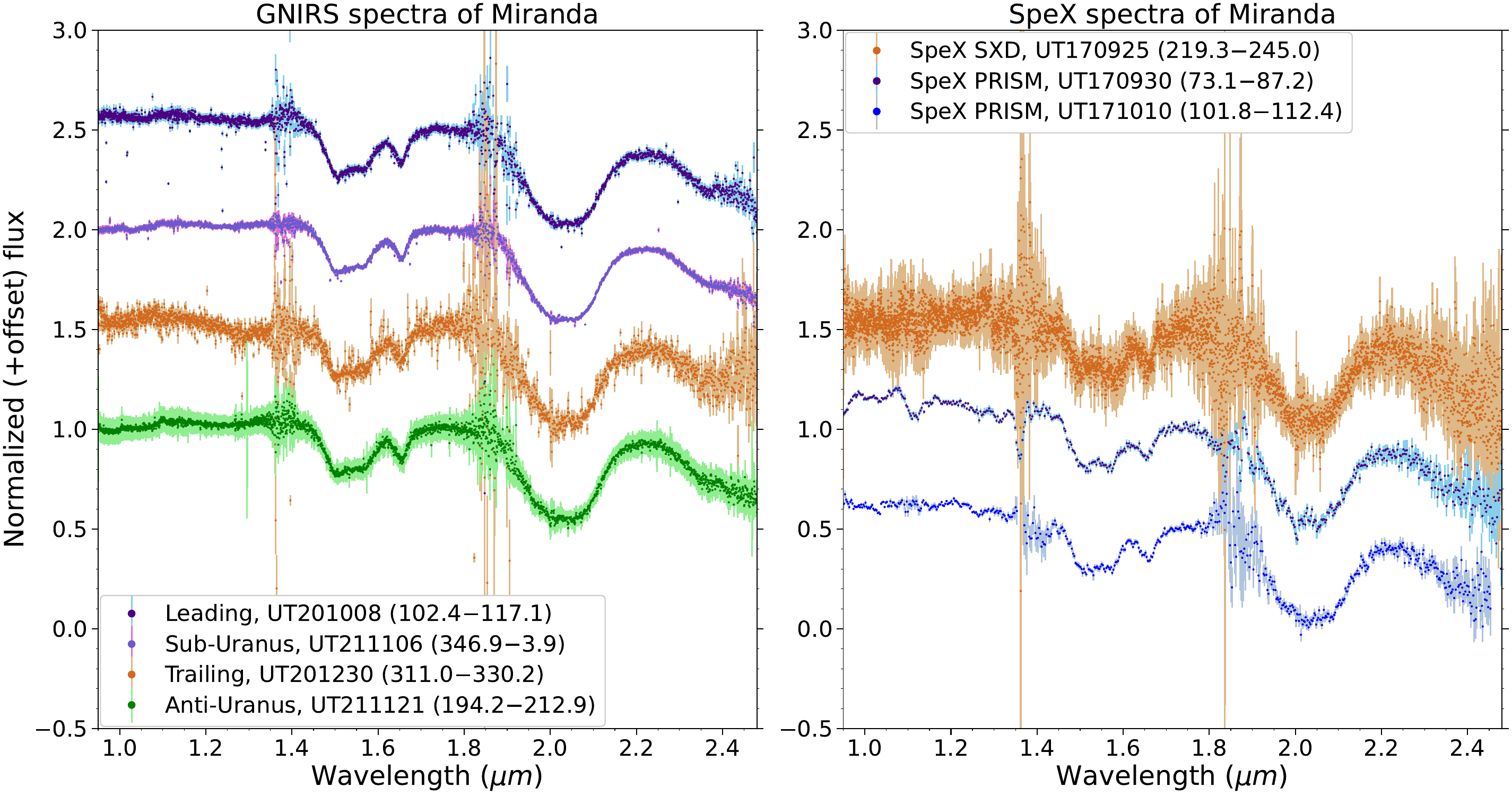}}
\vspace{-15pt}
\caption{\footnotesize \textit{(Left panel):} GNIRS spectra of Miranda. \textit{(Right panel):} The three previously unpublished SpeX spectra that we included in our analysis. Spectra in both panels are normalized to unity at 1.72 \um and offset in increments of 0.5. See Table \ref{tab:obstable} for observation details.}
\vspace{-5pt}
\label{fig:GNIRSSpeXExamples}
\end{figure}

We also acquired four near-IR spectra of Miranda using GNIRS \citep{Elias2006a,Elias2006b} on the 8.1-meter Gemini North telescope (Figure \ref{fig:GNIRSSpeXExamples}). We were awarded two hours on each of four nights (UT201008, UT201230, UT211106, and UT211121) to observe the leading, trailing, sub-Uranus, and anti-Uranus quadrants of Miranda. For the leading and trailing observations, GNIRS was configured with the 0.45$^{\prime\prime}$ slit with the 32.7 lines/mm grating, the SXD cross-dispersing prism, and the 0.15 $^{\prime\prime}$/pixel ``short blue'' camera. This resulted in a resolving power of R$\sim$1130 and coverage from 0.90--2.50 \um across five spectral orders, in a format similar to TripleSpec. For the anti-Uranus and sub-Uranus spectra, we instead used the 0.675$^{\prime\prime}$ slit, with a resolving power of R$\sim$750. We chose to increase the slit width for the sub-Uranus and anti-Uranus observations in order to increase the throughput. This increased the S/N achieved in a similar amount of allocated observing time, yet maintained sufficient resolving power for study of potential absorption features in the 2.2-\um range. We implemented the same slit position angle strategy as with our ARC 3.5m observations in order to minimize scattered light from Uranus. We observed the standard star (HD 16275) both before and after our Miranda observations.

Our leading, anti-Uranus, and sub-Uranus spectra were observed in excellent conditions, and represent some of the highest quality near-IR spectra of Miranda reported to date. The trailing hemisphere observation of Miranda with GNIRS was acquired in poor seeing conditions and at higher airmass, but this spectrum was still of sufficient quality to measure the 1.5-\um and 2.0-\um H$_2$O ice bands (Figure \ref{fig:GNIRSSpeXExamples}).

\subsection{IRTF / SpeX}

To compare our newly acquired observations to prior measurements, we included previously published near-IR spectra of Miranda in our analysis. These spectra were acquired with the SpeX spectrograph on NASA's 3-meter IRTF telescope on Maunakea \citep{Rayner2003}. The spectra obtained by PI Gourgeot were published in \citet{Gourgeot2014}. Spectra obtained by PI Rivkin and PI Cartwright were published in \citet{Cart2018}.  Three SpeX spectra, obtained in 2017 by PI Cartwright, are reported here for the first time (Figure \ref{fig:GNIRSSpeXExamples}). Unless otherwise noted, we refer the reader to these papers for details on the data reduction procedures. The procedure for reducing the three previously unpublished SpeX spectra was the same as that in \citet{Cart2018}.  The SpeX SXD spectrum from 2012/09/25 UT (mean longitude 256$\degr$E) possessed three spurious ($>5\sigma$) noise spikes at 1.5168, 1.6809, and 2.1624 \micron, which affected our band area and depth measurements of an otherwise high-quality spectrum. These noise spikes likely originated from either cosmic ray hits, bad detector pixels, or alpha particles from the radioactive ThF$_4$ anti-reflection coating on one of the lenses within the SpeX instrument. We interpolated over the affected pixels and replaced the flux and uncertainty values with ones drawn from a Gaussian distribution, based on the flux and uncertainty of nearby pixels. This was the only spectrum presented in this work that required this type of `by hand' fix, and none of the other SpeX spectra have been modified from their final reduced form that was provided to us.

\subsection{Data Reduction\label{ssec:datared}}
For the TripleSpec data, we used the IDL data reduction package TripleSpecTool, a modified version of Spextool \citep{Cushing2004}.
The TripleSpecTool pipeline uses raw spectrograph calibration and data frames as inputs. The pipeline performs flat-fielding, wavelength calibration, spectral extraction, stacking and combination, a basic telluric correction process, and merging of spectral orders. Wavelength calibration was performed on groups of frames using the numerous bright OH airglow lines in the near-IR. Spectral extraction used a variant of optimal extraction \citep{Horne1986} with background subtraction. Each extracted spectral frame was stacked and combined with other frames with a robust weighted mean statistic, using the TripleSpecTool \textit{xcombspec} software routine.

These groups of frames usually represented between two and four ABBA-pattern nod cycles (8--16 frames of 180 seconds each) on Miranda. Each group of frames was telluric-corrected by division by our G-type standard stars, using the TripleSpecTool \textit{xtellcor\_basic} software routine. The \textit{xtellcor\_basic} procedure divides the object spectrum by the telluric standard star spectrum, while allowing the user to apply sub-pixel wavelength shifts between the spectra on a per-spectral-order basis, which improves the correction of narrow, strong telluric absorption lines. In a minority of cases, a combination of standard stars provided the best correction. We did not attempt to individually correct for slope differences between these stars and the solar spectrum. We also applied a correction for the scattered light from Uranus to the same groups of spectral frames that we used for telluric correction; for details on this scattered light correction procedure, see \hyperref[adx:uranus]{Appendix A}. 
These telluric-corrected and Uranus-corrected groups of frames were again combined using a robust weighted mean statistic to create a single spectrum for each night. On some nights, there was a significant time interval between two contiguous Miranda observations. These were treated as separate spectra, denoted by an `a' and `b' suffix, in order to avoid the calculated mean longitude of that observation falling in a range that was not actually observed. We also constructed `grand average' spectra by combining (again using a robust weighted mean) every group of telluric-corrected, Uranus-subtracted TripleSpec spectral frames from our entire data set in which the central longitude fell within each defined quadrant (Figure \ref{fig:TSpecExamples}).

Finally, we merged the five spectral orders together using the TripleSpecTool \textit{xmergeorders} software routine. We note that this order-merging step apparently introduced a spurious `bump' between 0.95 and 1.15 \micron, the strength of which seems to be correlated to the contamination by Uranian scattered light. This effect is also discussed further in \hyperref[adx:bump]{Appendix A.3}. This region of the spectrum is not relevant to our measurements of the 1.5-\um and 2.0-\um H$_2$O ice bands, but it does introduce a localized spectral slope and some spurious spectral features at wavelengths $<$1.15 \micron.

Our Gemini North/GNIRS spectra were reduced using the Gemini IRAF package, following the reduction steps published on the GNIRS webpage. We found that we did not have satisfactory results from shifting and combining spectral frames into a single frame, as recommended in the reduction manual, and the signal-to-noise was high enough in each exposure that we decided to instead extract spectra from each spectral frame individually. We implemented an additional residual sky subtraction step with the IRAF task NSRESSKY to reduce the contamination from OH airglow lines. The extracted spectra from each frame were reformatted from the Gemini multi-extension FITS format to a format compatible with our other TripleSpecTool/Spextool FITS files. These individual spectra were then combined, telluric-corrected, and the spectral orders were merged, all in a manner identical to the methods described above for our TripleSpec data. Because of Gemini's superior PSF and narrower slit, the GNIRS spectra are not contaminated by scattered light from Uranus, and we therefore did not perform scattered light corrections on these data.

\section{Spectral Analysis\label{sec:analysis}}

Our spectra (Figures \ref{fig:TSpecExamples}, \ref{fig:GNIRSSpeXExamples}, \hyperref[adx:collection]{Appendix B}) are comparable to previous near-IR spectra of Miranda and the other Uranian satellites, with obvious evidence for the 1.5-\um and 2.0-\um H$_2$O ice absorption bands. The 1.65-\um feature associated with crystalline H$_2$O ice is also clearly visible. We do not detect any evidence of the 1.04-\um and 1.25-\um H$_2$O ice bands. This is consistent with the presence of a dark, spectrally neutral contaminant across the Uranian system, which even at low abundances can have a dramatic effect on otherwise reflective regoliths \citep{ClarkLucey1984}. In the case of the Miranda spectra presented here, additional complications are introduced by the strong contamination by scattered light from Uranus coinciding with the 1.04-\um and 1.25-\um bands, as the stray light could erase (or mimic) any weak absorption features.

\subsection{Band Area and Depth Measurements} \label{sec:measurements}

We measured the integrated band areas and fractional band depths of the 1.5-\micron, 1.65-\micron, and 2.0-\um H$_2$O ice absorption bands in all of the Miranda spectra reported here \citep[e.g.][]{ClarkRoush1984,Grundy1999,Emery2005,Cart2018}.  
In this work, we use the shorthand term ``1.5-\um band'' to refer to the entire absorption band complex between roughly $\sim$1.45--1.72 \micron, which includes the 1.65-\um band unless otherwise specified. The term ``2.0-\um band'' is used for the band between $\sim$1.90--2.20 \micron.
The integrated band areas and fractional band depths depend on the abundance of fresh H$_2$O ice exposed in the surface regolith, grain size variations \citep{Clark1983}, the presence of dark contaminants \citep{ClarkLucey1984}, and other factors like variations in crystallinity \citep{Grundy1998,Mastrapa2008}. All of these factors can play a significant role in determining the spectral signature of H$_2$O ice, and differences in band areas and depths are not linearly correlated with ice abundance.

In general, larger ice grains correspond to lower continuum albedo, a steeper spectral slope between 0.9--1.4 \um, and deeper absorption bands.
Increasing abundance of dark contaminants leads to lower albedo, a flatter spectral slope, and decreased absorption band depths. Crystallinity and temperature variations affect the shape of the 1.5-\um band and the presence of the 1.65-\um band, and combinations of these effects can lead to some absorption bands becoming deeper and others becoming shallower.

\subsection{Band Measurement Methods}\label{ssec:measmethods}
\begin{figure}[ht!]
\centering
\makebox[\textwidth][c]{\includegraphics[width=\textwidth]{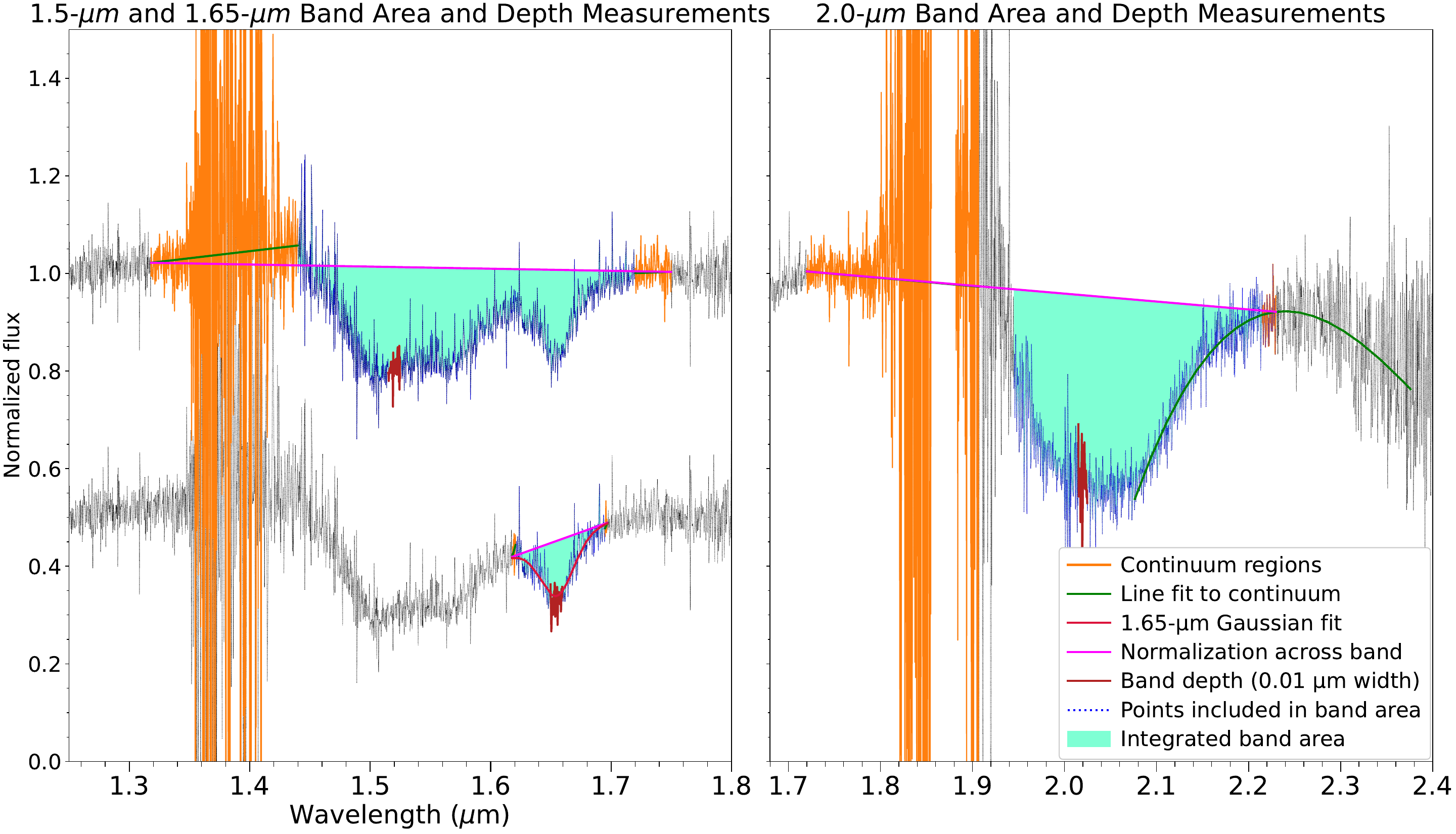}}
\vspace{-10pt}
\caption{\footnotesize A visual demonstration of the band area and depth measurement process, using the trailing hemisphere TripleSpec spectrum from UT201013. Error bars on the spectrum have been omitted for clarity. The gap between 1.85--1.88 \um is intrinsic to TripleSpec and those wavelengths are ignored in the continuum-fitting process. The right-side continuum for the 1.5-\um band overlaps with the left-side continuum for the 2.0-\um band.}
\vspace{-5pt}
\label{fig:banddefinitions}
\end{figure}
Our band area and depth measurements were conducted with a Python implementation of the technique described in \citet{Cart2015,Cart2018}, which used a modified version of the SARA band analysis routine originally developed for asteroid spectra \citep{Lindsay2015}. All of our analyses used unbinned, native resolution spectra.
The band analysis code first generated a sample spectrum, drawing from a Gaussian distribution that uses the measured flux and errors of the original spectrum in each pixel as the mean and standard deviation of the distribution. We fitted a line to the data points of the sample spectrum in two defined continuum regions on either side of the absorption feature of interest (Table \ref{tab:bandwaves}), weighting the points by their errors.
Using the endpoints of the continuum fit in the two regions on either side of the absorption band, we drew a continuum line across the entire bandpass, and normalized the spectrum by this line (Figure \ref{fig:banddefinitions}). We then integrated the area underneath it using the trapezoidal rule to calculate the band area. 
We implemented a Monte Carlo sampling method to estimate the uncertainties of our band area measurements. The above process was iterated 20,000 times, and the mean of the band area measurements was reported as the final result, with the standard deviation of the measured areas reported as the 1$\sigma$ errors.

We additionally measured the fractional band depth at the canonical band centers of 1.520 \um and 2.020 \um by calculating the mean reflectance of the points within $\pm$0.005 \um of the center (Table \ref{tab:bandwaves}). These reflectances were defined relative to the same continuum normalization line drawn across the band by the band area measurement procedure, and the mean reflectance was subtracted from unity to obtain a continuum-divided fractional band depth. The mean uncertainty was calculated by adding the errors of all individual spectral pixels in that wavelength range in quadrature and dividing by the number of pixels ($N$). We then calculated the standard error of the mean ($SEM = \sigma / \sqrt{N} = \sqrt{1/N(N-1) \sum{(x_i - \bar{x})^{2}}}$) to account for the point-to-point variation between spectral pixels, and finally added these two quantities together in quadrature to obtain the 1$\sigma$ error on the fractional band depth measurements.

We also measured the area and depth of the 1.65-\um band, as its association with crystallinity and temperature variations \citep[e.g.][]{Grundy1998,Grundy1999,Mastrapa2008} could provide useful information. In addition to the band area measurement, we fitted a Gaussian to the 1.65-\um band in order to identify the best-fit central wavelength of the absorption band, then measured the band depth using the data points within $\pm$0.005 \um of that wavelength. The 1.5-\um and 2.0-\um band complexes are composed of multiple overlapping overtones of longer-wavelength fundamental absorptions and cannot be easily described by a single Gaussian \citep{Grundy1998}, so we did not perform this step for these two H$_2$O bands.

\begin{deluxetable}{ccccc}
\tablecaption{H$_2$O ice band wavelengths\label{tab:bandwaves}}
\tablehead{\colhead{Band name} & \colhead{Left continuum} & \colhead{Band width} & \colhead{Right continuum} & \colhead{Center}\\ 
\colhead{} & \colhead{($\mu m$)} & \colhead{($\mu m$)} & \colhead{($\mu m$)} & \colhead{($\mu m$)} }
\startdata
1.5-$\mu m$ & 1.318--1.440 & 1.440--1.720 & 1.720--1.750 & 1.515--1.525\\
1.65-$\mu m$ & 1.618--1.621 & 1.621--1.695 & 1.695--1.697 & 1.65\tablenotemark{a} \\
2.0-$\mu m$ & 1.720--1.907 & 1.945--2.215\tablenotemark{b} & 2.215--2.230 & 2.015--2.025\\
\enddata
\tablenotetext{a}{\footnotesize The central wavelength of the 1.65-\um band was allowed to vary in our analysis, then we measured the band depth in a range $\lambda_{cen} \pm 0.005$ \micron.}
\tablenotetext{b}{\footnotesize The long-wavelength end of the 2.0-\um band is defined with a third-order polynomial via the method described in section \ref{ssec:measmethods}.}
\tablecomments{\footnotesize We tabulate the wavelengths we used to define the band continua, widths, and centers. Central wavelengths refers to the wavelength range we used to measure the fractional band depth.}

\end{deluxetable}

The 2.0-\um H$_2$O ice band required additional steps compared to the 1.5-\um band complex, as the nominal right-side continuum in the 2.2 \um region is curved. To generate our continuum, we first found the wavelength corresponding to the maximum flux in the data between 2.215 and 2.230 \micron. We then fitted a third-order polynomial to the flux in a range spanning $\pm$0.15 \um from that wavelength. The maximum value of the fit polynomial between 2.215 and 2.230 \um was then used as the endpoint of the continuum line across the entire bandpass used to normalize the data. 
We also chose to start our 2.0-\um band area integration at 1.945 \um instead of 1.907 \micron, as the presence of strong telluric noise between 1.90 and 1.94 \um can drastically affect the integrated 2.0-\um band area. 
This increased the reliability of our band area measurements, but with the caveat that this choice makes the measurements no longer directly comparable to the previous work in the field. For example, \citet{Cart2018} analyzed SpeX spectra of Miranda and set the left side of the absorption band at 1.907 \um, but telluric absorption is generally much less of an issue in the SpeX spectra due to the lower precipitable water vapor content above Maunakea (elevation 4200 m) compared to that at APO (2788 m). We re-measured the band areas of the same SpeX spectra in this work, but using the 1.945-\um wavelength definition for consistency with our TripleSpec and GNIRS spectra. We calculated band parameters for all of our spectra individually (Table \ref{tab:specbands}), although we completely omitted the night of UT191021 (leading quadrant) from our analysis as the S/N was too low to reliably measure any H$_2$O ice band parameters.

\subsection{H$_2$O Ice Temperature}

As described above, we measured the central wavelength of the 1.65-\um band for each spectrum, which shifts to longer wavelengths with decreasing temperature \citep{Grundy1998,Mastrapa2008}. We calculated the ice temperature of each spectrum following a procedure described in \citet{Holler2017}. We used the \textit{alpha\_h2o} IDL routine \citep[]{Grundy1998,Grundy1999}, which calculates synthetic absorption coefficients as a sum of Gaussians for the 1.65-\um band. We then constructed a lookup table for the wavelength of maximum absorption versus ice temperature between 20 and 120 K, and compared the central wavelengths we measured for the 1.65-\um band in our Miranda spectra. We report our results in section \ref{ssec:icetempresults}.

\subsection{Other Analyses}

Unlike the other major Uranian satellites, we do not see clear spectral evidence for CO$_2$ ice absorption features on Miranda's trailing hemisphere, which is consistent with the results of previous studies \citep{Gourgeot2014,Cart2018}. However, some of the spectra we report here may show low-level absorption features between 2.2 and 2.3 \um (Section \ref{ssec:surfcomp}). The exact compounds responsible for this feature are still undetermined, as many species have absorption features in this wavelength range. A detailed analysis of the potential presence of CO$_2$ ice and absorption features in the 2.2 \um region in our spectra of Miranda will be included in future work.

\begin{figure}[ht!]
\centering
\makebox[\textwidth][c]{\includegraphics[width=\textwidth]{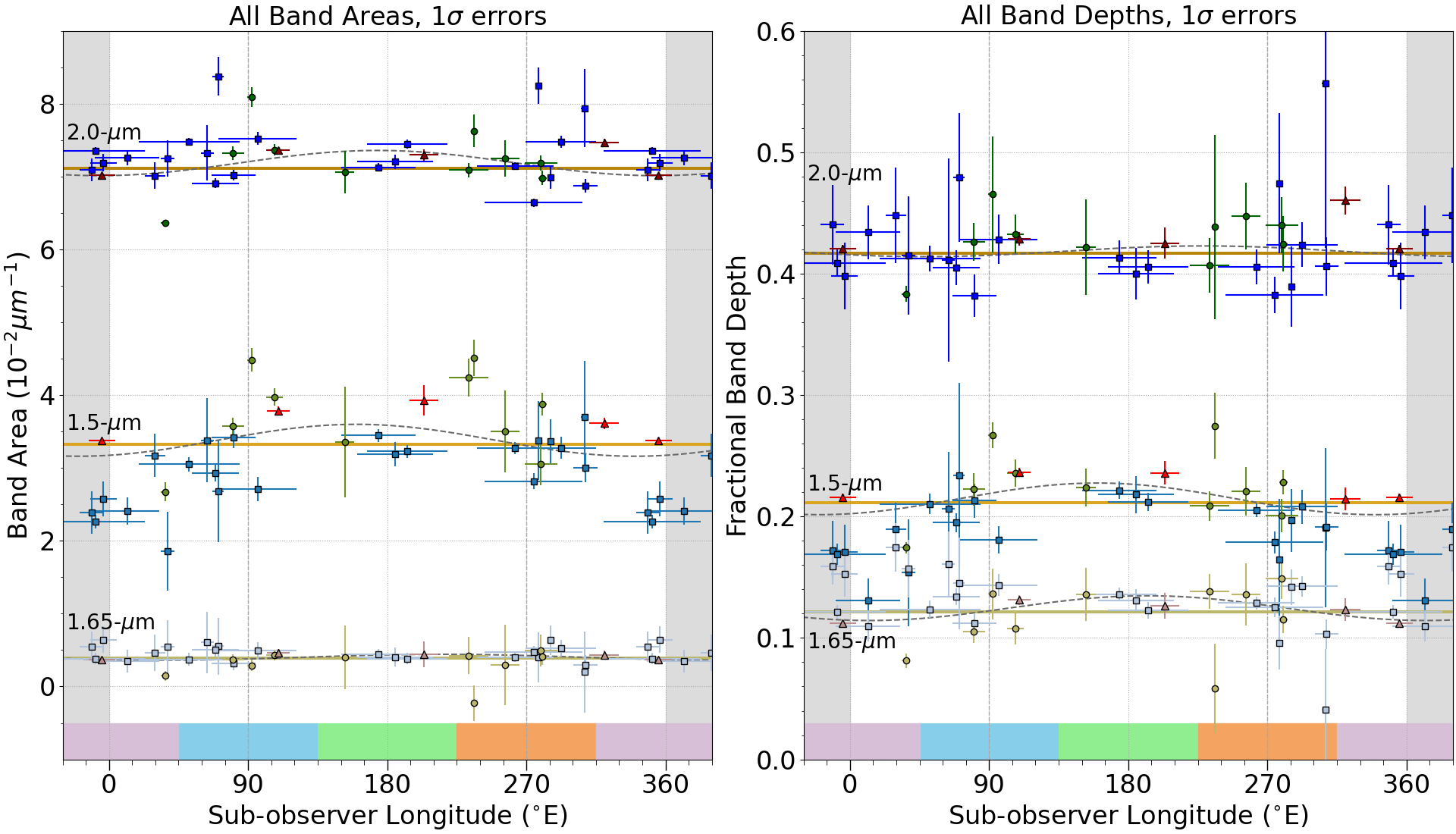}}
\vspace{-15pt}
\caption{\footnotesize A plot of the integrated band areas \textit{(left panel)} and fractional band depths \textit{(right panel)} versus longitude on Miranda. Blue squares are TripleSpec spectra, green circles are SpeX spectra, and red triangles are GNIRS spectra. Vertical error bars represent 1$\sigma$ errors in both panels. The lowest, lightest-colored set of measurements represents the 1.65-\um band, the middle set of measurements represents the 1.5-\um band, and the darker-colored upper set is for the 2.0-\um band. The golden horizontal lines represent the weighted mean band areas (and depths) and the gray dashed lines represent our best-fit sinusoidal models. Other features are the same as in Figure \ref{fig:locations}.}
\vspace{-5pt}
\label{fig:AllBandAreas}
\end{figure}

Northern/southern hemisphere differences are difficult to interpret with our data. The only spectrum in our data set that was acquired on the southern hemisphere (UT000907, latitude 35$\degr$S) lies at a mean sub-observer longitude of 36$\degr$, but the 1.5-\um band parameters are similar to our other sub-Uranus spectra at similar longitudes on the northern hemisphere (Figure \ref{fig:AllBandAreas}). For the following analyses, we did not account for latitudinal differences in the sub-observer locations of our spectra, and we cannot draw any strong conclusions about northern/southern hemispherical patterns on Miranda with only one spectrum at southern latitudes and no imaging coverage at northern latitudes.

\section{Results}\label{sec:results}
\begin{deluxetable}{lccCCCCCCcc}
\tabletypesize{\footnotesize}
\tablecaption{H$_2$O ice band measurements\label{tab:specbands}}
\tablehead{\colhead{UT Date} & \colhead{Long.} & \colhead{Lat.} & \multicolumn{3}{c}{Integrated band area} & \multicolumn{3}{c}{Fractional band depth} & \colhead{$\lambda_{cen}$} & \colhead{T$_{ice}$}\\ 
\colhead{} & \colhead{} & \colhead{} & \colhead{1.5-\um} & \colhead{1.65-\um} & \colhead{2.0-\um} & \colhead{1.5-\um} & \colhead{1.65-\um} & \colhead{2.0-\um} & \colhead{1.65-\um} & \colhead{}\\
\colhead{} & \colhead{(\degr)} & \colhead{(\degr)} & \colhead{(10$^{-2}\mu$m)} & \colhead{(10$^{-2}\mu$m)} & \colhead{(10$^{-2}\mu$m)} & \colhead{($\mu$m)} & \colhead{($\mu$m)} & \colhead{($\mu$m)} & \colhead{($\mu$m)} & \colhead{(K)}} 
\startdata
UT210115 & 11.8 & 47.2 & $2.41\pm0.19$ & $0.35\pm0.16$ & $7.26\pm0.10$ & $0.13\pm0.02$ & $0.11\pm0.01$ & $0.43\pm0.02$ & $1.6536\pm0.0006$ & 62$\pm$7 \\
UT201229b & 29.7 & 47.2 & $3.18\pm0.29$ & $0.46\pm0.25$ & $7.02\pm0.18$ & $0.19\pm0.02$ & $0.17\pm0.02$ & $0.45\pm0.04$ & $1.6539\pm0.0007$ & 58$\pm$9 \\
UT000907 & 36.3 & -35.4 & $2.68\pm0.13$ & $0.15\pm0.07$ & $6.37\pm0.04$ & $0.17\pm0.00$ & $0.08\pm0.01$ & $0.38\pm0.01$ & $1.6540\pm0.0005$ & 57$\pm$7 \\
UT191022a & 37.7 & 44.3 & $1.86\pm0.54$ & $0.54\pm0.37$ & $7.25\pm0.25$ & $0.15\pm0.04$ & $0.16\pm0.02$ & $0.42\pm0.05$ & $1.6558\pm0.0011$ & 33$\pm$15 \\
UT210101 & 51.8 & 47.2 & $3.05\pm0.10$ & $0.37\pm0.09$ & $7.48\pm0.05$ & $0.21\pm0.01$ & $0.12\pm0.01$ & $0.41\pm0.01$ & $1.6542\pm0.0003$ & 54$\pm$4 \\
UT191022b & 63.6 & 44.3 & $3.38\pm0.57$ & $0.61\pm0.42$ & $7.33\pm0.38$ & $0.21\pm0.05$ & $0.16\pm0.03$ & $0.41\pm0.08$ & $1.6581\pm0.0013$ & 1$\pm$19 \\
UT201104 & 68.7 & 48.7 & $2.93\pm0.11$ & $0.50\pm0.09$ & $6.91\pm0.07$ & $0.20\pm0.01$ & $0.13\pm0.01$ & $0.41\pm0.01$ & $1.6545\pm0.0004$ & 50$\pm$5 \\
UT200118 & 70.6 & 42.3 & $2.68\pm0.70$ & $0.55\pm0.39$ & $8.38\pm0.26$ & $0.23\pm0.08$ & $0.15\pm0.04$ & $0.48\pm0.05$ & $1.6518\pm0.0015$ & 83$\pm$17 \\
UT170930 & 80.2 & 36.7 & $3.58\pm0.11$ & $0.37\pm0.08$ & $7.32\pm0.09$ & $0.22\pm0.01$ & $0.11\pm0.01$ & $0.43\pm0.02$ & $1.6560\pm0.0006$ & 30$\pm$8 \\
UT191104 & 80.5 & 43.8 & $3.42\pm0.15$ & $0.31\pm0.09$ & $7.02\pm0.07$ & $0.21\pm0.01$ & $0.11\pm0.01$ & $0.38\pm0.02$ & $1.6552\pm0.0004$ & 41$\pm$5 \\
UT150912 & 92.1 & 30.4 & $4.49\pm0.16$ & $0.29\pm0.08$ & $8.10\pm0.14$ & $0.27\pm0.01$ & $0.14\pm0.02$ & $0.47\pm0.05$ & $1.6616\pm0.0018$ & -53$\pm$29 \\
UT200907 & 96.1 & 50.4 & $2.72\pm0.17$ & $0.49\pm0.12$ & $7.53\pm0.09$ & $0.18\pm0.01$ & $0.14\pm0.01$ & $0.43\pm0.02$ & $1.6538\pm0.0006$ & 60$\pm$7 \\
UT171010 & 107.1 & 36.1 & $3.97\pm0.12$ & $0.44\pm0.07$ & $7.37\pm0.08$ & $0.24\pm0.01$ & $0.11\pm0.01$ & $0.43\pm0.02$ & $1.6582\pm0.0007$ & 0$\pm$10 \\
UT201008g & 109.4 & 49.6 & $3.78\pm0.07$ & $0.46\pm0.06$ & $7.36\pm0.03$ & $0.24\pm0.00$ & $0.13\pm0.00$ & $0.43\pm0.01$ & $1.6530\pm0.0003$ & 69$\pm$4 \\
UT120926 & 152.6 & 21.4 & $3.36\pm0.76$ & $0.40\pm0.44$ & $7.06\pm0.29$ & $0.22\pm0.02$ & $0.14\pm0.02$ & $0.42\pm0.04$ & $1.6532\pm0.0004$ & 66$\pm$5 \\
UT201007 & 174.2 & 49.7 & $3.45\pm0.09$ & $0.44\pm0.07$ & $7.13\pm0.06$ & $0.22\pm0.01$ & $0.14\pm0.01$ & $0.41\pm0.01$ & $1.6534\pm0.0003$ & 65$\pm$4 \\
UT200913 & 184.9 & 50.3 & $3.19\pm0.17$ & $0.40\pm0.13$ & $7.21\pm0.09$ & $0.22\pm0.02$ & $0.13\pm0.01$ & $0.40\pm0.02$ & $1.6556\pm0.0005$ & 35$\pm$7 \\
UT200930 & 192.8 & 49.9 & $3.23\pm0.09$ & $0.38\pm0.08$ & $7.45\pm0.06$ & $0.21\pm0.01$ & $0.12\pm0.01$ & $0.41\pm0.01$ & $1.6546\pm0.0004$ & 50$\pm$5 \\
UT211121g & 203.7 & 53.3 & $3.93\pm0.20$ & $0.45\pm0.18$ & $7.30\pm0.08$ & $0.24\pm0.01$ & $0.13\pm0.01$ & $0.43\pm0.01$ & $1.6550\pm0.0003$ & 44$\pm$4 \\
UT170925 & 232.5 & 36.7 & $4.24\pm0.26$ & $0.43\pm0.25$ & $7.09\pm0.10$ & $0.21\pm0.01$ & $0.14\pm0.01$ & $0.41\pm0.02$ & $1.6551\pm0.0006$ & 42$\pm$8 \\
UT150911 & 236.2 & 30.4 & $4.52\pm0.25$ & $-0.23\pm0.24$ & $7.63\pm0.22$ & $0.27\pm0.03$ & $0.06\pm0.04$ & $0.44\pm0.08$ & $1.6571\pm0.0060$ & 15$\pm$76 \\
UT120925 & 256.1 & 21.4 & $3.50\pm0.56$ & $0.30\pm0.55$ & $7.25\pm0.25$ & $0.22\pm0.02$ & $0.14\pm0.03$ & $0.45\pm0.03$ & $1.6544\pm0.0004$ & 51$\pm$6 \\
UT201013 & 262.8 & 49.5 & $3.27\pm0.08$ & $0.40\pm0.07$ & $7.15\pm0.05$ & $0.21\pm0.01$ & $0.13\pm0.01$ & $0.41\pm0.01$ & $1.6549\pm0.0003$ & 45$\pm$4 \\
UT201206 & 274.6 & 47.7 & $2.82\pm0.11$ & $0.47\pm0.09$ & $6.64\pm0.06$ & $0.18\pm0.01$ & $0.13\pm0.01$ & $0.38\pm0.02$ & $1.6550\pm0.0005$ & 44$\pm$6 \\
UT191030 & 277.8 & 44.0 & $3.38\pm0.54$ & $0.40\pm0.34$ & $8.25\pm0.25$ & $0.16\pm0.05$ & $0.10\pm0.02$ & $0.47\pm0.06$ & $1.6590\pm0.0017$ & -12$\pm$24 \\
UT141130 & 279.4 & 24.7 & $3.05\pm0.29$ & $0.50\pm0.22$ & $7.19\pm0.11$ & $0.20\pm0.01$ & $0.15\pm0.02$ & $0.44\pm0.02$ & $1.6537\pm0.0008$ & 61$\pm$10 \\
UT150917 & 280.3 & 30.2 & $3.88\pm0.15$ & $0.41\pm0.12$ & $6.99\pm0.10$ & $0.23\pm0.01$ & $0.12\pm0.01$ & $0.42\pm0.02$ & $1.6543\pm0.0009$ & 52$\pm$12 \\
UT191013a & 285.7 & 44.6 & $3.37\pm0.30$ & $0.64\pm0.20$ & $7.00\pm0.16$ & $0.20\pm0.03$ & $0.14\pm0.01$ & $0.39\pm0.03$ & $1.6565\pm0.0012$ & 24$\pm$16 \\
UT200912 & 292.4 & 50.3 & $3.28\pm0.15$ & $0.53\pm0.11$ & $7.48\pm0.08$ & $0.21\pm0.01$ & $0.14\pm0.01$ & $0.42\pm0.02$ & $1.6554\pm0.0006$ & 38$\pm$7 \\
UT191013b & 307.5 & 44.6 & $3.71\pm0.77$ & $0.20\pm0.55$ & $7.94\pm0.53$ & $0.19\pm0.07$ & $0.04\pm0.05$ & $0.56\pm0.14$ & $1.6551\pm0.0038$ & 43$\pm$47 \\
UT191026 & 308.2 & 44.2 & $3.00\pm0.19$ & $0.30\pm0.13$ & $6.88\pm0.09$ & $0.19\pm0.02$ & $0.10\pm0.01$ & $0.41\pm0.02$ & $1.6550\pm0.0005$ & 44$\pm$7 \\
UT201230g & 320.4 & 47.2 & $3.62\pm0.08$ & $0.43\pm0.07$ & $7.47\pm0.05$ & $0.21\pm0.01$ & $0.12\pm0.01$ & $0.46\pm0.01$ & $1.6538\pm0.0006$ & 59$\pm$8 \\
UT201229a & 348.6 & 47.2 & $2.39\pm0.29$ & $0.55\pm0.21$ & $7.09\pm0.16$ & $0.17\pm0.02$ & $0.16\pm0.02$ & $0.44\pm0.03$ & $1.6558\pm0.0011$ & 33$\pm$15 \\
UT210105 & 351.5 & 47.2 & $2.27\pm0.10$ & $0.38\pm0.08$ & $7.36\pm0.05$ & $0.17\pm0.01$ & $0.12\pm0.01$ & $0.41\pm0.01$ & $1.6547\pm0.0004$ & 47$\pm$5 \\
UT211106g & 355.4 & 53.8 & $3.38\pm0.04$ & $0.37\pm0.04$ & $7.02\pm0.02$ & $0.22\pm0.00$ & $0.11\pm0.00$ & $0.42\pm0.00$ & $1.6546\pm0.0004$ & 49$\pm$5 \\
UT201212 & 356.4 & 47.5 & $2.58\pm0.24$ & $0.64\pm0.18$ & $7.19\pm0.12$ & $0.17\pm0.02$ & $0.15\pm0.02$ & $0.40\pm0.03$ & $1.6492\pm0.0009$ & 112$\pm$9 \\[0.25cm]
Leading & 79.1 & 47.2 & $3.22\pm0.06$ & $0.43\pm0.05$ & $7.26\pm0.04$ & $0.20\pm0.01$ & $0.13\pm0.00$ & $0.41\pm0.01$ & $1.6548\pm0.0002$ & 46$\pm$3 \\
Trailing & 280.7 & 48.1 & $3.25\pm0.05$ & $0.42\pm0.05$ & $7.01\pm0.03$ & $0.20\pm0.00$ & $0.12\pm0.00$ & $0.40\pm0.01$ & $1.6549\pm0.0002$ & 45$\pm$3 \\
Anti-Uranus & 184.1 & 50.0 & $3.28\pm0.06$ & $0.43\pm0.05$ & $7.28\pm0.04$ & $0.22\pm0.00$ & $0.13\pm0.00$ & $0.41\pm0.01$ & $1.6541\pm0.0002$ & 56$\pm$3 \\
Sub-Uranus & 3.8 & 47.0 & $2.47\pm0.07$ & $0.39\pm0.05$ & $7.31\pm0.03$ & $0.17\pm0.01$ & $0.12\pm0.00$ & $0.41\pm0.01$ & $1.6542\pm0.0003$ & 54$\pm$3 \\
\enddata
\tablecomments{\footnotesize The measured H$_2$O ice integrated band areas, fractional band depths, central wavelengths of the 1.65-\um band, and measured H$_2$O ice temperatures for each spectrum. The dates denoted with a `g' suffix indicate that the spectrum was obtained with GNIRS. All errors are 1$\sigma$ errors. The last four rows are for our `grand average' spectra; combined spectra that include all TripleSpec exposures in which the central longitude lies in a specific quadrant (Figure \ref{fig:TSpecExamples}).}

\end{deluxetable}

We describe the results of the aforementioned analyses in this section. As perhaps might be expected from comparing spectra acquired at three different telescopes, we found that our measured band areas did not always agree. We therefore split our data set into several subsets. Our most comprehensive set is the `All spectra' data set, which includes all of our individual spectra from TripleSpec, SpeX, and GNIRS. This set should be assumed as the default data set under discussion and that is plotted in figures unless otherwise specified. We also created a `TSpec + SpeX' data set, which omits the GNIRS spectra, and the `GNIRS Only' data set, which is self-explanatory. Finally, we also have a separate `TSpec Grand Average' data set, which consists of the quadrant grand average spectra, which were constructed only from our TripleSpec data (see subsection \ref{ssec:datared} and Figure \ref{fig:TSpecExamples}). 

\subsection{Mean Band Areas}
\begin{deluxetable}{ccccccccccc}
\tabletypesize{\footnotesize }
\tablecaption{Quadrant and hemisphere-averaged band area measurements\label{tab:meanbands}}
\tablehead{\colhead{Dataset} & \colhead{Loc.\tablenotemark{a}} & \colhead{Long.} & \multicolumn{2}{c}{1.5-$\mu m$ band} & \multicolumn{2}{c}{2.0-$\mu m$ band} & \multicolumn{3}{c}{1.65-$\mu m$ band} & \colhead{T$_{ice}$} \\ 
\cline{4-5} \cline{6-7} \cline{8-10}
\colhead{} & \colhead{} & \colhead{} & \colhead{Mean area} & \colhead{Mean depth} & \colhead{Area} & \colhead{Depth} & \colhead{Area} & \colhead{Depth} & \colhead{Mean $\lambda_{cen}$} & \colhead{} \\ 
\colhead{} & \colhead{} & \colhead{(\degr)} & \colhead{(10$^{-2}\mu m$)} & \colhead{($\mu m$)} & \colhead{(10$^{-2}\mu m$)} & \colhead{($\mu m$)} & \colhead{(10$^{-2}\mu m$)} & \colhead{($\mu m$)} & \colhead{($\mu m$)} & \colhead{(K)}} 
\startdata
All spectra & LH & 78.9 & $3.18\pm0.18$ & $0.21\pm0.01$ & $7.31\pm0.12$ & $0.42\pm0.01$ & $0.42\pm0.06$ & $0.13\pm0.01$ & $1.6550\pm0.0007$ & $44\pm8$ \\
  & TH & 280.4 & $3.33\pm0.14$ & $0.20\pm0.01$ & $7.28\pm0.09$ & $0.43\pm0.01$ & $0.40\pm0.07$ & $0.12\pm0.01$ & $1.6550\pm0.0006$ & $45\pm7$ \\
  & LQ & 82.0 & $3.40\pm0.21$ & $0.22\pm0.01$ & $7.48\pm0.15$ & $0.43\pm0.01$ & $0.44\pm0.07$ & $0.13\pm0.01$ & $1.6557\pm0.0010$ & $35\pm13$ \\
  & TQ & 274.5 & $3.50\pm0.18$ & $0.21\pm0.01$ & $7.29\pm0.15$ & $0.43\pm0.02$ & $0.36\pm0.10$ & $0.11\pm0.01$ & $1.6555\pm0.0008$ & $38\pm10$ \\
  & AQ & 181.6 & $3.43\pm0.21$ & $0.22\pm0.01$ & $7.23\pm0.09$ & $0.41\pm0.01$ & $0.41\pm0.10$ & $0.13\pm0.01$ & $1.6544\pm0.0005$ & $52\pm6$ \\
  & SQ & 5.3 & $2.71\pm0.21$ & $0.18\pm0.01$ & $7.12\pm0.11$ & $0.42\pm0.01$ & $0.43\pm0.08$ & $0.13\pm0.01$ & $1.6539\pm0.0007$ & $57\pm9$ \\[0.25cm]
TSpec + SpeX & LH & 76.9 & $3.14\pm0.19$ & $0.20\pm0.01$ & $7.30\pm0.13$ & $0.42\pm0.01$ & $0.42\pm0.07$ & $0.13\pm0.01$ & $1.6552\pm0.0007$ & $42\pm9$ \\
  & TH & 278.1 & $3.27\pm0.17$ & $0.20\pm0.01$ & $7.28\pm0.10$ & $0.43\pm0.01$ & $0.39\pm0.08$ & $0.12\pm0.01$ & $1.6550\pm0.0007$ & $43\pm9$ \\
  & LQ & 79.0 & $3.36\pm0.23$ & $0.22\pm0.01$ & $7.49\pm0.17$ & $0.43\pm0.02$ & $0.44\pm0.08$ & $0.13\pm0.01$ & $1.6560\pm0.0010$ & $31\pm14$ \\
  & TQ & 274.5 & $3.50\pm0.18$ & $0.21\pm0.01$ & $7.29\pm0.15$ & $0.43\pm0.02$ & $0.36\pm0.10$ & $0.11\pm0.01$ & $1.6555\pm0.0008$ & $38\pm10$ \\
  & AQ & 176.1 & $3.31\pm0.21$ & $0.22\pm0.01$ & $7.21\pm0.12$ & $0.41\pm0.01$ & $0.41\pm0.12$ & $0.13\pm0.01$ & $1.6542\pm0.0006$ & $54\pm8$ \\
  & SQ & 13.1 & $2.48\pm0.19$ & $0.17\pm0.01$ & $7.08\pm0.14$ & $0.42\pm0.01$ & $0.44\pm0.10$ & $0.14\pm0.01$ & $1.6539\pm0.0009$ & $58\pm11$ \\[0.25cm]
GN UT201008 & LQ & 109.4 & $3.78\pm0.07$ & $0.24\pm0.00$ & $7.36\pm0.03$ & $0.43\pm0.01$ & $0.46\pm0.06$ & $0.13\pm0.00$ & $1.6530\pm0.0003$ & $69\pm4$ \\
GN UT201230 & ``TQ'' & 320.4 & $3.62\pm0.08$ & $0.21\pm0.01$ & $7.47\pm0.05$ & $0.46\pm0.01$ & $0.43\pm0.07$ & $0.12\pm0.01$ & $1.6538\pm0.0006$ & $59\pm8$ \\
GN UT211121 & AQ & 203.7 & $3.93\pm0.20$ & $0.24\pm0.01$ & $7.30\pm0.08$ & $0.43\pm0.01$ & $0.45\pm0.18$ & $0.13\pm0.01$ & $1.6550\pm0.0003$ & $44\pm4$ \\
GN UT211106 & SQ & 355.4 & $3.38\pm0.04$ & $0.22\pm0.00$ & $7.02\pm0.02$ & $0.42\pm0.00$ & $0.37\pm0.04$ & $0.11\pm0.00$ & $1.6546\pm0.0004$ & $49\pm5$ \\
\enddata
\tablenotetext{a}{\footnotesize The sub-observer locations of the spectra were averaged over a given quadrant (Q) or hemisphere (H). L, T, A, and S refer to leading, trailing, anti-Uranus, and sub-Uranus longitudes, respectively. }
\tablecomments{\footnotesize Mean band area and depth measurements, averaged over quadrants and hemispheres and calculated for different subsets of our dataset. All errors are 1$\sigma$ errors. The GNIRS spectra are not quadrant/hemisphere averages, but were included for easier comparison to the mean values/averages of other datasets. The GNIRS UT201230 spectrum does not actually lie in the trailing quadrant, but it is the GNIRS spectrum closest to the center of the trailing hemisphere, and therefore it was given the ``TQ'' label for simplicity.}
\end{deluxetable}

\begin{figure}[ht!]
\centering
\makebox[\textwidth][c]{\includegraphics[width=\textwidth]{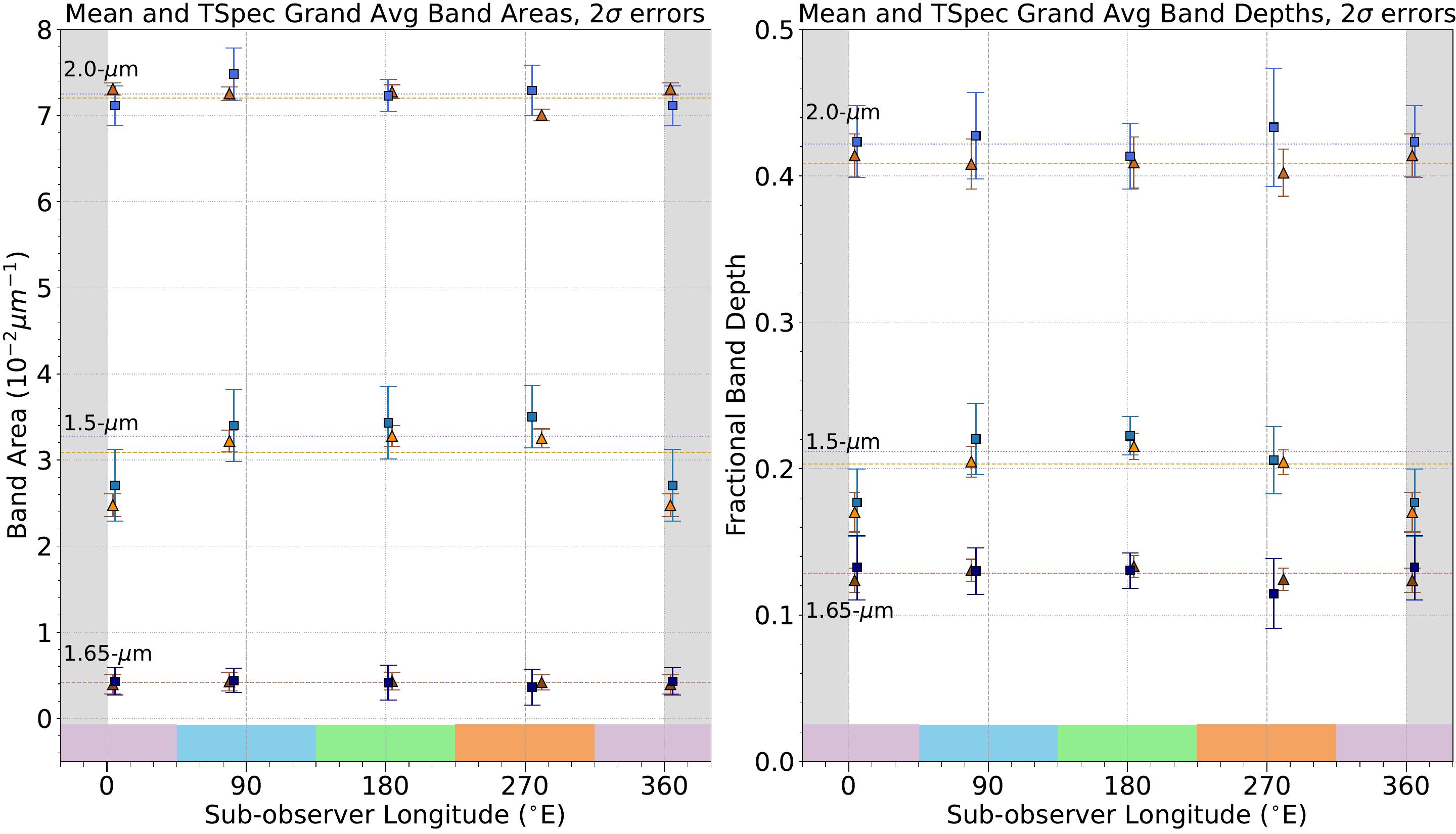}}
\vspace{-15pt}
\caption{\footnotesize The same as Figure \ref{fig:AllBandAreas}, but only showing the band measurements of our mean and TripleSpec grand average spectra. Blue squares are the mean band areas/depths for each quadrant, with purple horizontal lines representing weighted means. Orange triangles are the band areas/depths of the grand average spectra, with yellow horizontal lines representing weighted means. Vertical error bars represent 2$\sigma$ errors on both plots.}
\vspace{-5pt}
\label{fig:MeanGAvgBandAreas}
\end{figure}

As part of our analysis, we also calculated the mean band areas and depths of different longitudinal subsets of the data, split into the previously-defined (Section \ref{sec:obs}) quadrants and hemispheres. The mean is the arithmetic mean of the included band measurements, while the 1$\sigma$ error is calculated in a similar manner to the band depth measurement (i.e. the mean uncertainty and the standard error of the mean are added together in quadrature). We summarize our mean band parameters in Table \ref{tab:meanbands} and plot both the mean band parameters and the band parameters of the TripleSpec grand average spectra in Figure \ref{fig:MeanGAvgBandAreas}. 

While the band measurements of the individual GNIRS spectra are listed in Table \ref{tab:specbands}, we include the same measurements in Table \ref{tab:meanbands} for easier comparison to the mean values of the `All Spectra' and `TSpec + SpeX' data sets. We did not construct average quadrant and hemisphere measurements for the GNIRS spectra in the same manner as the other data sets for several reasons. The mean sub-observer longitude of the GNIRS “trailing” spectrum (320.4) technically lies in the sub-Uranus quadrant, while the mean sub-observer longitudes of the sub-Uranus (355.4) and anti-Uranus (203.7) quadrants are technically in the trailing hemisphere. Quadrant/hemisphere averages for this data set would therefore be presenting mean “trailing” quadrant values that are not technically located in the trailing quadrant, and mean trailing hemisphere values that are weighted towards the values from the anti-Uranus and sub-Uranus spectra (as they have higher S/N than the “trailing” spectrum). Finally, the GNIRS leading quadrant and leading hemisphere values would not be a “mean” at all, as both would only include one spectrum (longitude 109.4). We therefore decided to include the GNIRS values in Table \ref{tab:meanbands} for comparison, but with the caveat that they are not quadrant/hemisphere averaged values like the other values in the table.

\subsection{Sinusoidal Model and F-test Analysis}
\begin{deluxetable}{ccccccc}
\tablecaption{F-test and sinusoidal model fit\label{tab:Ftest}}
\tablehead{\colhead{Dataset} & \colhead{Measurement} & \colhead{N} & \colhead{Longitude} & \colhead{F-value} & \colhead{$p$-value} & \colhead{Reject null hypothesis?} \\ 
\colhead{} & \colhead{} & \colhead{} & \colhead{max absorption} & \colhead{} & \colhead{} & \colhead{} \\
\colhead{} & \colhead{} & \colhead{} & \colhead{(\degr)} & \colhead{} & \colhead{} & \colhead{} } 
\startdata
All spectra & 1.5-$\mu m$ area & 36 & N/A & 2.293 & 0.11684 & No \\
  & 1.5-$\mu m$ depth & 36 & 160.1 & 3.385 & 0.04599 & Yes \\
  & 1.65-$\mu m$ area & 36 & N/A & 1.313 & 0.28260 & No \\
  & 1.65-$\mu m$ depth & 36 & 192.4 & 4.292 & 0.02204 & Yes \\
  & 2.0-$\mu m$ area & 36 & N/A & 2.988 & 0.06415 & No \\
  & 2.0-$\mu m$ depth & 36 & N/A & 0.334 & 0.71862 & No \\
  & 1.65-$\mu m$ $\lambda_{cen}$ & 36 & N/A & 0.428 & 0.65534 & No \\[0.25cm]
TSpec + SpeX & 1.5-$\mu m$ area & 32 & 166.0 & 6.946 & 0.00343 & Yes \\
  & 1.5-$\mu m$ depth & 32 & 168.8 & 10.284 & 0.00042 & Yes \\
  & 1.65-$\mu m$ area & 32 & N/A & 1.082 & 0.35227 & No \\
  & 1.65-$\mu m$ depth & 32 & N/A & 1.832 & 0.17806 & No \\
  & 2.0-$\mu m$ area & 32 & N/A & 1.495 & 0.24101 & No \\
  & 2.0-$\mu m$ depth & 32 & N/A & 0.991 & 0.38330 & No \\
  & 1.65-$\mu m$ $\lambda_{cen}$ & 32 & N/A & 0.045 & 0.95652 & No \\
\enddata
\tablecomments{\footnotesize N refers to the number of data points (spectra) included in the analysis. The degrees of freedom for the sinusoidal model is N - 3, and the degrees of freedom for the constant model is N - 1. The longitude of maximum absorption is the longitude on Miranda for which the sinusoidal model predicts the largest integrated band areas/depths; entries are marked as `N/A' if the model was not statistically significant.  We reject the null hypothesis if the $p$-value calculated from the F-test is $p\leq$0.05.}
\end{deluxetable}

We conducted an additional test of the longitudinal distribution of the strengths of the H$_2$O bands by using an F-test, summarized in Table \ref{tab:Ftest}. We use the F-test to compare two nested models, in which one model is equivalent to a second model if certain parameters of the second model are restricted. In this study, we defined our null hypothesis that there is no discernible variation in band parameters with longitude (i.e., the band measurements are consistent with a constant, mean value.) The hypothesis we are testing is that a sinusoidal model provides a statistically significant better fit to the variation of band parameters with longitude. Our sinusoidal model was defined by the equation
\begin{equation}
y(x) = A sin(2\pi f x + 2\pi\phi) + C
\end{equation}
where $y$ is the integrated band area or fractional band depth, $A$ is the peak-to-peak amplitude, $x$ is the sub-observer longitude, $\phi$ is the phase shift of the sinusoid, and $C$ is a constant offset represented by a weighted mean. The frequency $f$ was held fixed at 1/2$\pi$, representing one rotation of Miranda. The sinusoidal model approach is well-suited to studying longitudinal asymmetries on Solar System bodies \citep{Grundy2006,Holler2016,Cart2018}. The constant, no-variation model against which we tested the sinusoid was defined by $y(x) = C$, where $C$ is a weighted mean of all band measurements (and functionally identical to the constant offset $C$ in the sinusoidal model). While this model fitting includes the uncertainties in the band area and depth measurements, we did not account for the `error bars' in longitude shown on the plots; we treated each spectrum as if it was located at its mean central longitude. We did not apply the sinusoidal model to the `GNIRS only' data set, the mean band areas, or the TripleSpec grand average spectra, given the limited number of data points. 

When all spectra were included, the sinusoidal model was a statistically significant better fit for the 1.5-\um and 1.65-\um band depths (Table \ref{tab:Ftest}). It was also a better fit for both the 1.5-\um areas and depths in the `TSpec + SpeX' dataset (omitting the GNIRS spectra). Any longitudinal variation was not statistically significant for the other band area or depth measurements. In Figure \ref{fig:AllBandAreas}, a sinusoidal pattern in band areas and depths is visually apparent for the 1.5-\um band, particularly in the TripleSpec and SpeX data, but is not clearly identifiable for the 1.65-\um or 2.0-\um band measurements. However, this pattern implies an asymmetry in band strengths between the sub-Uranus and anti-Uranus quadrants of Miranda's surface, \textit{not} the leading and trailing quadrants. The best-fit sinusoidal models for these measurements also placed the longitude corresponding to the strongest absorption bands as being located in the anti-Uranus quadrant.

\subsection{Quadrant and Hemisphere Ratios}
\begin{deluxetable}{cccccccc}
\tablecaption{Hemisphere and quadrant ratios\label{tab:ratios}}
\tablehead{\colhead{Dataset} & \colhead{Ratio\tablenotemark{a}} & \multicolumn{2}{c}{1.5-$\mu m$ band} & \multicolumn{2}{c}{1.65-$\mu m$ band} & \multicolumn{2}{c}{2.0-$\mu m$ band}\\ 
\colhead{} & \colhead{} & \colhead{Area ratio} & \colhead{Depth ratio} & \colhead{Area ratio} & \colhead{Depth ratio} & \colhead{Area ratio} & \colhead{Depth ratio} } 
\startdata
All spectra & LQ/TQ & $0.97\pm0.08$ & $1.07\pm0.08$ & $1.22\pm0.40$ & $1.13\pm0.14$ & $1.03\pm0.03$ & $0.99\pm0.06$ \\
  & LH/TH & $0.96\pm0.07$ & $1.01\pm0.07$ & $1.06\pm0.24$ & $1.08\pm0.09$ & $1.00\pm0.02$ & $0.99\pm0.04$ \\
  & AQ/SQ & $\mathbf{1.27\pm0.12}$ & $\mathbf{1.26\pm0.09}$ & $0.96\pm0.29$ & $0.98\pm0.09$ & $1.02\pm0.02$ & $0.98\pm0.04$ \\[0.25cm]
TSpec + SpeX & LQ/TQ & $0.96\pm0.08$ & $1.06\pm0.09$ & $1.21\pm0.41$ & $1.13\pm0.14$ & $1.03\pm0.03$ & $0.99\pm0.06$ \\
  & LH/TH & $0.96\pm0.08$ & $1.02\pm0.07$ & $1.06\pm0.27$ & $1.08\pm0.10$ & $1.00\pm0.02$ & $0.99\pm0.04$ \\
  & AQ/SQ & $\mathbf{1.33\pm0.13}$ & $\mathbf{1.32\pm0.10}$ & $0.93\pm0.34$ & $0.96\pm0.11$ & $1.02\pm0.03$ & $0.98\pm0.05$ \\[0.25cm]
GNIRS Only & LQ/TQ\tablenotemark{b} & $1.05\pm0.03$ & $\mathbf{1.10\pm0.05}$ & $1.06\pm0.21$ & $1.06\pm0.09$ & $0.99\pm0.01$ & $\mathbf{0.93\pm0.03}$ \\
  & AQ/SQ\tablenotemark{b} & $\mathbf{1.16\pm0.06}$ & $1.09\pm0.05$ & $1.22\pm0.50$ & $1.13\pm0.10$ & $\mathbf{1.04\pm0.01}$ & $1.01\pm0.03$ \\[0.25cm]
Grand Avg Only & LQ/TQ & $0.99\pm0.03$ & $1.00\pm0.03$ & $1.02\pm0.17$ & $1.05\pm0.04$ & $\mathbf{1.04\pm0.01}$ & $1.02\pm0.03$ \\
  & AQ/SQ & $\mathbf{1.33\pm0.04}$ & $\mathbf{1.26\pm0.06}$ & $1.09\pm0.20$ & $1.08\pm0.05$ & $1.00\pm0.01$ & $0.99\pm0.03$ \\[0.25cm]
Miranda & LH/TH & $0.94\pm0.18$ & \nodata & \nodata & \nodata & $0.95\pm0.09$ & \nodata \\
Ariel & LH/TH & $1.46\pm0.06$ & \nodata & \nodata & \nodata & $1.27\pm0.03$ & \nodata \\
Umbriel & LH/TH & $1.21\pm0.10$ & \nodata & \nodata & \nodata & $1.14\pm0.04$ & \nodata \\
Titania & LH/TH & $1.15\pm0.03$ & \nodata & \nodata & \nodata & $1.11\pm0.03$ & \nodata \\
Oberon & LH/TH & $1.09\pm0.05$ & \nodata & \nodata & \nodata & $1.09\pm0.02$ & \nodata \\
\enddata
\tablenotetext{a}{\footnotesize Abbreviations match those of Table \ref{tab:meanbands}.}
\tablenotetext{b}{\footnotesize The GNIRS LQ/TQ ratio only included the spectra from UT201008 and UT201230, and the AQ/SQ ratio only included the spectra from UT211121 and UT211106, representing the spectra closest to the centers of their respective quadrants/hemispheres.}
\tablecomments{\footnotesize Ratios of mean band measurements between specific quadrants and hemispheres. All errors are 1$\sigma$ errors, and ratios of measurements that differ from unity by $\geq2\sigma$ are displayed in a bold typeface. For comparison purposes, at the bottom of the table we include the leading/trailing hemisphere ratios for Miranda and the other Uranian satellites originally reported in Table 12 of \citet{Cart2018}, converted to 1$\sigma$ errors.}
\end{deluxetable}

\begin{figure}[ht!]
\centering
\makebox[\textwidth][c]{\includegraphics[width=\textwidth]{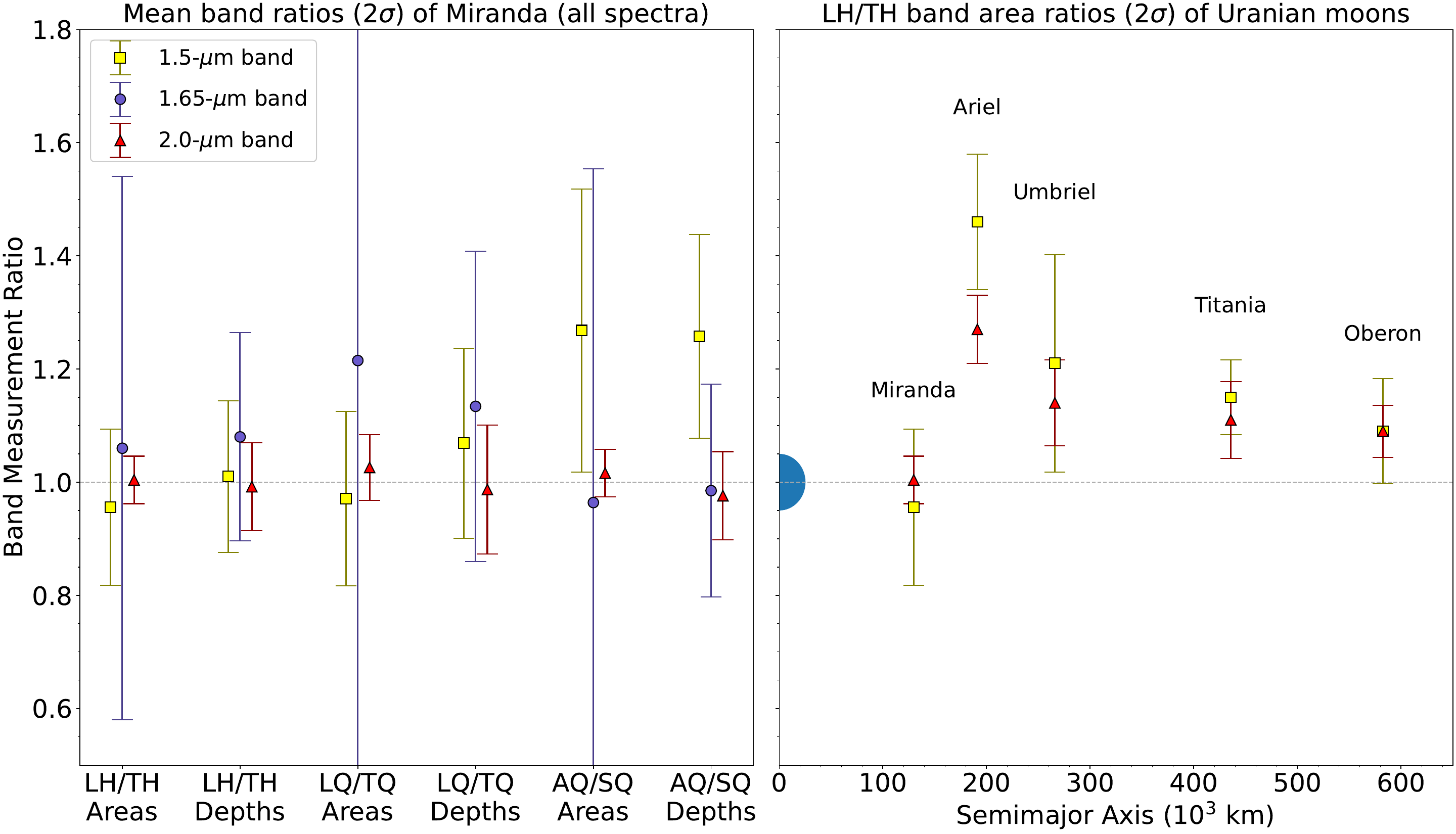}}
\vspace{-15pt}
\caption{\footnotesize \textit{(Left panel)}: Hemisphere and quadrant ratios for all three H$_2$O ice bands on Miranda, designated on the X-axis. These ratios are calculated using the mean band measurements with all spectra included. The yellow squares represent the 1.5-\um band, the purple circles are the 1.65-\um band, and the red triangles are the 2.0-\um band. \textit{(Right panel)}: Leading/trailing hemisphere ratios of H$_2$O ice band areas for the Uranian satellites versus planetocentric distance, based on Figure 9 in \citet{Cart2018}. Plotted errorbars represent 2$\sigma$ uncertainties. The mean radius of Uranus is included for illustrative purposes.}
\vspace{-5pt}
\label{fig:MeanRatios}
\end{figure}

Based on the leading/trailing trends seen on the other Uranian moons, we would expect a significant asymmetry in the strength of H$_2$O ice absorption bands between the leading and trailing quadrants (or hemispheres) of Miranda's surface; however, previous work has suggested that this is not the case for Miranda \citep{Cart2018,Cart2020IRAC}. When considering the results for all of the spectra reported here, we find that there is no statistically significant leading/trailing asymmetry ($>$2$\sigma$ difference) in the strength of Miranda's H$_2$O ice bands (summarized in Table \ref{tab:ratios}), supporting the results presented in prior work. When considering the different data sets separately, we do find a statistically significant leading/trailing asymmetry in some of the measurements for a particular H$_2$O ice band (bolded entries in Table \ref{tab:ratios}). However, none of these leading/trailing ratios are consistent across all data sets, and the band area and depth ratios within a data set for a single band often do not agree. In summary, we find that the leading/trailing hemispherical asymmetry in H$_2$O ice bands on Miranda is weaker than the hemispherical trends observed on the other Uranian satellites (Table \ref{tab:ratios}, Figure \ref{fig:MeanRatios}).

Although our results do not support a leading/trailing hemispherical asymmetry, many of our quadrant-averaged ratios for the 1.5-\um H$_2$O band provide evidence for a longitudinal asymmetry between the anti-Uranus and sub-Uranus quadrants of Miranda. These anti-Uranus/sub-Uranus quadrant ratios are apparent ($>2\sigma$ difference) for the 1.5-\um H$_2$O ice band area and depth measurements, when considering the combined TSpec, SpeX, and GNIRS results (`All spectra' entries in Table \ref{tab:ratios}). When considering the band area and depth ratios for these three datasets separately, they also show that the 1.5-\um H$_2$O ice band is stronger on the anti-Uranus quadrant compared to the sub-Uranus quadrant, albeit the magnitude of the asymmetry varies between each dataset. This is further supported by the sinusoidal model being a statistically significant better fit for the `TripleSpec + SpeX' 1.5-\um band measurements (compared to a model with no longitudinal variation in band area/depth), and that the longitude corresponding to the strongest bands in this sinusoidal model lies in the anti-Uranus quadrant (Table \ref{tab:Ftest}). 

The systematically lower band areas in the TripleSpec data (as compared to the GNIRS data) are likely due to a combination of factors. Not only are the TripleSpec data of lower S/N, but precipitable water vapor values above Apache Point are typically significantly higher than above Maunakea, and the higher spectral resolution of TripleSpec leads to more extreme mismatches in telluric correction between the narrow and strong atmospheric water vapor lines. This produces the high-frequency spectral noise clearly visible around 1.4 and 1.8 micron. These spectral noise ‘spikes’ can produce significant effects on the integrated band areas of the 1.5-um and 2.0-um bands; as mentioned in Section \ref{ssec:measmethods}, this was the rationale behind moving the short-wavelength edge of the wavelength range used for measuring the 2.0-um band area to 1.945 um. Even though this cuts out a significant part of the 2.0-um absorption band, it also excludes strong telluric noise that was significantly affecting the integrated band area. The telluric noise is also the likely reason behind the lower integrated band areas of the 1.5-um bands, as it could affect the continuum region determined on the short-wavelength side of the absorption band (which includes the region of strong telluric absorption around 1.4 \micron).

\subsection{Ice Temperatures \label{ssec:icetempresults}}
\begin{figure}[ht!]
\centering
\makebox[\textwidth][c]{\includegraphics[width=\textwidth]{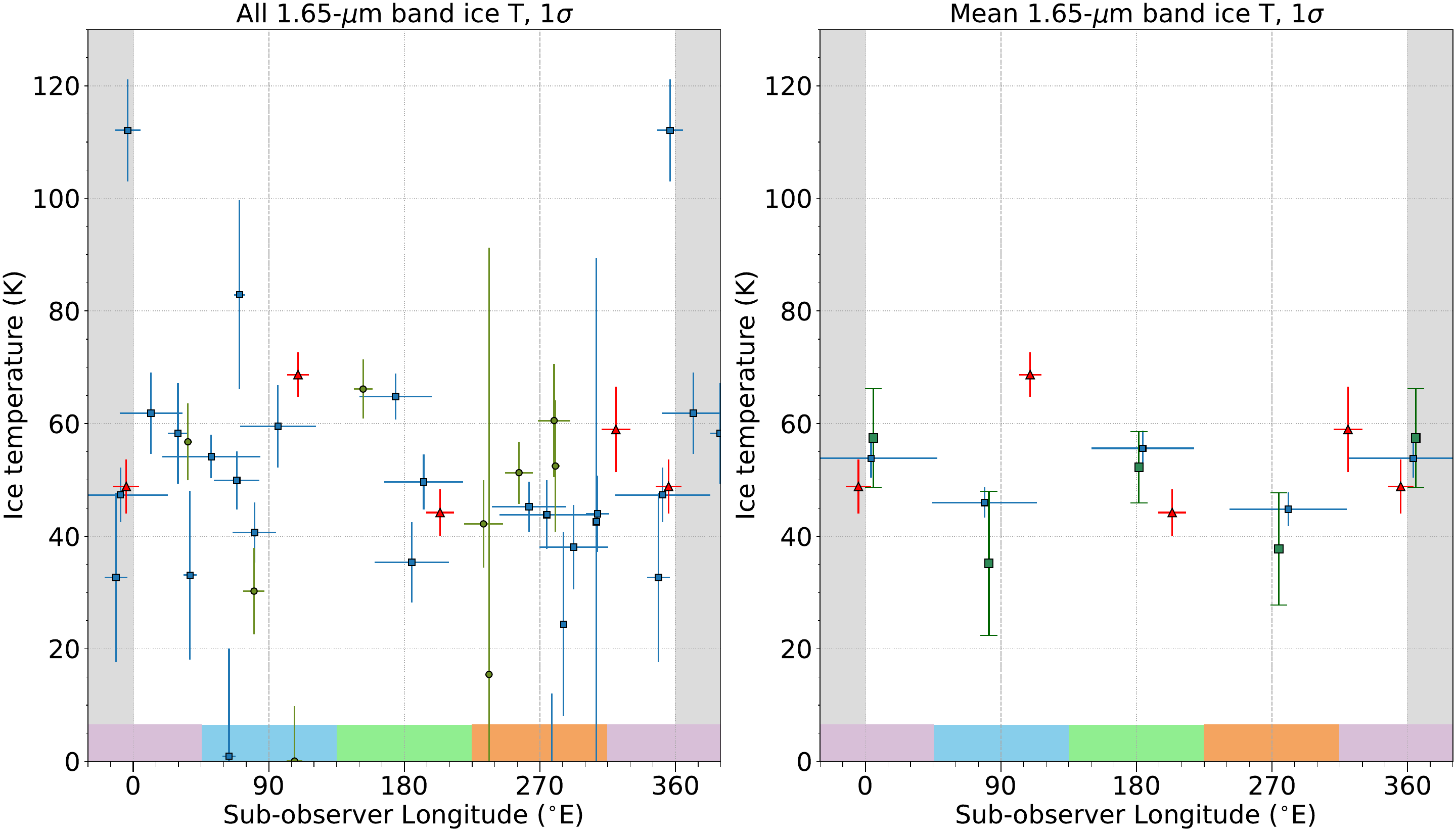}}
\vspace{-15pt}
\caption{\footnotesize \textit{(Left panel)}: A plot of the calculated H$_2$O ice temperatures for each spectrum versus longitude on Miranda. Blue squares are TripleSpec spectra, green circles are SpeX spectra, and red triangles are GNIRS spectra. \textit{(Right panel)}: The temperatures for the grand average TripleSpec spectra (blue squares), mean temperatures of all spectra for each quadrant (green squares), and temperatures of the GNIRS spectra (red triangles). Vertical error bars represent 1$\sigma$ errors in both panels. Other features are the same as in Figure \ref{fig:locations}.}
\vspace{-5pt}
\label{fig:IceTemps}
\end{figure}

Our high S/N spectra generally yield H$_2$O ice temperatures between 40--70 K, comparable to temperatures previously measured for the other Uranian satellites at comparable sub-observer latitudes on the southern hemisphere (37--50$\degr$ S) \citep{Grundy1999}. However, we also found conflicting results between our data sets, making it difficult to interpret. There is a wide spread of measured temperatures even among a single instrument, especially TripleSpec (Figure \ref{fig:IceTemps}, Table \ref{tab:specbands}). Additionally, some of the measured central wavelengths correspond to ice temperatures below 20 K, especially in low-S/N TripleSpec and SpeX PRISM spectra. The PRISM mode of SpeX only has a spectral resolution of R$\sim$95, and likely cannot detect these small wavelength shifts.

When only considering the GNIRS spectra, grand average TripleSpec spectra, and mean values of all spectra, the results are equally ambiguous (Figure \ref{fig:IceTemps}, Table \ref{tab:meanbands}). The GNIRS and TripleSpec grand average spectra show variation in temperature with longitude, but the datasets do not agree on the quadrants that show the highest/lowest temperatures.
However, this analysis assumes a simple relationship between ice temperature and the central wavelength of the 1.65-\um band, which is likely to be more complicated in disk-integrated spectra of planetary surfaces. Crystallinity variations induced by radiolytic processes, non-uniform spectral slopes, and the presence of other compounds with absorption features in the 1.6--1.7 \um range could affect the calculated band centers. Given the wide scatter between individual spectra and conflicting results between data sets, we find no convincing evidence for longitudinal variation in H$_2$O ice temperature on Miranda. 

\section{Discussion} \label{sec:discussion} 

Our results indicate that Miranda does not display a statistically meaningful leading/trailing hemispherical asymmetry in the strengths of H$_2$O ice absorption features, unlike the other major satellites in the Uranian system. In the cases in which a leading/trailing asymmetry is detected, it is only apparent in one or two measurements in a limited subset of our dataset. We do, however, find evidence for weaker 1.5-\um band H$_2$O ice bands in our spectra of the sub-Uranian quadrant of Miranda. This anti-Uranus/sub-Uranus asymmetry appears to be present to some degree in all of our data sets, but it is strongest in the TripleSpec spectra and weakest in the high S/N GNIRS spectra. The 1.65-\um and 2.0-\um band do not generally show statistically significant variations with longitude in either integrated band area or fractional band depth. In this section, we discuss potential causes for the lack of leading/trailing asymmetry, the anti-Uranus/sub-Uranus asymmetry, and differences between the 1.5-\um and 2.0-\um bands.

\subsection{Lack of Leading/Trailing Asymmetry} 

Previous studies have suggested that Miranda does not show as strong of a leading/trailing hemispherical asymmetry in H$_2$O ice band strengths compared to the other Uranian satellites \citep{Cart2018,Cart2020IRAC}. Leading/trailing asymmetries in H$_2$O ice band strengths have been identified in spectra on icy satellites in the Jovian, Saturnian, and Uranian systems \citep[e.g.][]{ClarkBrown1984,Calvin1995,Emery2005,Grundy2006}. As discussed in Section \ref{ssec:trends}, two processes are thought to be primarily responsible for this hemispherical dichotomy: heliocentric impactors and irradiation by charged particles in the host planet's magnetosphere. Over the age of the Solar System, a leading/trailing asymmetry should naturally develop on tidally-locked icy satellites. This asymmetry should be more pronounced on inner moons like Miranda, because these exogenic processes are likely to be more efficient closer to the host planet. 

Given that we do not see a leading/trailing asymmetry on Miranda, we must ask ``why not?'' Perhaps some of our base assumptions about the processes involved are incorrect. For example, multiple studies have suggested the possibility that Miranda has undergone polar reorientation (true polar wander) on one or more occasions, possibly driven by internal upwelling of low-density ice diapirs \citep{Plescia1988,Greenberg1991,Papp1993,Papp2013,Hammond2014,Bedd2020}. True polar wander by a similar process has been proposed to have occurred on Enceladus \citep{Nimmo2006,Matsuyama2007}. If such a reorientation occurred on Miranda, then its leading hemisphere may not have had a sufficient amount of time to build up a longitudinal asymmetry similar to the other Uranian moons. The reorientation process itself likely occurred on a timescale $\leq 10^5$ years \citep{Papp1994}, although not necessarily within $10^5$ years of Miranda's formation. Additionally, reorientation might have substantially shifted an existing leading/trailing asymmetry to Miranda’s sub-Uranus and anti-Uranus quadrants (see discussion in the next subsection).

\begin{figure}[ht!]
\centering
\makebox[\textwidth][c]{\includegraphics[width=0.6\textwidth]{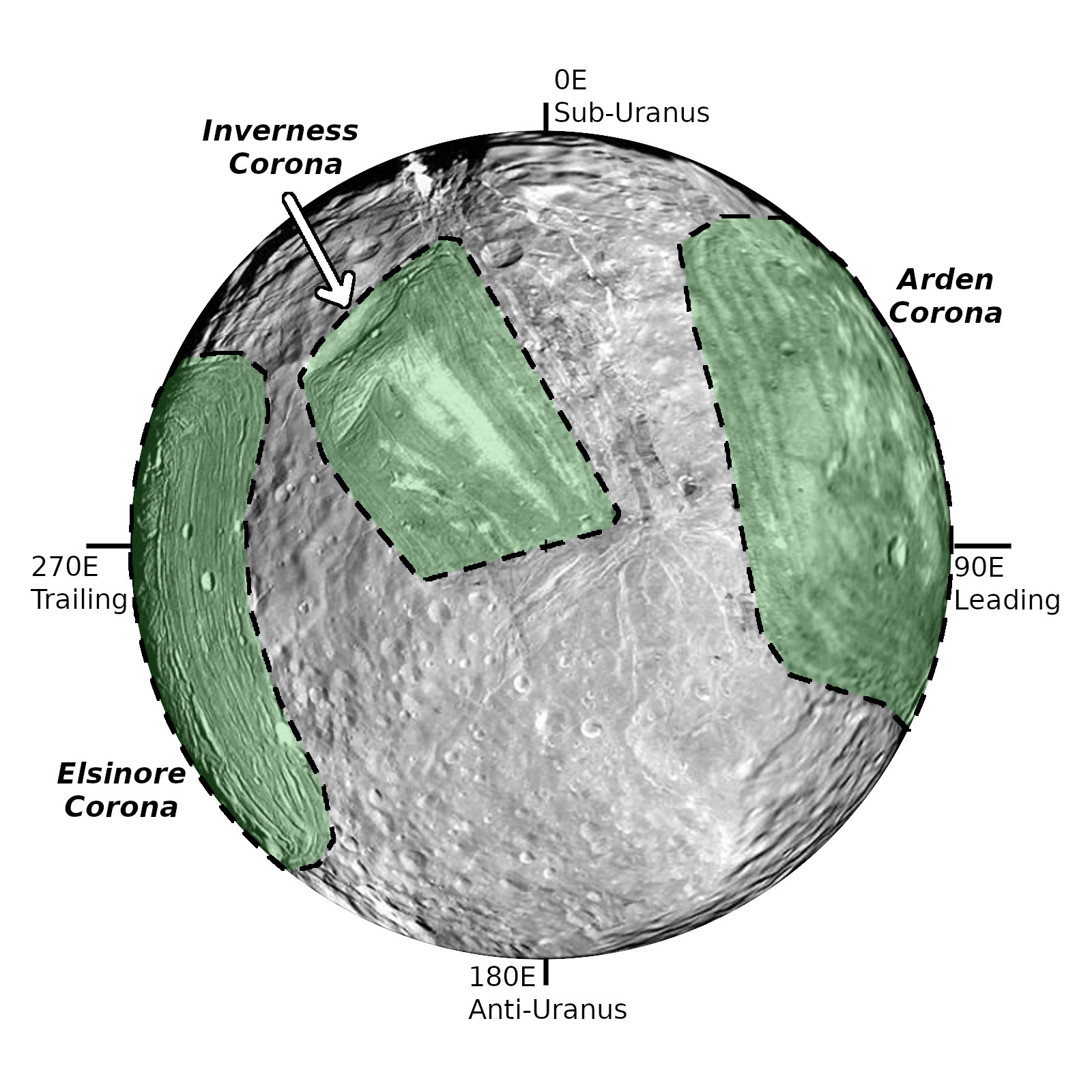}}
\vspace{-15pt}
\caption{\footnotesize \textit{Voyager 2}/ISS full-disk mosaic of the southern hemisphere of Miranda, including the three large coronae (highlighted in green and outlined with dashes), which were likely formed by tectonism and perhaps cryovolcanic processes as well. Miranda's south pole is at disk center. All but one of our spectra are located on the northern hemisphere, which was not imaged by \textit{Voyager 2}, so this figure is included as context for the discussion in Section \ref{sec:discussion}. Image credit: NASA/JPL/Caltech/USGS, PIA18185}
\vspace{-5pt}
\label{fig:MirMosaic}
\end{figure}

It is also possible that some other process or processes have reset or masked the leading/trailing asymmetrical signature. Many of the following processes we discuss are not longitudinally independent, and a combination of them may be required to fully explain the longitudinal variation (and lack thereof) that we see on Miranda. Consequently, this section is framed primarily as a discussion of possibilities and not as an authoritative explanation.

\subsubsection{Miranda's coronae}
The surface modification processes responsible for Miranda's coronae (Figure \ref{fig:MirMosaic}) are a plausible candidate for modifying any longitudinal asymmetry in H$_2$O ice, especially if they involved cryovolcanic emplacement of fresh ice or exposure of less-weathered ice. Processes that modify the distribution of ice grain sizes can produce differences in the depth of absorption bands \citep{Clark1983}. Arden and Elsinore Corona are both prominently located at leading and trailing longitudes of Miranda (respectively), an arrangement that matches both the predictions of low-order internal convection \citep{Papp1993,Papp1997,Hammond2014} and the positioning of heavily tectonized regions on Enceladus \citep{Papp2013}. 

Age estimates based on crater counts indicate that Elsinore Corona (trailing hemisphere) is likely the oldest of the three coronae, from $\sim$2 to $>$3.5 Gyr, while Arden Corona (leading hemisphere) and Inverness Corona (south pole) are much younger, between 100 Myr to 1 Gyr in age, with Arden Corona possibly being slightly older than Inverness Corona \citep{Zahnle2003,Kirchoff2022}. Although we cannot determine whether coronae are present on Miranda's northern hemisphere due to the lack of imaging coverage, Arden and Elsinore Corona should extend at least partway into the northern hemisphere. Perhaps the surface modification and/or resurfacing processes associated with Arden or Elsinore Corona refreshed the H$_2$O ice present on their respective hemispheres. However, Arden Corona's younger age implies that this would have strengthened the leading/trailing asymmetry, not erased it, if we assume that both coronae refreshed the spectral signature of H$_2$O ice as they formed. Furthermore, the cratering age estimates were made under the assumption that Miranda has maintained its current orientation for its entire history \citep{Zahnle2003}. Although Elsinore has fewer craters than Arden, its location on Miranda’s trailing hemisphere implies that it has experienced a much lower number of impact events over the age of the Solar System and is therefore much older than Arden. However, if Miranda underwent polar reorientation, then the relative age estimates between the coronae may not be correct, although all three coronae are clearly still younger than the surrounding cratered terrain in any scenario. 

\subsubsection{Mantling by fine-grained material}
In addition to geologic resurfacing \citep[e.g.][]{Smith1986,Greenberg1991,Schenk1991} and possible polar reorientation, Miranda's cratered terrain appears to have been partly buried by a large volume of fine-grained material, as shown by the mantled appearance of the subdued crater population identified on its surface \citep[e.g.][]{Smith1986,Greenberg1991}. The source of this material is unknown, and different hypotheses including burial by impact ejecta \citep{Croft1987}, ring particle accumulation \citep{Cart2018,Cart2020IRAC}, and burial by cryovolcanic plume material have been suggested \citep{Cart2020LPSC}. Depending on the age, volume, and distribution of the mantling material, these processes could help erase leading/trailing asymmetries on Miranda. However, mantling by impact ejecta is not a process unique to Miranda. Similar impact mantling should occur on the other classical Uranian satellites. The presence of leading/trailing asymmetries on those satellites has previously been established, so surface mantling by impacts is clearly not a major hindrance to developing such asymmetries. This suggests that if mantling with fine material does somehow contribute to the lack of leading/trailing asymmetry, that mantling process must be unique to Miranda.

Ring particles originating in Uranus's outermost and dusty $\mu$-ring could migrate outward and mantle Miranda. The $\mu$-ring is coincident with the orbit of the small (R$\sim$6--12 km) moon Mab, which is likely the primary source of $\mu$-ring material, possibly replenished from its surface in the form of impact ejecta \citep{Showalter2006}. The ring is likely composed primarily of fine-grained icy particles, and Mab may be an icy moon like Miranda \citep{dePater2006,Paradis2019}. If these ring particles -- whether dusty or icy -- were to become electrically charged, they could spiral outwards via the Lorentz force, and might reach Miranda’s orbit and collide with its surface. Depending on the velocity of these impacts, it is possible that $\mu$-ring grains could either overturn Miranda’s regolith and expose fresher ice, thereby enhancing H$_2$O ice bands, or mantle Miranda’s existing regolith, and possibly weaken H$_2$O absorption features \citep{Grundy2006}. If the ring particles are primarily composed of H$_2$O ice, then they could enhance Miranda’s H$_2$O ice bands via mantling as well. 
However, caution is warranted with this hypothesis, as the relationship between Miranda, the $\mu$-ring, and Mab is not well understood. If $\mu$-ring particles do migrate outwards sufficiently far to reach Miranda's orbit, their relative velocity is likely to determine their fate. Particles that have relative velocities similar to or greater than Miranda's orbital velocity will likely hit the moon's trailing side (and perhaps sub-Uranus side), while particles with slower relative velocities would preferentially hit Miranda's leading side (and perhaps also the sub-Uranus side).

Plume volcanism, similar to that seen on Enceladus, is another possible mechanism that could mantle Miranda’s craters in material and modify the spectral characteristics of H$_2$O ice on its surface. While Miranda is not known to be geologically active at the present time, the relatively young age of the coronae and Miranda's aforementioned similarities to Enceladus \citep[e.g][]{Papp2013,Bedd2020} raises the possibility that plume volcanism may have occurred in the past. For example, Inverness Corona at the south pole of Miranda may be as young as 100 Myr \citep{Zahnle2003}, and could be analogous to the currently-active south polar terrain on Enceladus. On Enceladus, plume deposition patterns dominate in the southern hemisphere, and trail off into the northern hemisphere in two swathes each offset about 40$\degr$ of longitude from the centers of the leading and trailing hemispheres \citep{Kempf2010}. Much less of this material would reach the mid-northern latitudes. This deposition pattern appears to be consistent with IR/UV albedo patterns on Enceladus's surface, likely indicating areas of larger grain size and therefore stronger water ice absorption \citep{Schenk2011,Scipioni2017}. 
More recent, updated modeling of the plume deposition pattern on Enceladus treated both discrete plumes and curtain-style eruptions. At a latitude of 45\degr N, the cumulative deposition rate ranges between 10$^{-6}$ and 10$^{-5}$ mm yr$^{-1}$, in primarily small, $\sim$1-\um grains \citep{Southworth2019}. While Miranda's surface does appear to be mantled in fine-grained water ice \citep{Cart2020IRAC,Cart2020LPSC}, the longitudes of peak deposition from Enceladus-style south polar plumes would correspond to longitudes of approximately 50$\degr$ and 230$\degr$ in Miranda's east-positive longitude system. There are no immediately apparent signatures of a similar pattern of deposition on Miranda, in terms of the strength of the H$_2$O ice band strengths presented here. However, any comparisons to Enceladus are only a rough guide, as plume activity on Miranda could have occurred at any, all, or none of the coronae. A detailed investigation of hypothetical plume deposition rates and locations on Miranda is needed to further explore this scenario, which is outside the scope of this study. 

\subsubsection{Lack of CO$_2$ ice}
An additional hypothesis was suggested by \citet{Cart2020IRAC}; that Miranda's {\it lack} of CO$_2$ ice deposits may be responsible for the lack of H$_2$O ice asymmetry. CO$_2$ molecules produced via radiolysis on Miranda and subsequently sublimated would follow a thermal velocity distribution, but Miranda's low mass (and therefore low escape velocity) implies that a majority of sublimated CO$_2$ molecules would be lost to Jeans escape before redeposition on the surface \citep{Sori2017}. The CO$_2$ ice detected on the trailing hemispheres of the other four Uranian satellites, particularly Ariel, appears to be present in thick, concentrated deposits and not as regolith-mixed CO$_2$ ice grains \citep{Grundy2003,Grundy2006,Cart2015,Cart2022}. These concentrated deposits of CO$_2$ ice could weaken the spectral signature of the underlying dirty H$_2$O ice on the trailing hemispheres of the other major Uranian moons. Consequently, the lack of CO$_2$ ice on Miranda could be influencing the strength of H$_2$O ice absorption bands between its leading and trailing hemispheres \citep{Cart2020IRAC}. This hypothesis is one of the more appealing options, as it supports the presence of leading/trailing asymmetries on the other Uranian moons, the decreasing asymmetry with increasing orbital distance, and the lack of asymmetry on Miranda. However, this does not rule out the presence of other processes (such as cratering and magnetospheric effects) that should produce leading/trailing asymmetries across all of the Uranian satellites, nor does it explain the anti-Uranus/sub-Uranus asymmetry on Miranda.

\subsection{Anti-Uranus/Sub-Uranus Asymmetry}
The anti-Uranus/sub-Uranus asymmetry hinted at by our spectra is more unusual, and it appears to be unique in the Uranian system. The aforementioned polar reorientation/true polar wander hypothesis could also play a role in the anti-Uranus/sub-Uranus asymmetry in H$_2$O band strengths on Miranda. The density of craters and the alignment of tectonic structures like rift systems is consistent with a 45--90$\degr$ longitudinal shift of the surface around the polar axis of Miranda at some point in its geologic history \citep{Plescia1988,Greenberg1991,Papp1993,Papp1994,Hammond2014}. Under this hypothesis, a leading/trailing asymmetry in H$_2$O band strengths could have formed on Miranda, but true polar wander subsequently relocated the formerly leading/trailing hemispheres to their present-day anti-Uranus/sub-Uranus positions.

\citet{Plescia1988} counted craters on Miranda's southern hemisphere in images collected by \textit{Voyager 2}, finding that Miranda shows an asymmetry in crater density (outside the coronae) similar to that expected from a typical leading/trailing distribution, but the ``leading'' point appears instead to be at anti-Uranian longitudes, which Plescia attributed to a $\sim$90$\degr$ reorientation about the rotation axis. \citet{Greenberg1991} found that the crater density on the sub-Uranian hemisphere is indeed less than the anti-Uranus hemisphere, but only for the younger, fresh craters and not the older, more subdued craters. \citet{Papp1993} and \citep{Papp1994} argue based on the analysis of tectonic features and the alignment of rift systems that Miranda underwent a polar reorientation similar to that suggested by \citet{Plescia1988}, but less drastic in that Miranda only rotated $\sim$45$\degr$ around the spin axis, with an additional minor poleward shift of $\sim$10$\degr$. This hypothesis still rotates the former apex towards modern anti-Uranian longitudes and the paleo-antapex towards modern sub-Uranian longitudes, which is consistent with our observations. The \citet{Plescia1988} and \citet{Papp1993} polar reorientation hypotheses are both potential explanations for both the lack of a leading/trailing asymmetry and the presence of an anti-Uranus/sub-Uranus asymmetry in H$_2$O ice bands. 

However, these hypotheses assume a significant longitudinal shift in the surface of Miranda relative to the rotational axis, with much smaller latitudinal components. In contrast, most models of true polar wander induced by density anomalies are instead characterized by primarily \textit{latitudinal} shifts, with little to no longitudinal changes \citep[e.g.][]{Nimmo2006,Matsuyama2007,Matsuyama2014}. True polar wander is driven by density anomalies, which align themselves to minimize the rotational moment of inertia. For an icy satellite, positive density anomalies would preferentially orient along the tidal axis towards the primary body (Uranus), while negative anomalies would align with the other two axes; if Miranda's coronae do represent the surface expression of low density diapirs, this would naturally drive them to the centers of the leading and trailing hemispheres and in the polar regions, consistent with where the coronae are located today. 

More recent work by \citet{Hammond2014} found that the positions of the observed coronae on Miranda are consistent with low-order convection driven by tidal heating of a satellite with no internal ocean. Their best-fit model suggests four roughly equally-spaced internal upwellings, which originally occurred at equatorial latitudes, may have formed the coronae. The density anomalies then cause Miranda to undergo true polar wander, in which the internal upwelling at the equatorial, sub-Uranian point rotated $\sim$60$\degr$ poleward to become what we now know as Inverness Corona. Under this hypothesis, the apex and antapex may not substantially change in longitude, but the equatorial regions of the anti-Uranus and sub-Uranus regions migrate to high latitudes in the northern and southern hemispheres, respectively. This prediction implies that there may be another, unobserved corona at high northern latitudes on Miranda’s anti-Uranus quadrant, possibly coincident with the anti-Uranus region that displays enhanced H$_2$O ice band strengths we report here. The current high northern latitude, sub-Uranian longitudes, meanwhile, would have originally been located near Miranda’s north pole, and may be old, cratered terrain. This model is more consistent with descriptions of primarily latitudinal true polar wander, yet could still potentially explain our anti-Uranus/sub-Uranus asymmetry; this then relies on the resurfacing of Arden and Elsinore Corona to explain the lack of leading/trailing asymmetry.

We therefore suggest that the anti-Uranus/sub-Uranus asymmetry we observe in our data may be spectral evidence of such a polar reorientation. Either the reorientations proposed by \citet{Plescia1988} and \citet{Papp1993,Papp1994} or the reorientation proposed by \citet{Hammond2014} could account for an anti-Uranus/sub-Uranus asymmetry and the lack of leading/trailing asymmetry. The hypotheses differ in the magnitude and the proposed axis of reorientation, and whether the anti-Uranus/sub-Uranus regions sensed by our spectra represent two heavily cratered regions or if the anti-Uranus region represents an additional corona on the northern hemisphere.

Another option to explain the enhanced band strengths at anti-Uranian longitudes invokes a large impact event on the northern hemisphere anti-Uranian quadrant. Studies of smaller, fresh craters on icy satellites elsewhere in the Solar System have shown that impacts can expose fresh H$_2$O ice, strengthening the 1.5-\um and 2.0-\um absorption features \citep[e.g.][]{Stephan2012}. The lack of imaging coverage of the northern hemisphere of Miranda means it is impossible to rule out a large impact crater with our current knowledge. 
However, an impact sufficient to produce the observed variation in our disk-integrated spectra would likely need to be very large or be composed of multiple smaller impact craters. A single impact origin would potentially be on the order of the impacts that produced Herschel on Mimas or even Odysseus on Tethys. There is some evidence for increased 1.5-\um or 2.0-\um band depths in these impact basins in \textit{Cassini} VIMS data, but it is complicated by insufficient spatial coverage in the VIMS data set and other hemispherical effects on the depths of H$_2$O absorption bands. \citet{Stephan2016} reported only minimal enhancement of band depths on the floor of the Odysseus basin on Tethys. \citet{Fila2022} found deeper 1.5-\um and 2.0-\um bands in both Herschel and Odysseus, while \citet{Scipioni2017DPS} reported that neither Herschel nor Odysseus showed evidence of enhanced absorption band strength. 
Both Herschel and Odysseus are likely ancient impacts dating to the early Solar System, so any ``fresh ice'' effect has been reduced by eons of space weathering. In order for a single impact to produce the observed effect on Miranda, it would therefore have to be both very large and relatively recent. 
This is certainly not out of the realm of possibility, but impacts of that size were most common early in the Solar System and the population of such impactors is generally expected to decrease with time. If we were to discover that Miranda's northern hemisphere was marked with a large, fresh impact basin, the implications for impactor distributions in the outer Solar System could be quite interesting. For now, we consider this option as possible, but unlikely.

Finally, another possibility is worth consideration: the Uranian magnetosphere is complex, and data from the {\it Voyager 2} flyby indicated that the dipole is both tilted from the Uranian spin axis by $\sim$60$\degr$ and offset from the center of the planet by $\sim$0.3 $R_U$ \citep{Ness1986}. Some models of the Uranian magnetosphere suggest that it has a proportionally much larger quadrupole moment than other magnetic fields in the Solar System \citep{Connerney1987}, and the offset tilted dipole model may not be sufficient to characterize magnetospheric effects at Miranda's orbital distance. However, no radiation dose maps of the Uranian satellites have been published, and it is currently difficult to investigate potential localized magnetospheric effects that could lead to a higher radiation dose in some latitude and/or longitude regions on Miranda. In the Jovian system, for example, detailed modeling of particle energies and ion species found that some ions preferentially bombard the poles of the moons instead of the trailing hemispheres \citep{Cassidy2013}. {\it Voyager 2} data indicates that protons are the only significant ions in the Uranian magnetosphere, along with a significant number of electrons \citep{Ness1986,Lanz1987}. The surface darkening time in the inner Uranian system may be as short as 10$^5$ years \citep{Lanz1987}, so any transient differences in localized radiation dose could become indistinguishable from the surroundings relatively quickly. More detailed modeling of moon-magnetosphere interactions are required to constrain if localized effects could be responsible for the anti-Uranus/sub-Uranus asymmetry.

\subsection{Differences Between 1.5-\um and 2.0-\um Absorption Bands \label{ssec:15v20}} 

Our results show that there is a statistically significant longitudinal variation in the strength of the 1.5-\um H$_2$O ice absorption band complex, but not in the 1.65-\um or 2.0-\um bands. Previous work \citep{Cart2018} showed that on the other Uranian moons, the leading/trailing asymmetry is stronger in the 1.5-\um band than in the 2.0-\um band (Figure \ref{fig:MeanRatios}), and this difference increases closer to the planet. The strengths of the 1.5-\um and 2.0-\um H$_2$O ice absorption bands are sensitive to grain size and contaminant mixing. In the idealized case of a dark contaminant, with a completely neutral spectrum, intimately mixed with H$_2$O ice, the 1.5-\um and 2.0-\um H$_2$O bands change in similar but distinct ways due to wavelength-dependent differences in the optical path lengths into H$_2$O ice. The 1.5-\um H$_2$O band samples longer optical path lengths than the 2-\um band and is therefore more sensitive to dark contaminant mixing and grain size effects \citep{ClarkLucey1984,ClarkRoush1984}. The mean optical path lengths for the Uranian satellites have previously been calculated to be $\sim$0.2 mm for the 1.5-\um band, and $\sim$0.1 mm for the 2.0-\um band \citep{Cart2018,Cart2022}.  Previous work showed that the regoliths of the Uranian satellites may be compositionally stratified (mantled with a layer of tiny H$_2$O ice grains) based on long-wavelength NIR spectra from SpeX and Spitzer IRAC photometry, which sample shallower ($<$0.05 mm) depths \citep{Cart2015,Cart2018,Cart2020IRAC}. As the 1.5-\um band is more sensitive to grain size effects than the 2.0-\um band, surface level changes (such as a coating of fine ice grains) may therefore confuse attempts at interpretations based on depth. It is possible that the H$_2$O ice band differences are influenced by changes in the surface or subsurface, but the lack of long-wavelength (3--5 \micron) spectra of Miranda makes this more difficult to constrain.

The 1.04-\um and 1.25-\um H$_2$O ice bands are even more sensitive to grain size effects and the presence of dark contaminants than longer wavelength bands, and in conjunction with the other bands therefore provide an important constraint on grain size and contamination, as has been used for the Saturnian and Jovian moons \citep[e.g.][]{Calvin1995,Emery2005}. However, in the Uranian system, the 1.04-\um and 1.25-\um ice bands are absent; this implies that a dark and neutral absorber is present in sufficient abundance to effectively suppress these bands \citep{Brown1983,BrownClark1984}. 

Grain size also plays a role in the relative depths of H$_2$O ice absorption bands. For H$_2$O grains with diameters between 10 \um and 1 mm, the relationship between grain size and the relative depths of the 1.5-\um and 2.0-\um bands is roughly linear \citep[e.g.][]{ClarkLucey1984,Fila2012}. This relationship is valid as long as the grains are much larger than the wavelength of light, and are therefore in the geometric optics regime. For larger grains and deeper, longer-wavelength absorption bands, a band can become `saturated' (optically thick) and no longer proportional to other bands \citep{ClarkBrown1984}. The morphology of the 1.5-\um and 2.0-\um H$_2$O bands in the Uranian system do not indicate that they are saturated, although low reflectance in the 3.0-\um region on the larger Uranian satellites suggests that the 3.0-\um H$_2$O ice band may be saturated \citep{Cart2018}. 

However, previous work showed that $\sim$1 \um (or smaller) grains are widespread on the surfaces of the Uranian satellites \citep{Avramchuk2007,Afanasiev2014}, with further evidence that at least some of these grains are composed of H$_2$O ice \citep{Cart2015,Cart2018,Cart2020IRAC}. Temperatures are low enough in the Uranian system that small ice grains should not experience efficient thermal sintering into larger grains \citep{Clark1983}. The effects of diffraction in grains on the order of or smaller than 1 \um can introduce spectral variability not seen in H$_2$O ice surfaces dominated by larger grains. \citet{Clark2012} investigated the presence of sub-micron ice grains (and sub-micron contaminants) on the surface of Iapetus, and discussed the general characteristics in the 1.0--2.5 \um region. The presence of sub-micron H$_2$O ice grains tends to reduce the strength of the 1.5-\um band, whereas the 2.0-\um band gets slightly stronger. Furthermore, the 2.0-\um band shape becomes asymmetrical, with the deepest part of the band center shifting to slightly longer wavelengths, and the 1.65-\um band is also deeper than usually seen in crystalline H$_2$O ice samples composed of larger grains. \citet{Scipioni2017} focused on mapping the presence of sub-micron ice grains on the surface of Enceladus, finding that the regions with the most plume deposits had the fewest spectral indicators associated with sub-micron grains.

Crystallinity and temperature variations can also affect the depths of some H$_2$O ice bands \citep{Grundy1998,Grundy1999}. Crystalline water ice at lower temperatures will have a stronger 1.65-\um absorption band. Our measured ``1.5-\micron'' band includes the 1.65-\um band, so variations in temperature or crystallinity could have an effect on our measured 1.5-\um band areas. However, these variations would not substantially affect our 1.5-\um depths, which are measured between 1.515--1.525 \micron. 

We found minimal evidence to suggest that the 1.65-\um band depths may show a similar anti-Uranus/sub-Uranus longitudinal pattern to that of the 1.5-\um band, based on the `all spectra' sinusoidal model F-test, but this is not strongly supported by the mean band measurement ratios. Our measured 1.5-\um band areas include the area of the 1.65-\um band, so this may imply that the process responsible for the 1.5-\um hemispherical asymmetry is limited to the 1.5-\um band itself, not the 1.65-\um band. Additionally, the asymmetry is seen in the 1.5-\um depth measurements, which are independent of the 1.65-\um band. It is possible that the anti-Uranus/sub-Uranus asymmetry in the 1.5-\um band represents an effect of scattered light contamination from Uranus which is not accounted for by our correction procedure. The sub-Uranus and anti-Uranus quadrants are more heavily affected by scattered light, as Miranda's projected angular separation from Uranus is smaller when these quadrants are visible from Earth (Figure \ref{fig:slitviewer}). Nevertheless, the lack of leading/trailing asymmetry on Miranda is a robust result whether the 1.5-\um band is included or not, as scattered light from Uranus drops off sharply at wavelengths longer than 1.6 \micron, and would have minimal effects on the 1.65-\um and 2.0-\um bands.

\section{Conclusions} \label{sec:conclusion}
Miranda's near-infrared spectrum resembles those of the other Uranian satellites, dominated by crystalline H$_2$O ice and a dark, spectrally neutral constituent. However, unlike the other Uranian satellites and many other satellites in the Solar System, Miranda does not show evidence for an asymmetry in the strength of the H$_2$O ice absorption bands between the leading and trailing quadrants or hemispheres. With our expanded longitudinal coverage of Miranda's northern hemisphere, our analysis found that Miranda shows hints of a longitudinal asymmetry in the strength of H$_2$O ice absorption in the 1.5-\um band, but this asymmetry is instead between the anti-Uranus and sub-Uranus quadrants, a situation that has not been previously observed on the other classical Uranian moons. 

We reviewed many mechanisms potentially responsible for controlling the spectral signature of H$_2$O ice with longitude. While it is difficult to disentangle the many possible contributing factors, one explanation may involve polar reorientation (true polar wander). Previous investigations of cratering and tectonic structural features provided supporting evidence for at least one occurrence of true polar wander on Miranda, which resulted in a primarily longitudinal rotation of $\sim$45--90$\degr$ of the paleo-apex (leading hemisphere) towards the anti-Uranian point, and the paleo-antapex (trailing hemisphere) towards the sub-Uranian point \citep{Plescia1988,Papp1993,Papp1994}. Alternatively, the proposed true polar wander of \citet{Hammond2014} would imply that our anti-Uranus/sub-Uranus asymmetry is the result of spectral differences between a northern-hemisphere corona at anti-Uranian longitudes and old cratered terrain at sub-Uranian longitudes.

Even with the polar reorientation hypothesis, there are still unresolved questions. Why does the detected anti-Uranus/sub-Uranus longitudinal asymmetry occur in the 1.5-\um band, but not the 1.65-\um or 2.0-\um band? Are NH$_3$/NH$_4$-bearing species or other volatiles like CO$_2$ ice present on Miranda? How much does the Uranian magnetosphere play a role in creating longitudinal asymmetries in H$_2$O ice in the Uranian system? Some other icy bodies in the Solar System (like Europa and Enceladus) have been proposed to have undergone a similar reorientation event, but they still show leading/trailing asymmetries even with very young surfaces. If the reorientation hypothesis is correct, does this imply that Miranda's reorientation was recent, that the surface modification processes in the Uranian system are much slower, or some combination thereof?

In the long term, a deeper understanding of Miranda's unusual characteristics can only be achieved with a return to the Uranian system with an orbiter mission. An orbiter equipped with a high-resolution camera and (V)NIR imaging spectrometer would enable thorough, spatially-resolved investigations of geology and surface composition of the Uranian satellites, while a particle and plasma instrument package and magnetometer measurements would provide insight into the Uranian radiation environment, moon-magnetosphere interactions, and constrain the possible presence of internal oceans \citep[e.g.][]{Hofstadter2019,Cart2021,Leonard2021}. In order to study the northern hemispheres of the Uranian satellites, such a mission must arrive in the next few decades, as the Uranian system passes through northern summer solstice in 2030 and towards the next equinox in 2050. Without a spacecraft mission to Uranus, ground-based and space-based telescopes will remain our only method of studying the Uranian system, inherently limiting our potential understanding of these moons.
\pagebreak
\section*{Acknowledgements}
\begin{acknowledgments}
We wish to extend our gratitude to the observing, engineering, and administrative staff at Apache Point Observatory, Gemini North, and the IRTF, without whom this project would not have been possible, and to the anonymous reviewers whose comments improved this manuscript. In no particular order, we also would like to thank Jon Holtzman, Jodi Berdis, Ted Roush, Amanda Townsend, John Wilson, and Michael Cushing for advice and useful discussions. Matthew Varakian and Audrey Dijeau observed Miranda for us on UT 2019/10/21 and UT 2019/10/26, respectively.

This work is funded under NASA FINESST grant 80NSSC20K1378, and parts of it were previously funded under the NMSU Astronomy Department's William Webber Voyager Graduate Fellowship. This work is primarily based on new observations obtained with the ARC 3.5-meter telescope at Apache Point Observatory, which is owned and operated by the Astrophysical Research Consortium. This work also incorporates observations previously obtained at the Infrared Telescope Facility (IRTF), which is operated by the University of Hawaii under contract 80HQTR19D0030 with the National Aeronautics and Space Administration. 

Finally, this work is also based in part on new observations obtained under program IDs GN-2020B-FT-205 and GN-2021B-FT-210 at the international Gemini Observatory, a program of NSF’s NOIRLab, which is managed by the Association of Universities for Research in Astronomy (AURA) under a cooperative agreement with the National Science Foundation on behalf of the Gemini Observatory partnership: the National Science Foundation (United States), National Research Council (Canada), Agencia Nacional de Investigaci\'{o}n y Desarrollo (Chile), Ministerio de Ciencia, Tecnolog\'{i}a e Innovaci\'{o}n (Argentina), Minist\'{e}rio da Ci\^{e}ncia, Tecnologia, Inova\c{c}\~{o}es e Comunica\c{c}\~{o}es (Brazil), and Korea Astronomy and Space Science Institute (Republic of Korea). The Gemini North and IRTF telescopes are located within the Maunakea Science Reserve and adjacent to the summit of Maunakea. We are grateful for the privilege of observing the Universe from a place that is unique in both its astronomical quality and its cultural significance, and wish to emphasize our respect for the Native Hawaiian community's cultural and historical ties to Maunakea.

This research has made use of the SIMBAD database, operated at CDS, Strasbourg, France.

\end{acknowledgments}

\vspace{5mm}
\facilities{ARC(TripleSpec), Gemini:Gillett(GNIRS), IRTF(SpeX)}

\software{alpha\_H2O IDL routine \citep[][\url{http://www2.lowell.edu/users/grundy/abstracts/ice/alpha_H2O.pro}]{Grundy1998}, AstroPy \citep{Astropy2013}, Astroquery \citep{astroquery}, NumPy \citep{numpy}, SciPy \citep{scipy}, Matplotlib \citep{Matplotlib}, IRAF \citep{IRAF}, JPL Horizons Online Ephemeris Service (\url{https://ssd.jpl.nasa.gov/horizons/}), Spextool \citep{Cushing2004,Vacca2003}, SIMBAD \citep{SIMBAD}, SpectRes \citep{Spectres}, }

\appendix

\section{Uranus scattered light correction} \label{adx:uranus}
Miranda's proximity to Uranus leads to significant contamination from scattered light in the spectrograph slit. This contamination is strong in the region between 1.5 and 1.6 \micron, overlapping the 1.5-\um H$_2$O ice absorption band complex. Below, we discuss our scattered light removal procedure.

\subsection{The spectrum of Uranus}
\begin{figure}[ht!]
\centering
\makebox[\textwidth][c]{\includegraphics[width=\textwidth]{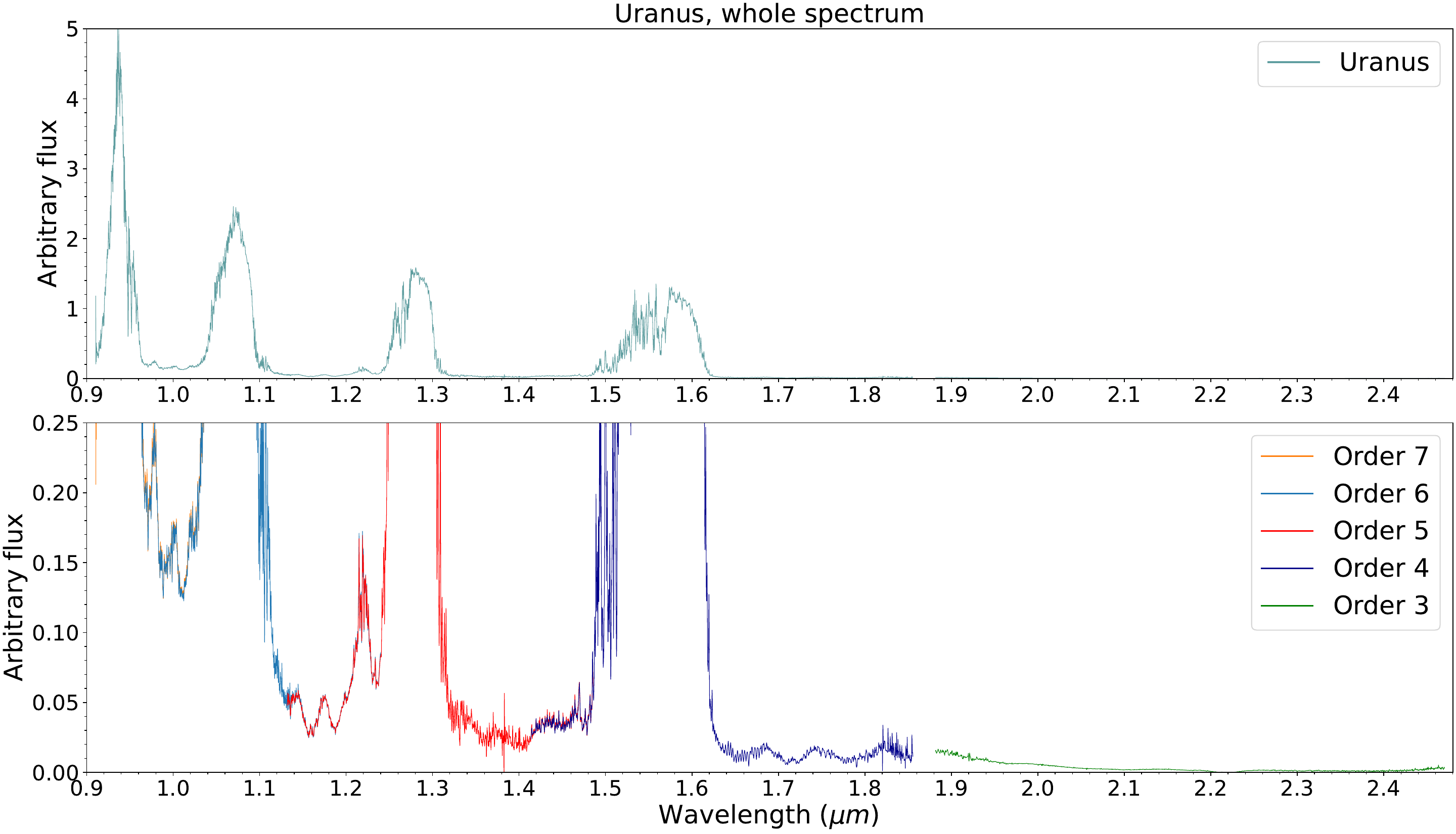}}
\vspace{-10pt}
\caption{\footnotesize The spectrum of Uranus as observed with TripleSpec, demonstrating the complex structure of methane absorption both within and outside of atmospheric windows. Note the change in scale on the y-axis between plots, the strong contamination between 1.48--1.63 \um and the minimal flux in K-band (order 3).}
\vspace{-5pt}
\label{fig:uspectrum}
\end{figure}

Scattered light from Uranus primarily contaminated our Miranda spectra through four `peaks' in the Uranus spectrum, as shown in Figure \ref{fig:uspectrum}. These `peaks' are in reality atmospheric windows between strong methane absorption in the Uranian atmosphere. This contamination was strongest between 0.90--0.965 \micron,
1.02--1.12 \micron, 1.24--1.32 \micron, and 1.48--1.63 \micron. The last peak coincides with the H$_2$O ice band complex between 1.5 and 1.7 \micron. At longer wavelengths than this, there was little obvious contamination in our Miranda spectra; the atmospheric methane window at 2.04 \um is almost completely absent on Uranus due to the pressure-induced H$_2$ dipole \citep{FinkLarson1979}. Consequently, Uranus is about a factor of 100 fainter in K-band compared to its 1.5-\um peak, and in K-band imaging the Uranian rings rival the brightness of the planet itself \citep{Baines1998}. This is easily seen in the TripleSpec K$_s$-band slitviewer image in Figure \ref{fig:slitviewer}.

In general, the peaks at shorter wavelengths are stronger, and in the `troughs' between them, there is residual light about a factor of 10--20 fainter, with a complex and non-uniform structure to the spectrum. The residual light in the troughs also has a blueward slope and is stronger in shorter wavelength orders. This slope is dominated by Rayleigh scattering, with the `wavy' structure being due to H$_2$ pressure-induced absorption \citep{FinkLarson1979}.

\subsection{Scattered light correction}

Our TripleSpec spectra of Miranda resisted simple attempts at correction of scattered light by subtraction of a single scaled spectrum of Uranus. Subtracting a spectrum scaled by a factor determined for any one `peak' of contamination produced significant residuals on the other peaks. 

\begin{deluxetable}{lcccc}
\tablecaption{Uranus correction wavelengths\label{tab:uranuswaves}}
\tablehead{\colhead{Spectral order} & \colhead{Left continuum} & \colhead{Right continuum} & \colhead{``Peak'' range} & \colhead{Scale fit range}\\ 
\colhead{} & \colhead{($\mu m$)} & \colhead{($\mu m$)} & \colhead{($\mu m$)} & \colhead{($\mu m$)} }
\startdata
4 (H-band) & 1.455--1.465 & 1.708--1.725 & N/A\tablenotemark{a} & 1.480--1.625 \\
5 (J-band) & 1.185--1.190 & 1.320--1.330 & 1.2750--1.2805 & 1.235--1.330 \\
6 (yJ-band) & 1.005--1.015 & 1.155--1.163 & 1.0687--1.0767 & 1.020--1.110 \\
7 (y-band) & 0.988--0.989 & 1.008--1.012 & 0.933--0.940 & 0.922--1.063 \\
\enddata
\tablenotetext{a}{\footnotesize The peak of order 4 was not used to generate a first guess scale factor; the first guess from order 5 was used for this purpose.}
\tablecomments{\footnotesize For each of the spectral orders in the TripleSpec data, we tabulate the wavelengths we used to define the continuum regions and the strength of peak contamination, used for calculating a first guess scale factor for Uranus correction. We also include the wavelength range that we used for the linear least squares scale fitting, which approximates the wavelength ranges in which scattered light from Uranus is noticeable in the Miranda spectra. Order 3 (K-band) is omitted, as contamination is negligible in this wavelength range.}
\end{deluxetable}

To achieve an adequate correction across the entire spectrum, we fit and subtracted a spectrum of Uranus separately from each spectral order of the TripleSpec data (Table \ref{tab:uranuswaves}). We independently corrected multiple groups of Miranda exposures per night (the same groups of 4--16 exposures used for telluric correction). These telluric-corrected, Uranus-subtracted groups are then combined with a weighted average to create the final full-night spectrum. This largely mitigates the effects of an over-subtracted or under-subtracted Uranus correction on any one individual group. We determined a `first-guess' scale factor by measuring a `continuum' level in the Miranda spectra on either side of the regions of strongest contamination. We then measured the flux in both the Uranus and Miranda spectra at the wavelengths with the strongest contamination, at the `summit' of the `peak'. We determined our first-guess scale factor $s_1$ with
\begin{equation} 
s_{1} = \frac{P_M - C_M}{P_U}  
\end{equation}
where $P_M$ is the flux observed from Miranda at the `peak', $P_U$ is the flux observed from Uranus at the same peak, and $C_M$ is the average of the flux in the two continuum regions of Miranda's spectrum on either side. This process was applied for each order separately. Spectral order 4 (H-band) includes the 1.5-\um H$_2$O ice band, and determining a `continuum' was more difficult; we therefore used the initial guess from order 5. The scattered light contamination was minimal in order 3 (K-band), and we therefore also used the guess from order 5 for consistency.

\begin{figure}[ht!]
\centering
\makebox[\textwidth][c]{\includegraphics[width=0.7\textwidth]{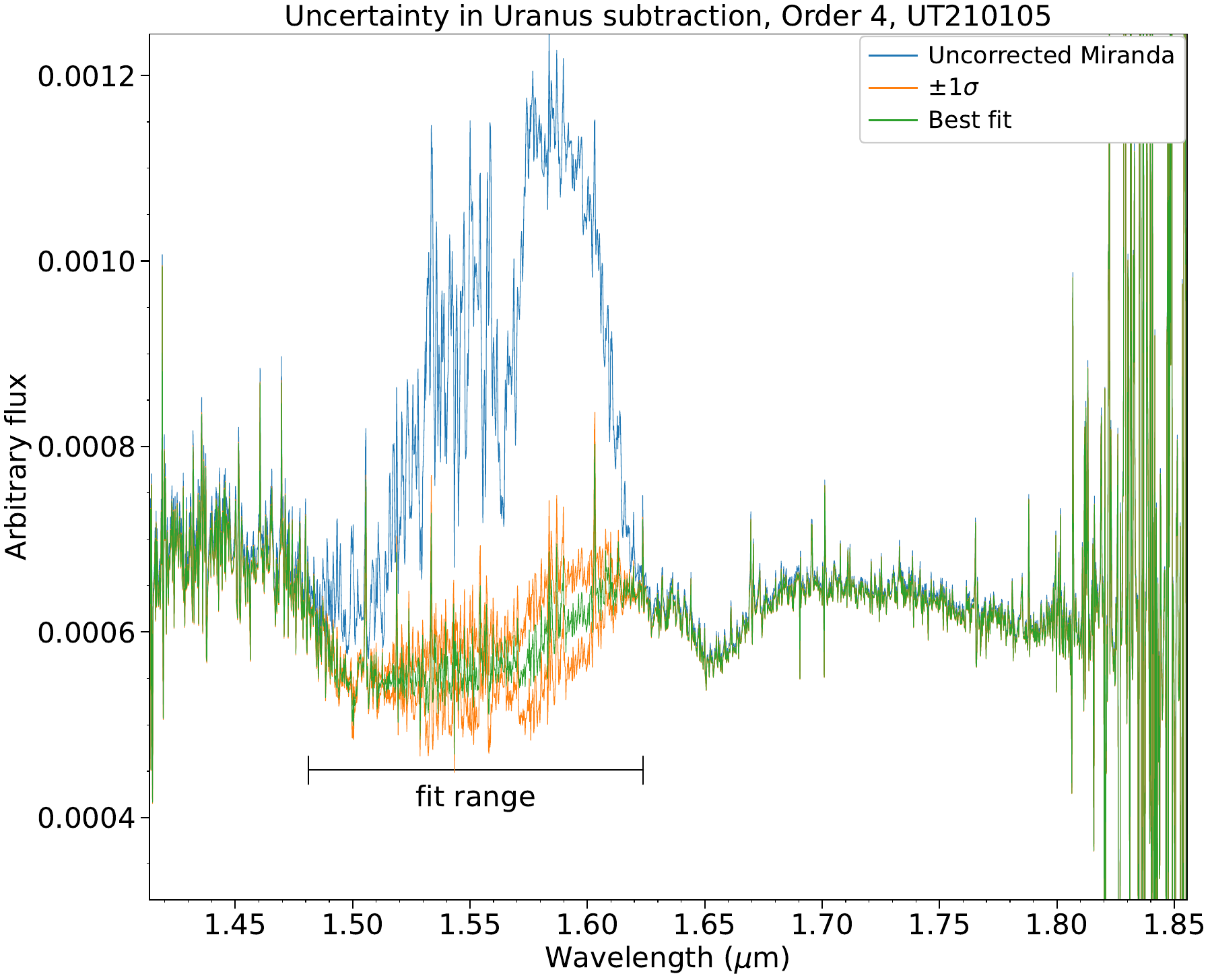}}
\vspace{-10pt}
\caption{\footnotesize We demonstrate the uncertainty in the derived scale factor in order 4 (H band) and its effect on the resulting Uranus-subtracted spectrum. The blue spectrum is the `observed' Miranda spectrum from UT210105, the green spectrum is the Miranda spectrum after subtracting Uranus (best-fit scale factor), and the orange spectra represent the $\pm$1$\sigma$ uncertainty in the scale factor. We note that we corrected a full-night spectrum in this figure for clarity, but corrections for the data used for analysis in this work were applied to multiple groups of telluric-corrected spectra per night.}
\vspace{-5pt}
\label{fig:scalefactorunc}
\end{figure}

Using a preliminary subtracted spectrum with these first guesses for all orders, we then found sub-pixel wavelength shifts of the Uranus spectrum relative to the Miranda spectrum, minimizing the standard deviation in the regions of interest. We then applied a linear least squares algorithm to find the best fit for the scale factor, assuming that 
\begin{equation}
M = O - sU
\end{equation}
where $M$ is the `true' spectrum of Miranda, $O$ is the `observed' spectrum of Miranda, $U$ is the spectrum of Uranus, and $s$ is the best-fit scale factor. The scale factors found in this method were often close to the first guesses. 
The error in this derived scale factor was estimated with:
\begin{equation}
\sigma_{s}^2 = \frac{RSS}{N - 3}
\end{equation}
where $RSS$ is the sum of squared residuals and $N$ is the number of data points in the fit region. The uncertainty in this scale factor is demonstrated in Figure \ref{fig:scalefactorunc}. The wavelength-dependent error arrays in the output, Uranus-corrected spectra were defined by:
\begin{equation}
\sigma_{M}^{2} = s^{2}\sigma_{U}^{2} + U^{2}\sigma_{s}^{2} + \sigma_{O}^{2}
\end{equation}
where $\sigma^{2}$ represents the variance in each factor. The variances for the input spectra are the variance arrays output from TripleSpectool.
This method of accounting for potential error in the scale factor was represented in the wavelength-dependent variance arrays as a decrease in the S/N in the regions where the Uranus contamination was strong, and with effectively no change in regions where it was weak.

\begin{figure}[ht!]
\centering
\makebox[\textwidth][c]{\includegraphics[width=\textwidth]{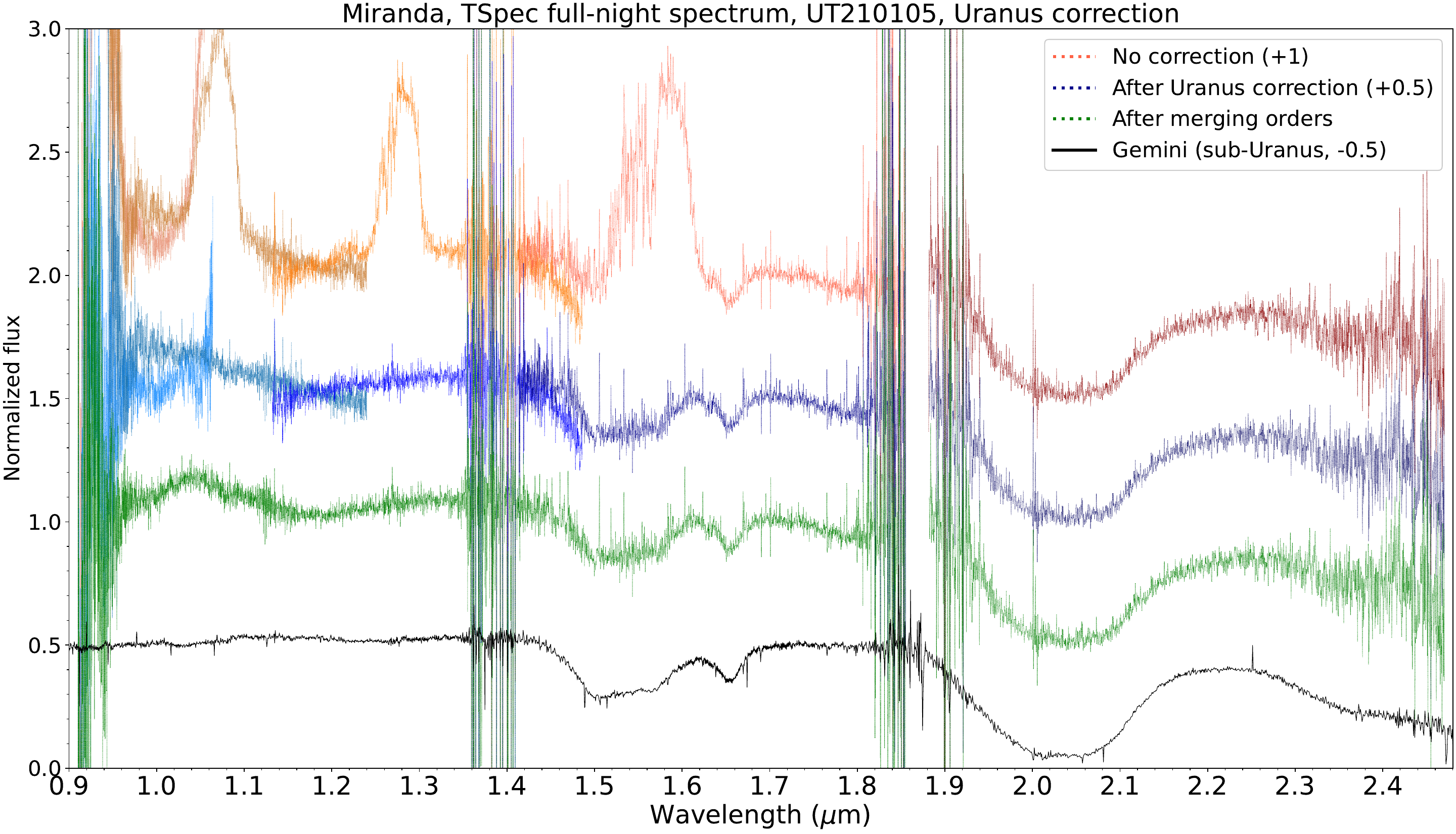}}
\vspace{-10pt}
\caption{\footnotesize We use the TripleSpec spectrum of Miranda from UT210105 to demonstrate our ability to correct the scattered light from Uranus and retrieve the shape of the 1.5-\um H$_2$O ice band. From top to bottom, we plot the different spectral orders in the uncorrected and unmerged spectrum (in shades of orange), in shades of blue the Uranus-corrected and unmerged spectrum, in green the Uranus-corrected and merged spectrum, and in black the GNIRS sub-Uranus spectrum, which was acquired at similar sub-observer longitudes. While the `peaks' of contamination are effectively removed, some spurious spectral slopes remain between different spectral orders ($<$1.25 \micron), which propagate into the final, merged spectrum and create the 0.95--1.15 \um `bump'. All spectra are normalized to unity at 1.72 \um and offset in increments of 0.5.}
\vspace{-5pt}
\label{fig:uranuscorrection}
\end{figure}

The resulting correction was successful at retrieving the shape of the 1.5-\um H$_2$O ice band complex and minimizing the contamination in other spectral orders (Figure \ref{fig:uranuscorrection}). These final output spectra were then merged together with the TripleSpecTool {\it xmergeorders} software routine. 

\subsection{The short-wavelength `bump'\label{adx:bump}}
As previously noted, our corrected TripleSpec spectra retained a wide `bump' between 0.95--1.15 \micron. The direct cause of this spurious bump was due to merging overlapping spectral orders with differing spectral slopes. Many of our Miranda spectra showed alternating positive and negative spectral slopes in spectral orders 5, 6, and 7. When we merged the regions of overlapping spectral coverage, the difference in slopes between orders was interpreted as a bump with inflection points where these orders overlapped. These spectral slopes were apparent both before and after application of the scattered-light procedure discussed above, and they were not visibly affected by that procedure (Figure \ref{fig:uranuscorrection}).

However, the underlying cause of these alternating slopes in the spectral orders is not clear.
The strength of this bump and the underlying spectral slopes appears to be correlated with the level of contamination of our spectra by scattered light from Uranus. These spectral slopes were present in observations of Miranda (and to a lesser extent Ariel), but were not apparent in our spectra of other objects, including the other Uranian satellites, asteroids, or standard stars. On one occasion, we angled the TripleSpec slit such that we could observe Titania simultaneously with Miranda. The bump was apparent in that spectrum of Titania, but was not present in any of the other spectra we obtained of Titania without Miranda, in which Uranus was typically much farther from the slit. The apparent connection between the strength of scattered light from Uranus and the presence of the bump suggests that it is not a real feature in the spectrum of Miranda, an assertion supported by the absence of the bump in previously-published Miranda spectra or our new GNIRS spectra.

Possible causes for this effect could include low-level nonlinearity in the response of the detector, although we took care not to observe any of our targets or telluric standard stars with exposures long enough to reach a highly nonlinear regime, as there is no nonlinearity correction algorithm tailored for TripleSpec. Slit losses from imperfect guiding or atmospheric dispersion are known to induce spurious slopes or spectral mismatches, but are unavoidable for our observations due to the difficulty of maintaining the slit at a parallactic angle in the proximity of Uranus. The spatial location of Miranda's spectrum in the slit could potentially be traced incorrectly during the spectral extraction procedure, possibly biased by the presence of scattered light. It is also possible that the spectral slopes may be a manifestation of cross-order contamination by scattered light. Similar scattered light crossing spectral orders can be seen in other echelle-format spectrographs observing bright objects, but TripleSpec's orders are spaced much farther apart on the detector than in a typical high-resolution echelle spectrograph. As we did not understand the underlying cause of this `bump' and it did not affect the wavelength ranges that were the primary concern for this work, we did not attempt to correct for it. 

\section{Telluric Standards\label{adx:telluric}}
\begin{deluxetable*}{lccccll}
\tablecaption{TripleSpec Telluric Standards\label{tab:telluric}}
\tablewidth{0pt}
\tablehead{
\colhead{Name} & \colhead{Alt. Name} & \colhead{Sp. Type} & \colhead{B - V} &
\colhead{V - K} & \colhead{A0 star} & \colhead{Notes} \\
\colhead{} & \colhead{} & \colhead{} & \colhead{mag} &
\colhead{mag} & \colhead{} & \colhead{}
}
\startdata
Sun & \nodata & G2 V & 0.62 & 1.49 & \nodata & values included for comparison \\
HD 16275 & HIP 12198 & G5 & 0.65 & 1.51 & HD 15983 & also used for Gemini observations\\
HD 15942 & HIP 11941 & G0 & 0.66 & 1.48 & HD 15983 & \\
HD 13545 & HIP 10287 & G0 & 0.61 & 1.42 & HD 15983 & \\
HD 16017 & SAO 110618 & G2 V & 0.66 & 1.46 & HD 17544 & \\
HD 19061 & BD+14 507 & G2 V & 0.60 & 1.63 & HD 20086 & \\
HD 28099 & Hyades 64 & G2 V & 0.66 & 1.57 & HD 27761 & \\
BD+15 4915 & U Peg & G2 V & 0.63 & 1.57 & HD 225001 & W UMa-type eclipsing binary \\
HD 224251 & HIP 118013 & G2 V & 0.60 & 1.59 & HD 27761 & \\[0.25cm]
\enddata
\tablerefs{SIMBAD - \citet{SIMBAD}}
\tablecomments{Designations, spectral types, and colors were referenced from the SIMBAD database. The A0-type stars listed were used for correction of the G-type stars for display in Figure \ref{fig:telluricstandards}.}
\end{deluxetable*}

We used a variety of G-type stars as combined solar and telluric standards for our TripleSpec observations. Theoretically, the spectrum of an observed Solar System object can be divided by a spectrum of an early G-type star observed at a similar time and airmass, which should correct atmospheric absorption, the instrumental throughput curve, and solar spectral features all at once and result in a clean spectrum of the target object. Unfortunately, it can be difficult to locate well-characterized G2V stars of the appropriate brightness and proximity to a target in the sky. We prioritized correction of telluric absorption over correction of solar spectral features, and we therefore selected stars as close as possible to Uranus in the sky, with spectral types in the literature ranging from G0 to G5 (Table \ref{tab:telluric}). We observed our telluric standards at least once an hour, so the airmass difference between a group of Miranda frames and one or more standard star spectra was typically $<$0.1.

During the fall 2019 observing season, our telluric standards were primarily BD+15 4915, HD 16017, and HD 19061. BD+15 4915 was discarded from our standard list upon realizing it was a variable star. On one night we observed HD 224251, and we observed HD 28099 on two nights, although HD 28099 was much farther from Uranus in the sky. We noted that while HD 28099 is known to be a reasonably good solar analog star, using it for telluric correction also produced the strongest mismatches to solar absorption lines of any of our standard stars, possibly due to flexure of the instrument between significantly different pointing positions. While a sub-pixel wavelength shifting routine can correct flexure-induced shifts in spectral lines, doing so would also cause greater mismatches between the target object spectrum and telluric absorption features, which we judged to be of higher priority. 

Between our 2019 and 2020 observations, we changed our selection of standard stars to three G-type stars that were closer to Uranus in the sky, aiming to improve our atmospheric corrections. While we allowed a wider range of spectral types (G0 to G5), we chose stars with B-V and V-K color indices similar to those of the Sun in an attempt to minimize any spurious spectral slopes that might be introduced by spectral type mismatches. In general, these stars (HD 13545, HD 15942, and HD 16275) were suitable analogs, but we noticed spectral slope differences in HD 13545 and HD 15942 during data reduction. Partway through our fall 2020 observations, we switched to solely using HD 16275 as our primary correction star. HD 16275 was our preferred choice because it has near-solar colors and spectral slope, relatively weak non-solar spectral features, was only a few degrees from Uranus during that season, and was not bright enough to saturate the GNIRS detector. However, it is bright enough that observations as a telluric standard could be completed quickly and frequently throughout the night without spending significant amounts of observing time that could otherwise be used for Miranda. 

As part of our observing campaign, on a few nights we also observed nearby A0 stars. A0 stars are more commonly used for telluric absorption correction in non-Solar System astronomy, as they have clean spectra with few metal lines. One example of this correction process is implemented in the \textit{xtellcor} routine of the Spextool/Triplespectool software package \citep[see][]{Vacca2003,Cushing2004}. In brief, \textit{xtellcor} models the intrinsic spectrum of an A0 star, scaling by given B and V magnitudes to approximate the effective temperature and received flux. The routine then divides the spectrum of the observed A0 star by the model, which results in an atmospheric absorption spectrum (with instrumental throughput effects) without any stellar features. Dividing the spectrum of the target by this atmospheric spectrum retrieves the intrinsic spectrum of the target object. 

\begin{figure}[ht!]
\centering
\makebox[\textwidth][c]{\includegraphics[width=0.8\textwidth]{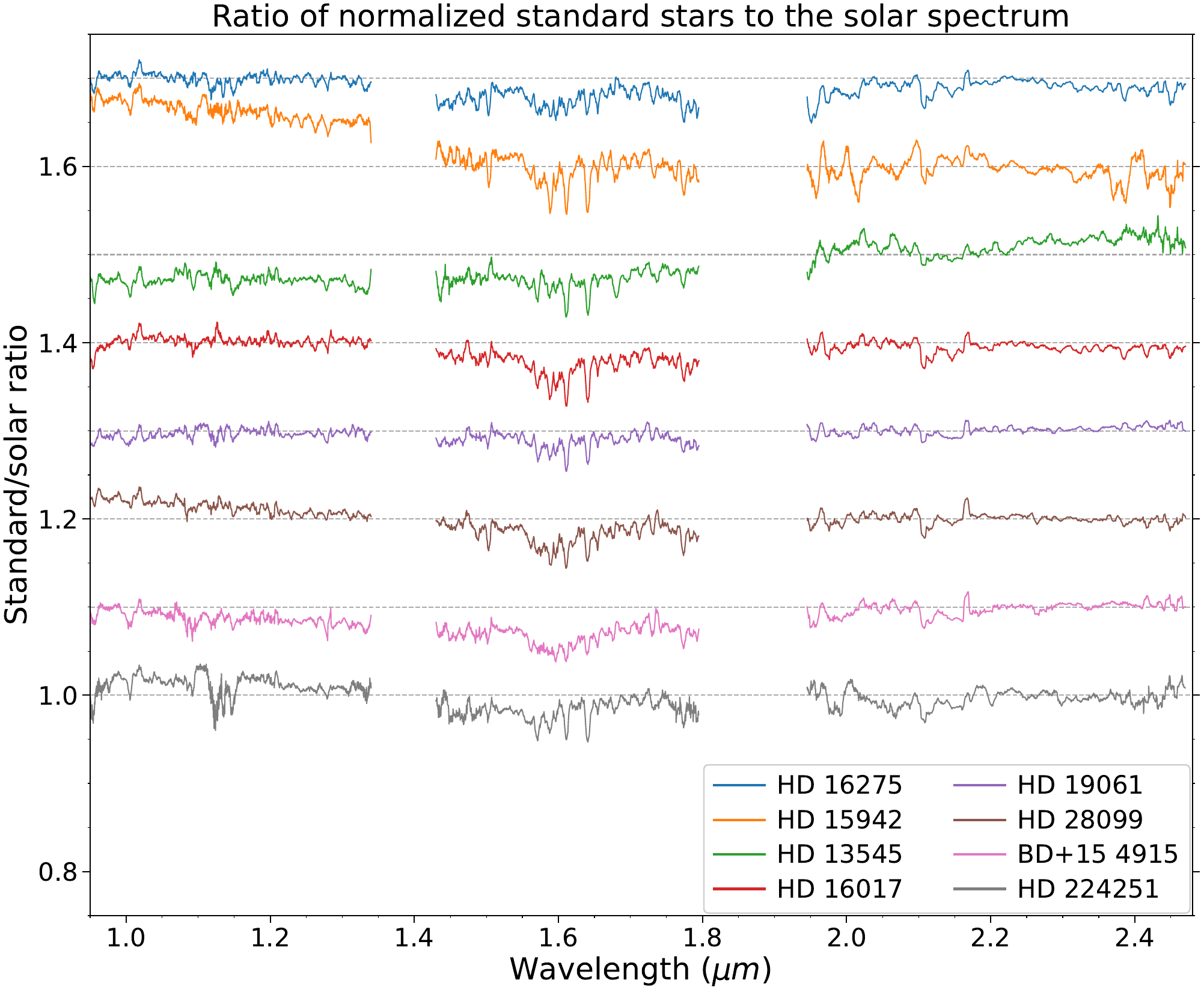}}
\vspace{-15pt}
\caption{\footnotesize This figure plots TripleSpec spectra of our G-type telluric standard stars, divided by the spectrum of the Sun. All spectra are normalized to unity between 2.21--2.225 \micron, and are offset in increments of 0.1. For the purposes of clarity, the spectra have been substantially smoothed by a 25-pixel boxcar kernel.}
\vspace{-5pt}
\label{fig:telluricstandards}
\end{figure}

We used our A0 star observations to correct our observations of our G-type standards on the same night. Our goal was to check their similarity to the solar spectrum and therefore their suitability for use as telluric standards for Solar System observations. While this introduces additional uncertainty (as the observed A0 star is not necessarily a perfect match to the model), it serves as a reasonable first-order check of the spectral properties of our standards. We plot reduced, A0-corrected spectra of our G-type standard stars in Figure \ref{fig:telluricstandards}. Observations of each standard star were corrected using an A0 star observed close in time and proximity in the sky. We then normalized the spectrum to unity between 2.21--2.225 \um and divided by a similarly normalized spectrum of the Sun, using a sub-pixel wavelength shifting routine to minimize spectral line mismatches.

While we were not aware of this work until after our observations were complete, \citet{Lewin2020} ranked a large number of northern hemisphere stars based on their suitability for use as solar analog stars in near-infrared observations of Solar System objects. Although we note that this determination was generally made using SpeX PRISM spectra of much lower resolution (R$\sim$200) than our TripleSpec data, three of the G-type stars we chose for our observations are also listed in Lewin et al.'s catalog. HD 28099 (Hyades 64) was first identified as a solar analog in \citet{Hardorp1978} and has frequently been used as a near-IR telluric standard for Solar System observations. Lewin et al. lists HD 28099 as one of their 17 ``primary'' (most accurate) solar analogs. HD 13545 is listed as a ``rank 1'' star, with differences from the solar spectrum across the 0.8--2.5\um range of $<$10\% in slope, while HD 15942 is a `'rank 2'' star, with slope differences between 10--20\%. These characterizations generally match our experience with these two stars, although we discontinued use of both of them during our 2020 observations in favor of HD 16275.

In general, we find that our stars were reasonably well-suited to their purpose as telluric/solar standards. However, some discrepancies between our standards and the solar spectrum are apparent in Figure \ref{fig:telluricstandards}. The most noticeable discrepancies are mismatches in the strength of \ion{H}{1} lines throughout the spectrum, especially between 1.55--1.70 \micron. Characterization of this mismatch is complicated by uncertainty regarding modeling the strength of the same \ion{H}{1} lines in the A0 stars used to correct the G-type standards. Other line strength mismatches in the wavelength range covered by our spectra appear to be mostly due to either \ion{Mg}{1} or \ion{Al}{1}. However, these line strength mismatches have minimal effects on our corrected spectra of Miranda or the derived band measurements, and the S/N of our TripleSpec data is simply too low for such effects to become apparent.

The other noticeable discrepancies are slope differences between the solar and standard spectra. Once again, any uncertainties in the spectral slopes/colors of the A0 stars will propagate into the final result, but even the G-type standards that apparently deviate most strongly from the solar spectrum (HD 15942 and HD 13545) differ by no more than 10\% between 1.0--2.5 \micron. As we normalized by a line drawn across the width of each individual H$_2$O ice absorption band during the band measurement process anyway, these spectral slopes are unlikely to affect the final results, especially compared to other sources of uncertainty in our observations. 

\pagebreak
\section{Spectra collection\label{adx:collection}}
\begin{figure}[ht!]
\vspace{-10pt}
\centering
\makebox[\textwidth][c]{\includegraphics[width=\textwidth]{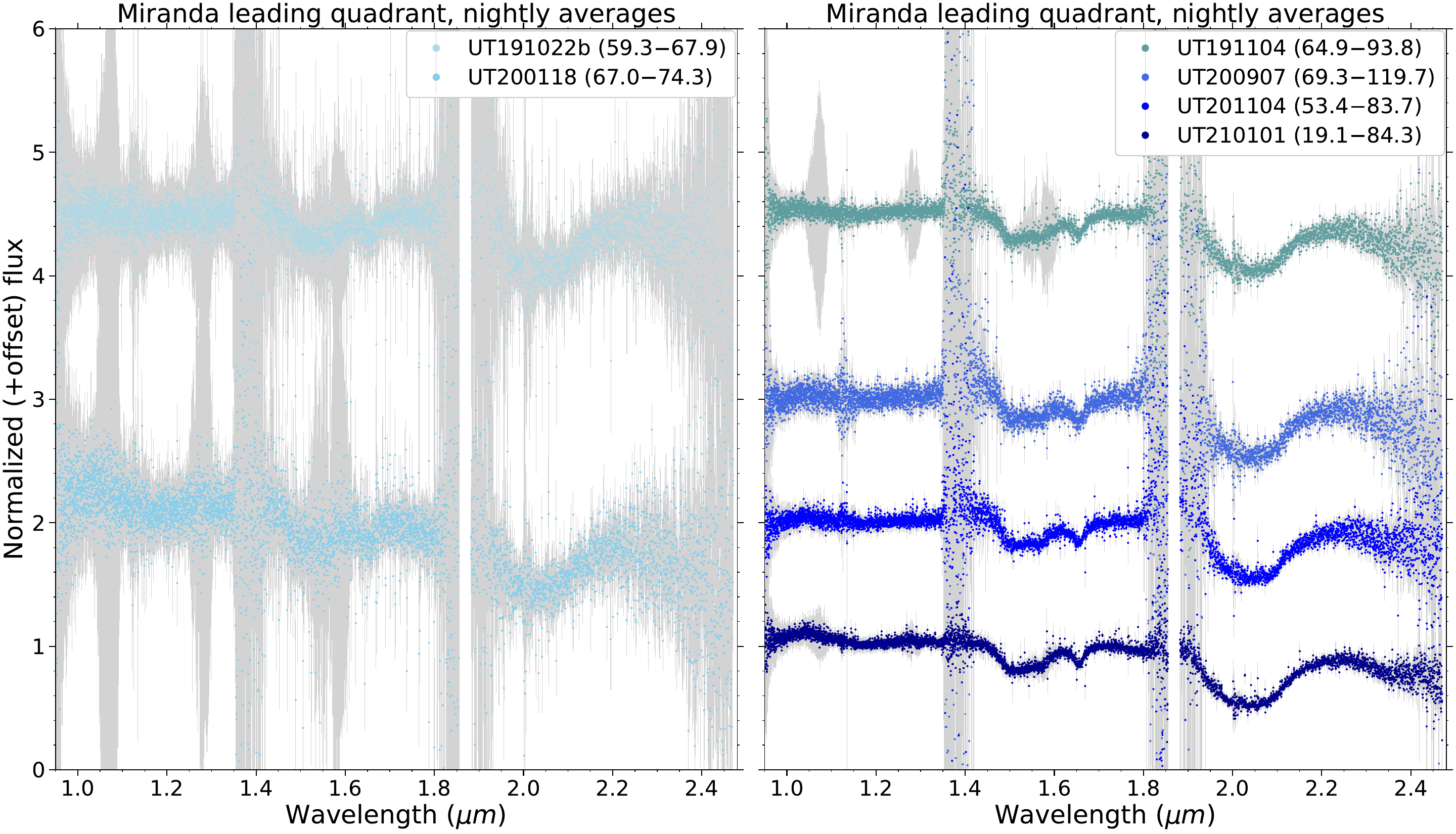}}
\vspace{-15pt}
\caption{\footnotesize Leading quadrant TripleSpec nightly-average spectra of Miranda, at native resolution. All spectra are normalized to unity at 1.72 \micron. Spectra are offset vertically in increments of +3.5 and +1 in the left panel, and +3.5, +2, +1, and +0 in the right panel.}
\vspace{-15pt}
\end{figure}

\begin{figure}[h!]
\vspace{-5pt}
\centering
\makebox[\textwidth][c]{\includegraphics[width=\textwidth]{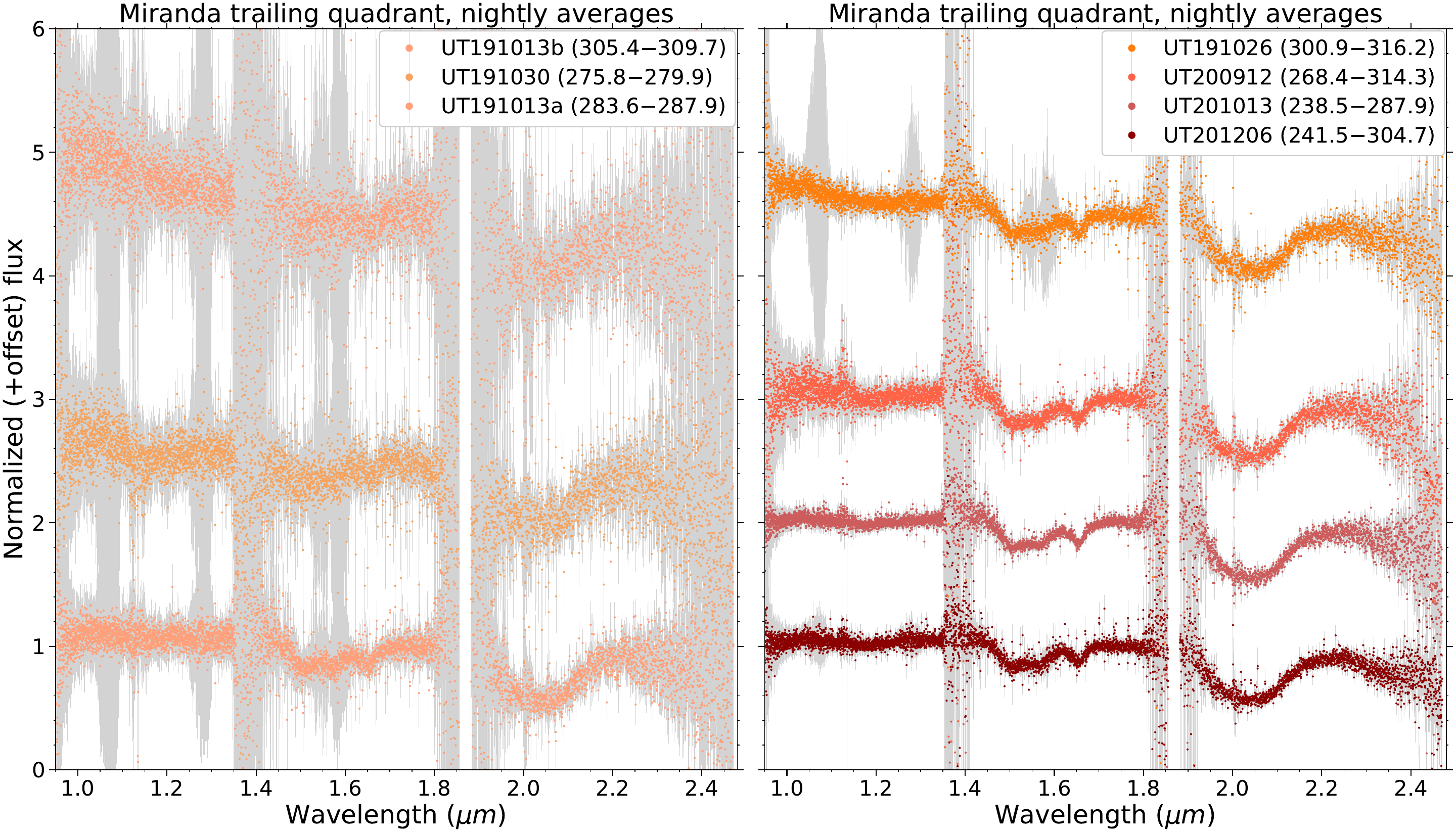}}
\vspace{-15pt}
\caption{\footnotesize Trailing quadrant TripleSpec nightly-average spectra of Miranda. Spectra are offset vertically in increments of +4, +2, and +0 in the left panel, and +3.5, +2, +1, and +0 in the right panel. }
\vspace{-15pt}
\end{figure}

\begin{figure}[ht!]
\centering
\makebox[\textwidth][c]{\includegraphics[width=\textwidth]{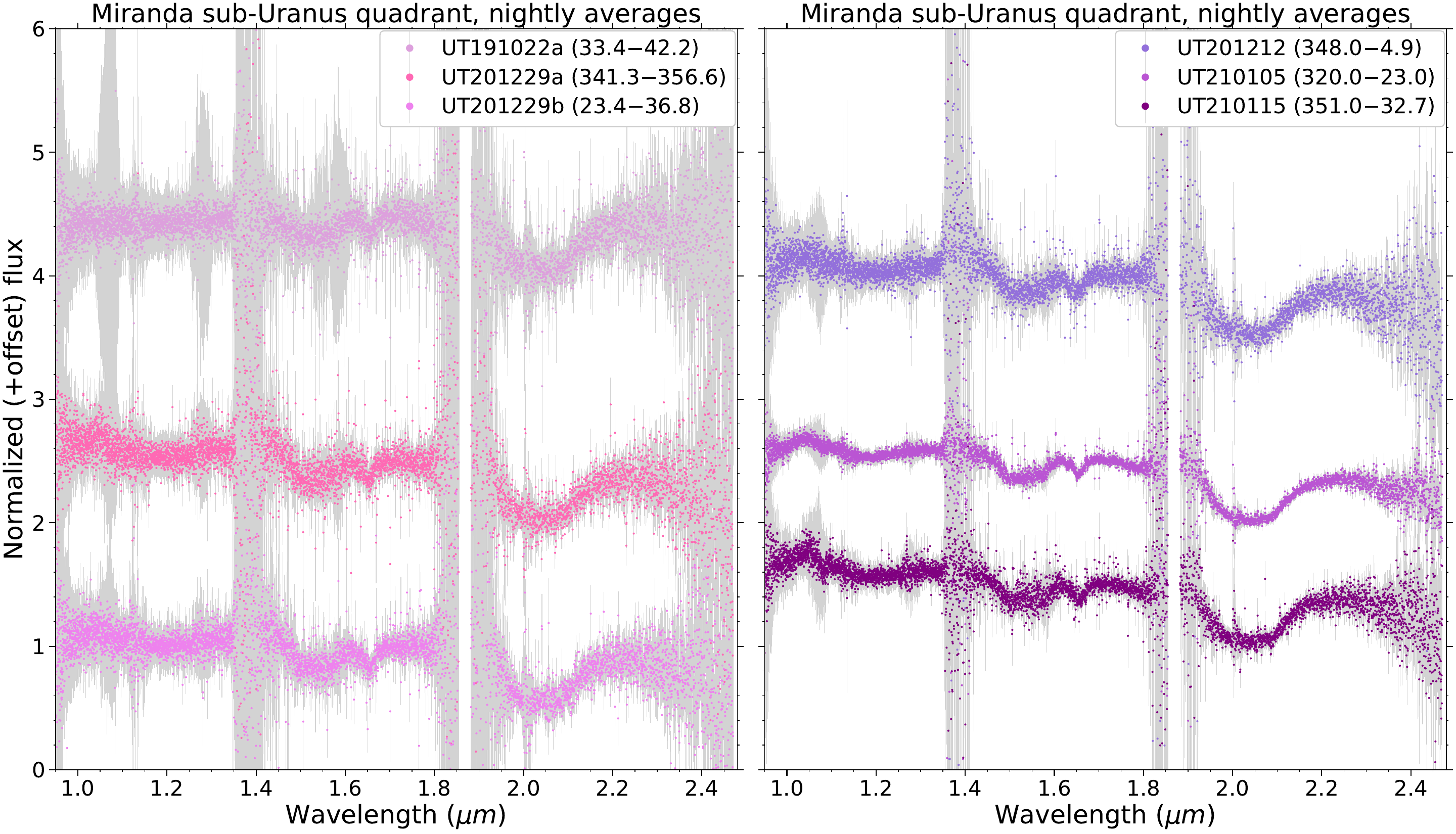}}
\vspace{-15pt}
\caption{\footnotesize Sub-Uranus quadrant TripleSpec nightly-average spectra of Miranda. Spectra are offset vertically in increments of +3.5, +1.5, and +0 in the left panel, and +3, +1.5, and +0.5 in the right panel.}
\vspace{-5pt}
\end{figure}
\nopagebreak
\begin{figure}[ht!]
\centering
\makebox[\textwidth][c]{\includegraphics[width=\textwidth]{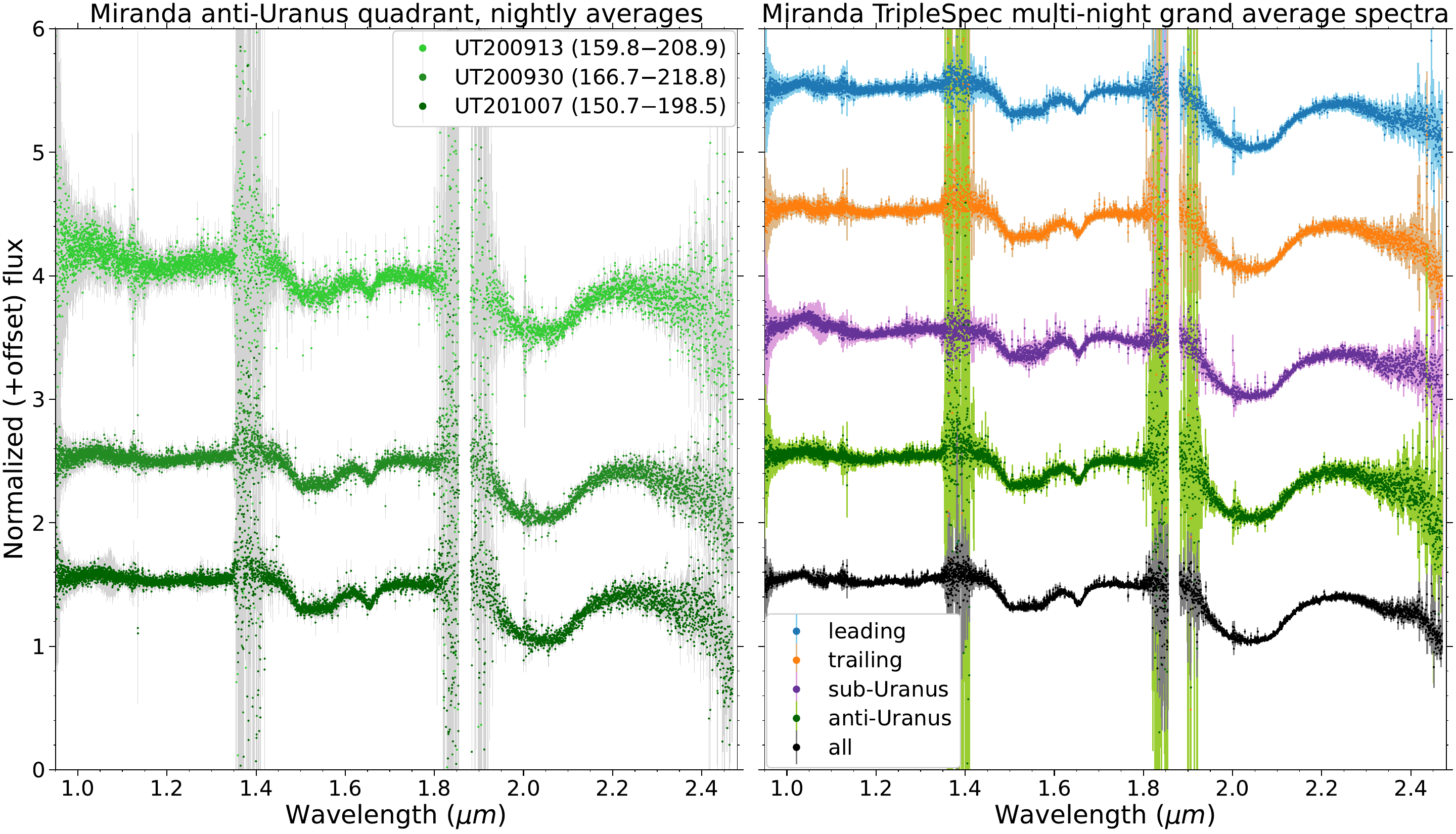}}
\vspace{-15pt}
\caption{\footnotesize Anti-Uranus quadrant nightly-average (left) and grand-average (right) TripleSpec spectra of Miranda. Spectra are offset vertically in increments of +3, +1.5, and +0 in the left panel, and +4.5, +3.5, +2.5, +1.5, and +0.5 in the right panel. }
\vspace{-5pt}
\label{fig:GrandAverages}
\end{figure}
\pagebreak

\newpage
\bibliography{MirBib}{}
\bibliographystyle{aasjournal}


\end{document}